\documentclass[fleqn,usenatbib]{mnras}
\usepackage{fix-cm}
\usepackage{newtxtext,newtxmath}
\usepackage[T1]{fontenc}
\usepackage{bm} 
\DeclareRobustCommand{\VAN}[3]{#2}
\let\VANthebibliography\thebibliography
\def\thebibliography{\DeclareRobustCommand{\VAN}[3]{##3}\VANthebibliography}

\usepackage{graphicx}	
\usepackage{amsmath}	
\usepackage{subcaption} 

\title[]{Conditions for mass-loading: Investigating jet-disk material exchange in strongly magnetised accretion flows} 

\author[S. Praharaj et al.]{Subhrat Praharaj,$^{1, 2}$\thanks{E-mail: s.praharaj@uva.nl}
Matthew T.P. Liska,$^{3}$
Sera Markoff,$^{1,2,5}$
Koushik Chatterjee,$^{4}$
Bart Ripperda,$^{4,6,7,8}$
\and Lorenzo Sironi,$^{9,10}$
Aylecia S. Lattimer$^{1,2}$
\\
$^{1}$Anton Pannekoek Institute, University of Amsterdam, Science Park 904, 1098 XH, Amsterdam, The Netherlands\\
$^{2}$Gravitation Astroparticle Physics Amsterdam (GRAPPA) Institute, University of Amsterdam, Science Park 904, 1098 XH Amsterdam, The Netherlands\\
$^{3}$Center for Relativistic Astrophysics, Georgia Institute of Technology, Howey Physics Bldg, 837 State St NW, Atlanta, GA 30332, USA\\
$^{4}$Canadian Institute for Theoretical Astrophysics, 60 St. George Street, Toronto, Ontario M5S 3H8, Canada\\
$^{5}$ Institute of Astronomy, University of Cambridge, Madingley Road, CB3 0HA, Cambridge, United Kingdom\\
$^{6}$Department of Physics, University of Toronto, 60 St. George St, Toronto, ON M5S 1A7, Canada\\
$^{7}$ David A. Dunlap Department of Astronomy \& Astrophysics, University of Toronto, 50 St. George St,Toronto, ON M5S 3H4, Canada\\
$^{8}$Perimeter Institute for Theoretical Physics, 31 Caroline St. North, Waterloo, ON N2L 2Y5, Canada\\
$^{9}$Department of Astronomy and Columbia Astrophysics Laboratory, Columbia University, New York, NY, 10027, USA\\
$^{10}$Center for Computational Astrophysics, Flatiron Institute, 162 5th Avenue, New York, NY 10010, USA
}

\date{Accepted XXX. Received YYY; in original form ZZZ}

\pubyear{\the\year{}}
\begin{document}
\label{firstpage}
\pagerange{\pageref{firstpage}--\pageref{lastpage}}
\maketitle

\begin{abstract}
Relativistic jets are produced by accreting black holes across a wide range of masses, from stellar-mass X-ray binaries to supermassive active galactic nuclei (AGN). In low-luminosity AGN, these jets are believed to be powered by strongly magnetised accretion flows, a picture supported by Event Horizon Telescope and VLBI observations of M87. The jets often exhibit high-energy flaring emission which, if partly hadronic, implies that baryons must be loaded into the jet and accelerated despite strong magnetic barriers. However, the mechanism behind this remains unclear. We present high-resolution, three-dimensional General Relativistic Magnetohydrodynamic (GRMHD) simulations of a high-spin Magnetically Arrested Disk (MAD), a strongly magnetised accretion regime. Using passive Eulerian tracers, we track the exchange of material between the jet and disk to quantify mixing. We find that magnetic flux eruptions — a hallmark of MAD accretion — expel magnetic flux tubes initially filled with jet material (expected to be pair-dominated in real systems). Through magnetic reconnection during these eruptions, this material is mixed into the disk. Crucially, the same flux tubes reconnect the disk back to the jet, allowing disk material — expected to be primarily hadronic — to infiltrate the jet sheath and boundary. This disk-to-jet injection offers a natural mass-loading mechanism that does not require the jet to be globally baryon-dominated. The resulting ejection-injection cycle may instead drive episodic hadronic acceleration, providing a viable pathway for high-energy emission in relativistic jets.
\end{abstract}



\section{Introduction} \label{sec:intro}

Relativistic jets represent one of the most ubiquitous phenomena across astrophysical objects and scales, first unwittingly discovered in the nearby galaxy M87 over a century ago \citep{Curtis1918} and later found to be polarised (observed at optical wavelengths;  \citealt{Baade1956}), a key step towards establishing them as structures originating due to ordered  magnetic fields. The recognition of quasars as distant, luminous galaxies \citep[e.g.,][]{Schmidt1963} and the theoretical framework for accretion-powered systems \citep[e.g.,][]{Lynden1971} established jets as fundamental components of black hole physics. Much later, Very Large Baseline Interferometry (VLBI) observations confirmed relativistic bulk motion through proper superluminal motions \citep[e.g.,][]{Reid, 1989ApJ...342..128B, Biretta}. Over subsequent decades, jets have been identified across a remarkable range of systems harboring or forming black holes: including stellar-mass X-ray binaries (XRBs) where the jets exhibit superluminal motion \citep[e.g.,][]{Mirabel1994, Fender2004} to the most luminous active galactic nuclei  \citep[AGN; e.g.,][]{Bridle1984, Begelman1984}, and more recently in neutron star mergers \citep[e.g.,][]{Mooley2018} and tidal disruption events \citep[e.g.][]{Bloom2011, cendes2024ubiquitouslateradioemission}. 

Within this larger topic, there is a long-standing question regarding the origin of high-energy accelerated, hadronic particles called cosmic rays, that requires a better understanding of jet composition and potential sites for acceleration. Current instruments cannot directly measure the jet composition, and broadband Spectral Energy Distribution (SED) modeling alone remains highly degenerate between leptonic and hadronic scenarios \citep[e.g.][]{Boettcher2013, Cerruti2015, Madejski2016, Romero_2017, Blandford_2019}. To make progress on this question, it is important to better constrain how and where  hadronic material manages to enter the jets, and is the loading continuous, or does it occur in episodic events? Some progress has been made on measuring composition via source-specific observations where a few black hole jets show direct signatures of baryonic material. For example, the black hole XRB 4U 1630‑47 displays Fe XXVI lines at $0.66c$ \citep{Trigo_2013, Trueba_2019}; another black hole XRB SS~433 also displays Doppler-shifted Balmer, HeI, Fe XXV/XXVI consistent with baryonic jets moving at $\approx 0.26c$ \citep{Margon_79,Abell_79,Marshall_2002}, where $c$ is the speed of light. However linking these results to acceleration sites and mechanisms remains challenging.

Historically, the prime candidate site proposed for extreme particle acceleration in AGN jets is a putative recollimation shock that occurs around the Bondi radius ($R_{\rm Bondi} = 2GM/c_s^2$; where $c_s$ is the speed of sound, and, $M$ is the mass of the black hole). The Bondi radius (hereafter $R_{\rm B}$) is the gravitational capture radius of the black hole's ambient medium, where the inflow transitions from quasi-spherical to disk-dominated, depending on its residual angular momentum \citep[e.g.][]{Homan_2015, Zech_2021, Nalewajko_2014}. In the nearby low-luminosity AGN M87, the prominent HST-1 knot at $\sim 60$ pc ($\sim 10^5 r_g \sim R_B$) has long been considered to be associated with a reconfinement shock \citep{Stawarz_2006}. HST-1 is observed to be a site of particle re-acceleration, with a major multiwavelength outburst peaking in 2005 and exhibiting correlated variability from radio through X-rays \citep{Harris2009, Stawarz_2006}. If protons are entrained into these acceleration zones and sufficiently accelerated, they can potentially be a source of neutrinos. The association of AGN jets with high-energy neutrinos is supported by candidate neutrino events from sources such as TXS~0506+056 \citep{Aartsen2018multimessenger}, while recent work has identified other radio galaxies, including 3C~403, as candidate neutrino-emitting sources \citep{bruni20263c403candidateneutrinoemittingradio}. More broadly, AGN contributions to the extragalactic $\gamma$-ray background \citep{Ackermann2016cosmic} and the diffuse neutrino flux \citep{Aartsen2017lowenergy} depend sensitively on the fraction of jets that carry sufficient baryons for hadronic acceleration \citep{Stecker2013cosmic, Sahakyan2018}, and where that acceleration occurs.

 Recent observations in the very‑high‑energy (VHE) $\gamma$‑rays ($>100$ TeV) with the Large High Altitude Air Shower Observatory (LHAASO; \citealt{Cao_2025}) and the IceCube neutrino detector \citep{Abbasi2022time} have begun to challenge the canonical picture of predominantly leptonic (pair-dominated) jets. LHAASO, and now also H.E.S.S., have somewhat unexpectedly detected VHE $\gamma$‑rays from several XRBs (SS433, GRS 1915+105, V4641 Sgr, MAXI J1820+070; \citealt{Cao_2025}), sources that are fed by Roche-lobe overflow from a companion star rather than by a quasi-spherical ambient medium, so that a recollimation shock near $R_{\rm B}$ is not an obvious acceleration site. Several low-luminosity AGN including M87, whose jet is not pointed towards us, are also VHE $\gamma$‑ray bright \citep{Acciari_2009}, showing that VHE emission does not require a blazar-like jet closely aligned with our line of sight. Interestingly, the VHE $\gamma$-ray flare detected during the 2018 EHT campaign on M87 \citep{EHTMWL2024} was not accompanied by emission from the extended jet. 
 
 IceCube's detection of events from a few blazars such as TXS~0506+056 \citep{Aartsen2018multimessenger}; and association of high‑energy neutrinos with weakly jetted/choked AGN (NGC 1068, NGC 4151; \citealt{abbasi2024searchneutrinoemissionhard}) further indicate hadronic acceleration mechanisms that likely require significant baryon loading in the jet launching region, or in X-ray emitting coronal regions close to the black hole due to the lack of corresponding $\gamma$-ray detections. Consequently there is a growing consensus that hadronic acceleration can favourably occur in close‑in regions of AGN that are opaque to GeV–TeV $\gamma$-rays but allow neutrinos and X-rays to escape—such as AGN coronae or their inner jets \citep[e.g.][]{Murase2016hidden, Murase2020AGNcores, Inoue2019coronae, Eichmann_2022, Fiorillo24a, Fiorillo24b}.

 Furthermore, understanding mass-loading may help reconcile the unrealistically high power requirements found in several 
 hadronic models for  simultaneous $\gamma$‑ray and neutrino emission, such as those invoked for the blazar TXS 0506+056 flares (\citealt{IceCube2018neutrino, Ansoldi2018magic}, see also, e.g., \citealt{Petropoulou2015hadronic, Lucchini_2023}. Finally,  minute‑scale $\gamma$-ray variability in blazars like 3C 279 \citep[e.g.][]{Shukla2020} and rapid morphological changes in VLBI imaging \citep{Hada2016, Kim2018} suggest dynamic material exchange at jet–disk boundaries on short timescales (although they can also be framed through internal shock-turbulence, magnetic reconnection, or geometric effects). Together, all of these results provide the motivation to explore the conditions for mass mixing and loading within the highly magnetised inner regions of accreting black holes. 

 Ideal general relativistic magnetohydrodynamics (GRMHD) offers a framework to study the global accretion and outflow dynamics, where typical time and length scales can be expressed in units of gravitational radius $r_g = GM/c^2$ and dynamical time $t_g = r_g/c$, where $M$ is black hole mass, $G$ is the gravitational constant, and $c$ speed of light. In regimes when  radiative losses do not affect the dynamics, GRMHD simulations allow us to compare the essential accretion and jet-launching physics across all black holes masses \citep[e.g.,][]{Gammie2003, Jiang2014, Porth2019event}. The current generation of GRMHD simulations has robustly produced low-density, Poynting-flux dominated outflows via the Blandford-Znajek mechanism \citep{BlandfordZnajek1977, McKinney2006, Tchekhovskoy2011}, and also reveal that jets exist in dynamic equilibrium with their surrounding accretion flows, with steep density ($\rho$) gradients ($\rho_{\rm disk}/\rho_{\rm jet} \sim 10^4$) and velocity shear at jet-disk boundaries. These conditions can drive mixing in the inner jets through magnetohydrodynamic instabilities \citep[e.g.,][]{Mizuno2012, Mbarek2019, figueiredo2024compositionasymmetricshearedrelativisticmagnetic}. Kelvin-Helmholtz instabilities can develop in shear layers \citep[e.g.,][]{Chatterjee_2019J, Sironi_2021, Chow2023}, while Rayleigh-Taylor instabilities can grow at density interfaces during pressure-driven acceleration \citep{Stehle2001, Mbarek2022, Ripperda2022}. Furthermore, magnetic reconnection in equatorial current sheets enables cross-field transport of material through plasmoid formation and ejection \citep{Ripperda2020, Ripperda2022, Hakobyan2023, Hakobyan2025, chow2026baryoncontentmagneticallyarrested}, potentially injecting disk material directly into the jet, or jet material into the disk.

Despite these advances, quantifying material exchange between jets and disks remains challenging for GRMHD simulations due to insufficient resolution to resolve non-local mixing at jet-disk boundaries in 3D \citep{salas2024resolutionanalysismagneticallyarrested}, and the inability to distinguish between physically initialised disk material and numerically injected "floor" material required for computational stability in low-density regions \citep{Ryan2017, Liska2022, Chatterjee_2019J}. This last limitation is particularly critical because floors become dynamically important in at least the base of the Poynting-flux dominated jets.

Here we explore mass-loading and mixing in a high-resolution 3D GRMHD simulation where we have added density tracers to distinguish disk material from floored jet material, following the approach of \citet{chow2026baryoncontentmagneticallyarrested}. We choose a prograde, rapidly spinning black hole surrounded by a magnetically arrested disk (MAD; \citealt{Narayan2003, McKinney2007, Tchekhovskoy2011}). MADs launch powerful jets, which makes them well suited to study jet composition and possible mechanisms of mixing. 
MADs are thought to be relevant for systems with powerful jets: in M87, the EHT resolves a dynamically strong magnetic field with ordered azimuthal and helical structure in 1.3\,mm polarised images, favouring MAD models \citep{eht2021, EHT2021magnetic}. Importantly, such magnetically dominated flows often experience flux eruptions—quasi-periodic events where accumulated magnetic field violently reorganises near the event horizon \citep{McKinney2007, Tchekhovskoy2011, Ripperda2022}. These eruptions push magnetic flux tubes out into the disk, making them a natural candidate for episodic exchange of material between the disk and the jet. The ability to trace disk and jet material separately enables us to begin addressing critical questions such as: What fraction of disk baryons manage to enter the jet, and how does this fraction vary with different accretion states? Can we find evidence in our simulations for mechanisms that could potentially energise baryons in the base of the jets, as invoked in semi-analytical models of hadronic jets \citep[e.g.,][]{Kantzas2021, Lucchini_2023}?

This paper is organised as follows. Section \ref{sec:methods} describes our numerical approach, and the eulerian tracer algorithm (with validation tests demonstrating tracer conservation and mixing accuracy in the appendix). Section \ref{sec:results} presents our systematic analysis of disk-jet mixing driven by flux eruption cycles in a MAD simulation. We discuss possible links between our simulation and observed phenomena from some astrophysical sources in Section \ref{sec:discussion}, before summarizing our conclusions and future directions in Section \ref{sec:conclusions}.

\section{Numerical setup}\label{sec:methods}

We perform our simulation using \texttt{H-AMR} \citep{Liska2022}, which evolves the ideal GRMHD equations on a fixed Kerr spacetime. The code employs the Harten-Lax-van Leer-Riemann (HLLR) solver \citep{Harten1983} for flux calculations, a staggered grid configuration for magnetic fields \citep{Gardiner2005}, and the piecewise parabolic method (PPM; \citealt{Colella1984}) for second-order spatial reconstruction. To avoid polar singularities inherent in standard spherical coordinates, we implement a 90° coordinate rotation that aligns the black hole spin axis with the grid equatorial plane. The simulation utilises adaptive mesh refinement (AMR; see appendix \ref{sec:amrapp}) coupled with local adaptive timestepping (LAT) to achieve high resolution in dynamically important regions while maintaining computational efficiency.

We adopt geometrised Lorentz–Heaviside units where $G = M = c = 1$ and a factor $\sqrt{4\pi}$ is absorbed into $B$ ($B$ being the magnetic field strength), setting characteristic length and time scales to the gravitational radius $r_g = GM/c^2$ and light-crossing time $t_g = r_g/c$, respectively. Our computational domain is set on a spherical grid that spans $r \in [0.7r_H, 10^4 r_g]$, $\theta \in [0, \pi]$, and $\phi \in [0, 2\pi]$, where $r_H = (1 + \sqrt{1-a^2})r_g$ is the event horizon radius for dimensionless spin parameter $a$. In our studies, we consider only the high spin case of $a=0.9375$. The initial base grid employs logarithmic radial spacing with $N_r \times N_\theta \times N_\phi = 512 \times 256 \times 256$ cells. This ensures that the magnetorotational instability (MRI) driving the initial accretion is adequately resolved \citep{Porth2019event}. Beyond the base grid, we employ AMR: blocks are refined and de-refined dynamically during the run according to physics-driven criteria that target the jet and the jet sheath (based on the magnetisation (range 0.1-1) and the Bernoulli parameter(>1.02)) and regions of strong magnetic field gradients (>50\% change) (see Appendix~\ref{sec:amrapp} for exact numerical constraints). We allow up to 4 levels of refinement above the base grid, with all criteria triggering the same maximum level, and do not refine cells below a minimum size of $\Delta x = 0.002\,r_g$. With derefinement techniques within the constraints defined we record maximum grid resolution of $2688 \times 1280 \times 1152$ cells in the jet and its boundary.

The initial conditions consist of a Fishbone-Moncrief torus \citep{Fishbone1976} with inner edge at $r_{\rm in} = 20r_g$ and pressure maximum at $R_{\rm max} = 41r_g$. The torus is threaded by an initial magnetic field configuration designed to produce a MAD state ($A_\phi \propto \max \left[\rho / \rho_{\max }\left(r / r_{\mathrm{in}}\right)^3 \sin ^3 \theta \exp (-r / 400)-0.2,0\right]$, where $\rho_\mathrm{0}$ is the disk midplane density at t=0 (set for a corresponding plasma $\beta$; $2p_g/b^2$ where $p_g$ is the gas pressure; of 100) and $\rho_\mathrm{max} = 1$ in code units \citep{Narayan2003, Igumenshchev2003, Porth2019event}. We set the adiabatic index to a constant value of 5/3. We implement outflow boundary conditions at the inner and outer radial boundaries, transmissive conditions at the polar boundaries of the rotated grid (which, owing to the $90^\circ$ rotation, lie in the black hole equatorial plane rather than along the spin axis; see Appendix~\ref{flipGR}), and periodic boundary conditions in the azimuthal direction of the rotated grid. We describe in the following our tracer implementation, for a more detailed description please see Appendix \ref{sec:inittracers}.

\subsection{Tracer implementation}\label{subsec:tracer}
To investigate mass loading and mixing processes in our black hole jet-disk system, we implement a tracer scheme that tracks the distinct origins of plasma throughout the simulation domain \citep[as in][]{chow2026baryoncontentmagneticallyarrested}. We employ two complementary density tracer species that follow the mass fraction of disk-origin plasma (assumed to be fully ionised, charge-neutral electron-ion plasma) and jet-origin plasma (assumedly pair-dominated), typically injected as density floors.

\subsubsection{Tracking of jet and disk material}
During evolution, maintaining numerical stability in highly magnetised regions requires injection of density floors when $\rho$ falls below a critical threshold, or the magnetization, $\sigma = b^2/\rho$, becomes too high \citep{Gammie2003}. These floor injections occur predominantly in the magnetically-dominated jet base along the Black Hole spin axis interior to the stagnation surfaces, where charge separation and strong electric fields are expected to drive pair production \citep{Beskin1992, Levinson2000, Hirotani2016}. Theoretical works have demonstrated that diverging mass flow near stagnation surfaces leads to charge separation and subsequent pair cascade formation \citep{Parfrey2019, Crinquand2020, Crinquand2021, medvedev2026electronpositroncascademagnetospheressupermassive}. The injected floor density thus seems to serve as a reasonable proxy for pair plasma created through these microphysical processes \citep{McKinney2012}, which are not captured in fluid GRMHD simulations. Numerical tracers have been used to follow the origin of injected floor plasma in GRMHD simulations \citep[e.g.,][]{Wong2021tracers, chow2026baryoncontentmagneticallyarrested}.

The density floor criterion employed is the standard \citep{Porth2019event}:
\begin{equation}
\rho_{\mathrm{fl}}=\max\left[\frac{b^2}{25\rho_0 c^2}, 10^{-7} \rho_0 r^{-2}, 10^{-20}\rho_0\right],
\end{equation}
where $b^2$ is the squared magnitude of the magnetic field in the fluid's comoving-frame, while $\rho$ is the rest-mass density. The first term prevents the magnetization $\sigma = b^2/(\rho c^2)$ from exceeding 25, the second term provides a power-law floor, and the third term sets an absolute minimum \citep{Liska2018,Liska2022}. 

\subsubsection{Tracer evolution equations}
Following the approach of \citet{Keppens2014}, we model the tracers as advected scalar quantities. We define the conserved tracer density for each species as:
\begin{equation}
D_{\mathrm{tr}} = \rho \theta_{\mathrm{tr}}
\end{equation}
which evolves according to the continuity equation:
\begin{equation}
\frac{\partial (\sqrt{-g}D_{\mathrm{tr}})}{\partial t}+\nabla \cdot\left(\sqrt{-g}\mathbf{v} D_{\mathrm{tr}}\right)=0,
\end{equation}
where $\nabla = \frac{\partial}{\partial x^i}$ and $g$ is the metric determinant in non-rotated frame, ensuring that $D_{\mathrm{tr}}$ is advected with the flow velocity $\mathbf{v}$ in the same manner as the bulk density $\rho$. The primitive tracer fraction is recovered as $\theta_{\mathrm{tr}} = D_{\mathrm{tr}}/\rho$, which satisfies:
\begin{equation}
\frac{\partial (\sqrt{-g}\theta_{\mathrm{tr}})}{\partial t}+\mathbf{v} \cdot \nabla (\sqrt{-g}\theta_{\mathrm{tr}})=0
\end{equation}

Moreover, in our tilted spherical Kerr-Schild coordinate system, the tracer advection equation retains its covariant form:
\begin{equation}
\frac{\partial}{\partial t'}(\sqrt{-g'}\rho u^{t'} \theta_{\mathrm{tr}}) + \frac{\partial}{\partial x'^i}(\sqrt{-g'}\rho u^i \theta_{\mathrm{tr}}) = 0
\end{equation}
where primed quantities refer to the rotated coordinate system, $g'$ is the metric determinant, and the four-velocity components transform according to:
\begin{equation}
u'^{\mu'} = \frac{\partial x'^{\mu'}}{\partial x^\mu} u^\mu
\end{equation}

\subsubsection{Tracer update during floor injection}

The highest magnetization regions primarily occur near the black hole.  When density floors are triggered, the tracer fractions must be updated to account for the injected mass. If the pre-floor density is $\rho_{\mathrm{pre}}$ and the post-floor density is $\rho_{\mathrm{post}}$ (same subscripts for $\theta$), the tracer fractions are updated as

For the disk tracer (species 0):
\begin{equation}
\theta_{\mathrm{disk}}^{\mathrm{post}} = \frac{\rho_{\mathrm{pre}} \theta_{\mathrm{disk}}^{\mathrm{pre}}}{\rho_{\mathrm{post}}}
\end{equation}

For the floor tracer (species 1):
\begin{equation}
\theta_{\mathrm{floor}}^{\mathrm{post}} = \frac{\rho_{\mathrm{pre}} \theta_{\mathrm{floor}}^{\mathrm{pre}} + (\rho_{\mathrm{post}} - \rho_{\mathrm{pre}})}{\rho_{\mathrm{post}}}
\end{equation}

In words, a floor injection leaves the mass density of disk-origin material, $\rho_{\rm disk}\equiv\rho\,\theta_{\mathrm{disk}}$, unchanged, so the disk fraction is merely diluted by the factor $\rho_{\mathrm{pre}}/\rho_{\mathrm{post}}$, while the mass density of jet-origin (floor) material, $\rho_{\rm jet}\equiv\rho\,\theta_{\mathrm{floor}}$, grows by exactly the injected amount, $\rho_{\mathrm{post}}-\rho_{\mathrm{pre}}$. If $\theta_{\mathrm{disk}}+\theta_{\mathrm{floor}}=1$ before the injection, this therefore still holds afterwards. This prescription thus ensures mass conservation (see Fig.~\ref{fig3}) while attributing all injected mass to the floor tracer species. Throughout this paper, the subscript `jet' on tracer quantities (e.g. $\rho_{\rm jet}$, $\dot{M}_{\rm jet}$) refers to this jet-origin (floor) material, and `disk' to disk-origin material. The tracers thus evolve dynamically with the flow.

The described implementation allows us to quantify mass exchange between the disk and jet components, track the penetration of disk material into the jet-disk boundary (the jet sheath) and the main jet itself, as well as jet material into the disk. For a prograde MAD around a high-spin black hole ($a=0.9375$), we quantify mixing into the jet as a function of time relative to flux eruption phases, as a function of height in the jet and across the co-existing jet, sheath (jet disk boundary layer) and disk components.

\section{Results}\label{sec:results}

\subsection{Temporal Phase Classification}

\begin{figure}
\centering
\includegraphics[width=\columnwidth]{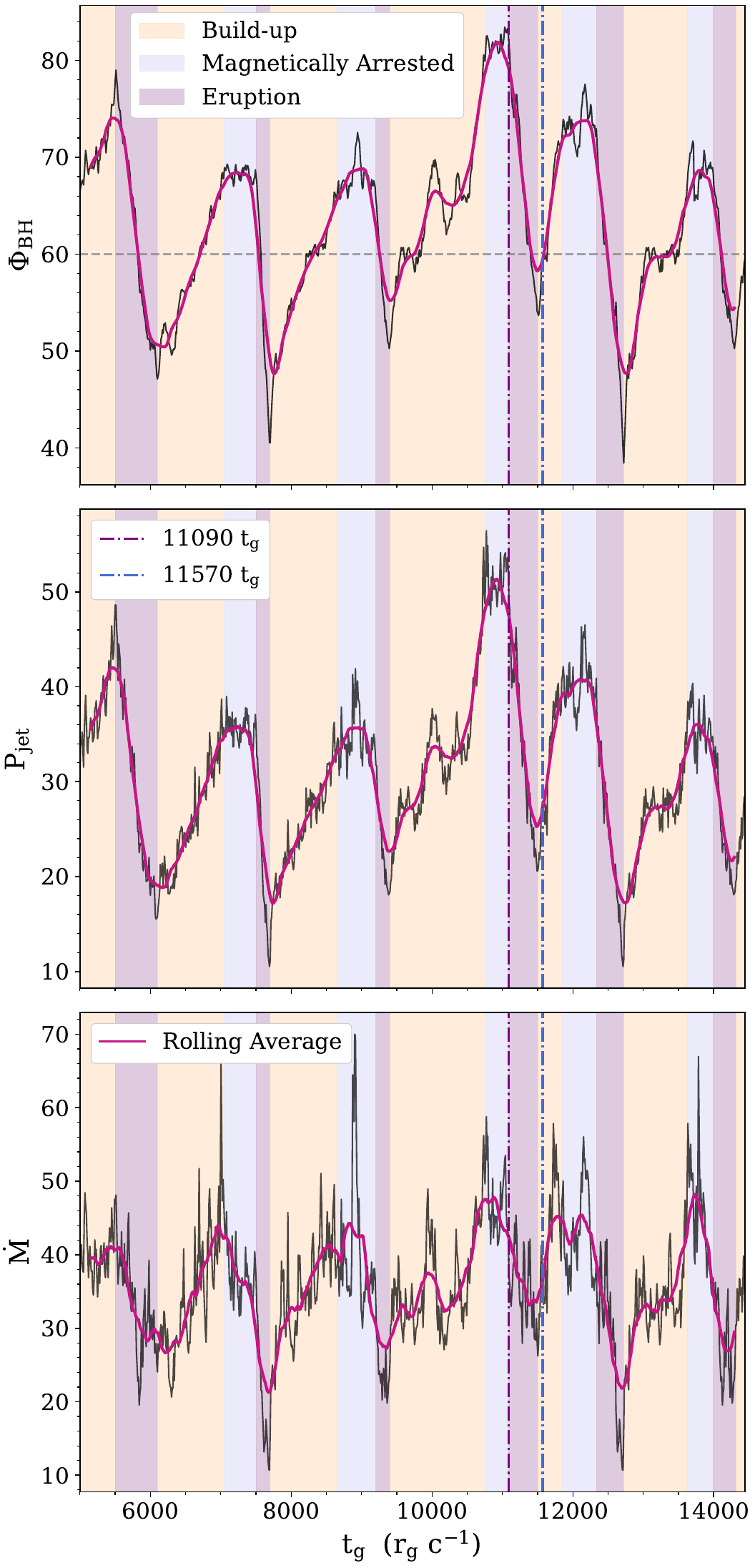}
\caption{
\textbf{Time evolution of magnetic flux, jet power, and mass accretion rate, showing flux eruption cycles in a MAD.}
\textbf{Buildup phases} (peach) where magnetic flux accumulates near the horizon as field lines are advected inward by accretion ($\mathrm{d}\Phi_{\rm BH}/\mathrm{d}t > 0$);
\textbf{MAD phases} (light blue) where flux reaches saturation and exhibits high variability as further flux accumulation is inhibited ($\Phi_{\rm BH} > 60$ with elevated variance);
and \textbf{Eruption phases} (lavender) where accumulated flux is expelled through magnetic reconnection ($\mathrm{d}\Phi_{\rm BH}/\mathrm{d}t < 0$).
\textbf{Top panel}: Dimensionless magnetic flux threading the black hole horizon $\Phi_{\rm BH}$, showing a characteristic saw-tooth pattern with gradual buildup followed by sudden drops during eruptions. The dashed gray line marks the MAD saturation level measured in GRMHD simulations \citep{Tchekhovskoy2011}.
\textbf{Middle panel}: Jet power revealing strong modulation with eruption cycles. 
\textbf{Bottom panel}: Mass accretion rate $\dot{M}$ (horizon-averaged), which also correlates with $\Phi_{\rm BH}$: accretion rates gradually increases during buildup states, saturates around a fixed value during the MAD state, then drops during eruptions.
}
\label{fig:phase_classification}
\end{figure}

Our fiducial GRMHD simulation of a high‑spin ($a=0.9375$), prograde MAD evolves over $\sim 15000\,t_g$, revealing three distinct dynamical phases: build-up to a MAD, the phase where the magnetic pressure in the inner region is saturated, i.e. MAD, and the eruption phase where magnetic flux is ejected through reconnection \citep{Ripperda2022, najafi_2024}. These phases are identified through the evolution of the magnetic flux across the event horizon $\Phi_{\rm BH}$, jet power $P_{\rm jet} = \dot{M}c^2 - \dot{E}$ (where $\dot{M}c^2$ is the rest-mass energy flux into the black hole and $\dot{E}$ is the total energy flux through the event horizon, both taken positive for inflow into the black hole \citep[e.g.,][]{Tchekhovskoy2011}, so that $P_{\rm jet}>0$ is the net power extracted from the black hole, which in a MAD is dominated by the electromagnetic Blandford–Znajek contribution), and mass accretion rate $\dot{M}$ at the black hole event horizon (Fig.~\ref{fig:phase_classification}). The characteristic saw‑tooth pattern of $\Phi_{\rm BH}$, gradual accumulation followed by sudden drops, marks the MAD eruption cycle, with periods of $\sim 2000\,t_g$.

\subsubsection{Buildup Phase}

The buildup phase is the longest and most variable, lasting $723\pm422\,t_g$ (a coefficient of variation, CV, i.e. standard deviation divided by mean, of $58\%$). Magnetic flux gradually accumulates from post‑eruption lows to near‑MAD levels, while jet power rises by a factor $\sim2$ from eruption minima to saturation. The disk–jet material accretion ratio ($\dot{M}_{disk}/\dot{M}_{jet}$, with the numerator and denominator being the accretion rates of disk and jet tracer masses respectively) recovers from $1.4$ in the eruption phase we see later, to $2.9$ – a greater than $100\%$ increase – signalling the re‑establishment of a disk‑dominated accretion flow. This phase also exhibits the largest fluctuations in jet power ($72\%$ CV).

\subsubsection{MAD Phase}

The MAD phase is the most dynamically settled state, lasting $427\pm85\,t_g$ ($\sim20\%$ variation). Here the magnetic flux saturates at a value $\sim20\%$ above the eruption threshold, driving the strongest Blandford–Znajek jet power (a factor $\sim2.2$ higher than during the subsequent eruptions) and the highest mass accretion rate ($\sim1.5\times$ the eruption value). 

\subsubsection{Flux Eruption Phase} \label{sec:eruptions}

During eruptions, the magnetic flux drops by $\sim15\%$ on a timescale of $343\pm151\,t_g$ (high cycle‑to‑cycle variation, $44\%$). This is consistent with \citet{Ripperda2022}, who show that the duration of this phase is set predominantly by the universal fast magnetic reconnection rate, which is independent of the Lundquist number in the plasmoid-dominated regime. The jet power decreases by $\sim55\%$ relative to the MAD phase, and the accretion rate falls by $\sim35\%$. Most strikingly, the disk–jet material accretion ratio drops to $1.4$ ($58\%$ disk, $42\%$ floor) – a $30\%$ reduction from MAD levels. This demonstrates that eruptions preferentially disrupt the disk component while enhancing wind/outflow contributions, consistent with flux expulsion via magnetic reconnection \citep{Ripperda2020, Ripperda2022, Chatterjee_2022}.

The fourth eruption cycle (the strongest in our simulation) is chosen for further analysis because it exhibits the highest pre‑eruption flux and the most dramatic post‑eruption reorganisation, offering the clearest signature of baryon loading into the jet funnel. We restrict the mass-flux and $dM/d\ln\sigma$ analysis below to this eruption and have not analysed the other cycles in the same detail.

\subsection{Quantifying eruption-induced mass redistribution}
\label{sec:dmdsigma}

\begin{figure*}
  \centering
  \includegraphics[width=0.32\linewidth]{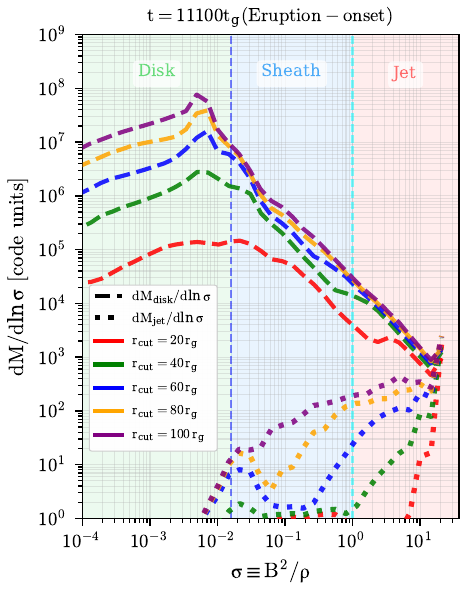}
  \includegraphics[width=0.32\linewidth]{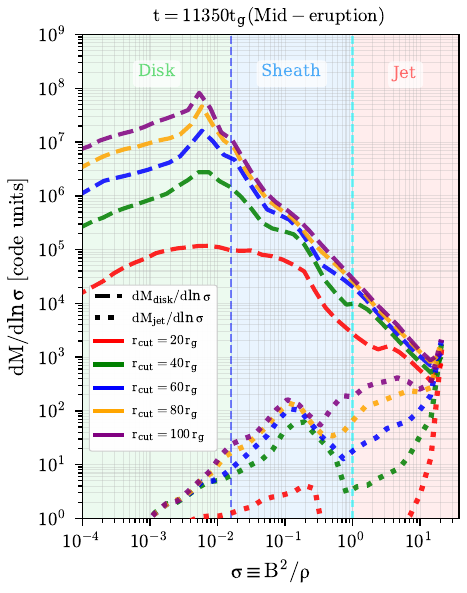}
  \includegraphics[width=0.32\linewidth]{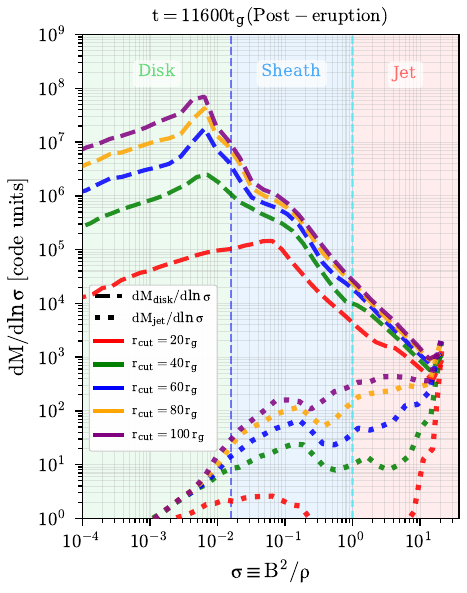}
  \caption{Magnetisation-resolved mass distributions
    $dM/d\ln\sigma$ at $t=11090$, $11340$, and $11590\,t_g$ for multiple radial
    cuts $r_{\rm cut}=20$--$100\,r_{\rm g}$.  Dashed curves show
    disk-origin material (tracer~1) and dotted curves show jet/floor
    material (tracer~2).  The eruption window corresponds to the
    transition from the onset ($11090\,t_g$) to the mid-eruption ($11340\,t_g$)
    and post-eruption ($11590\,t_g$) snapshots.}
  \label{fig:dMdlnsigma_snaps}
\end{figure*}

To quantify how mass is distributed over magnetisation, we compute $dM/d\ln\sigma$: the tracer-weighted mass within a radius $r_{\rm cut}$ per logarithmic interval of the magnetisation $\sigma$ (Appendix~\ref{appendixdM}). This allows us to track how disk and jet plasma are redistributed in magnetisation and radius as the MAD undergoes a flux eruption: it shows how much disk material reaches the magnetisations of the sheath and the jet, and conversely how much jet material ends up at disk-like magnetisations. In Fig.~\ref{fig:dMdlnsigma_snaps} we show $dM/d\ln\sigma$ at times 11090, 11340, and 11590 $t_g$ for multiple radial cuts $r_{\rm cut}=20$--$100\,r_{\rm g}$, with three shaded bands in $\sigma$: a ``disk'' band at $\sigma<\sigma_{\rm sh}$, a ``sheath'' band $\sigma_{\rm sh}<\sigma<1$, and a ``jet'' band at $\sigma>1$. The detailed calculations, definitions and considerations for this analysis are laid out in Appendix \ref{appendixdM}.

At the onset of eruption (11090 $t_g$) the disk tracer is concentrated in the disk physical region (green band on the panels), with only a modest high-$\sigma$ tail in the sheath and jet regions. The jet/floor tracer, by contrast, contributes relatively little in the disk band but shows a clear peak in the jet band. There is some separation between $r_{\rm cut}=60$--$100\,r_{\rm g}$ for jet material in the sheath marking the remnant of an earlier eruption trapped in the sheath, structures relating to which possibly cause the separation between $r_{\rm cut}=20$--$40\,r_{\rm g}$ for material originating in the disk. 

In the middle of the eruption (11340 $t_g$) the $dM/d\ln\sigma$ curves change qualitatively in the sheath band.  For both tracers, the curves at $r_{\rm cut}=40$--$100\,r_{\rm g}$ become nearly coincident in the disk and sheath bands, indicating that material in this radial range is all contained in the inner $40 r_{\rm g}$.  As the eruption develops, the $r_{\rm cut}=20\,r_{\rm g}$ curve for the disk tracer remains at roughly the same level, while the curves for $r_{\rm cut}\ge40\,r_{\rm g}$ move upward in the sheath band.  This behaviour – near overlap of the 40–100 $r_{\rm g}$ curves, followed by a co-rise of the cuts towards the end of the eruption (11590 $t_g$) – shows that disk material between 20 and 40 $r_g$ is shifted to sheath-like magnetisations during the eruption. On its own, $dM/d\ln\sigma$ does not show whether this material enters the jet; we address this with the mass fluxes across the jet boundary below. 

The jet/floor tracer also responds to the eruption.  Over the
eruption window its $dM/d\ln\sigma$ curves increase in the disk band (green region) and the sheath, indicating that floored material is pushed into the flux tubes through reconnection in the equatorial current sheet, and is carried into the disk as the tubes are ejected. In the middle of the eruption ($11340\,t_g$), we see no separation between $r_{\rm cut}=40$--$100\,r_{\rm g}$ for the jet tracer in the sheath, whereas towards the end of the eruption ($11590\,t_g$) widening gaps develop between the cuts. This is consistent with former jet material, carried within the ejected flux tubes, moving to larger radii as the tubes move outwards.

\begin{figure}
  \centering
  \includegraphics[width=\linewidth]{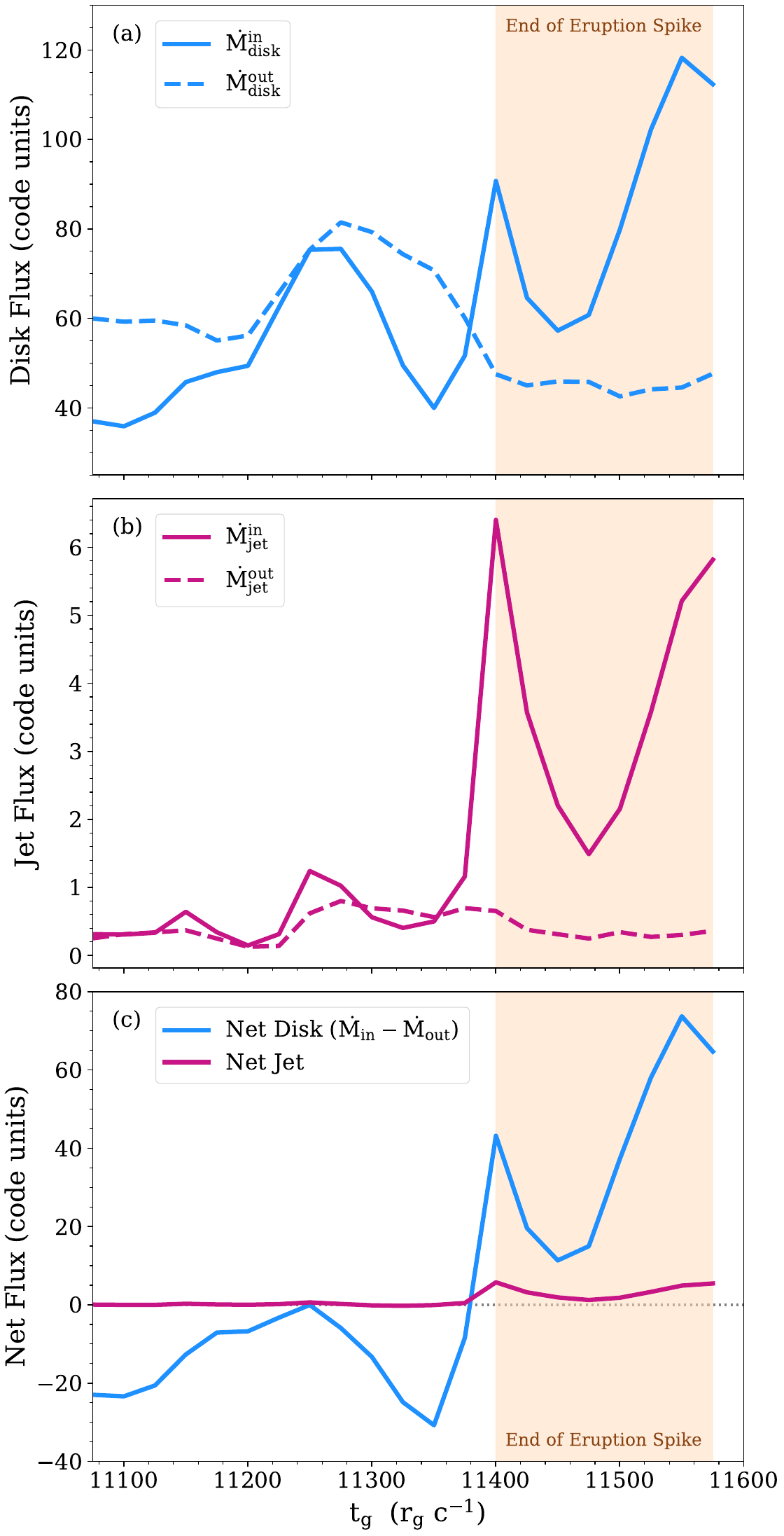}
  \caption{Tracer-weighted mass fluxes across the magnetisation-defined
    jet boundary ($\sigma=1$) as a function of time over the fourth eruption period.  Panel~(a): disk tracer inflow
    and outflow, $\dot{M}_{\rm disk}^{\rm in}$ (solid) and
    $\dot{M}_{\rm disk}^{\rm out}$ (dashed).  Panel~(b): jet/floor tracer
    inflow and outflow, $\dot{M}_{\rm jet}^{\rm in}$ (solid) and
    $\dot{M}_{\rm jet}^{\rm out}$ (dashed), on a logarithmic $y$-axis.
    Panel~(c): net fluxes (in--out) for disk (blue) and jet/floor
    (red).  The shaded band marks the period in the eruption window where the flux tubes bring entrained disk material across the jet boundary.}
  \label{fig:funnel_flux}
\end{figure}

Because most of the radial motion seen in the $dM/d\ln\sigma$ curves
occurs in the $20$--$80\,r_{\rm g}$ range for both tracers, we focus our analysis of the mass fluxes across the jet boundary on this band.  The three-panel flux figure
(Fig.~\ref{fig:funnel_flux}) summarises the time evolution of the
tracer-weighted mass flux across the jet boundary, defined as the $\sigma=1$ surface:
panel~(a) shows disk tracer inflow and outflow,
$\dot{M}_{\rm disk}^{\rm in}$ and $\dot{M}_{\rm disk}^{\rm out}$;
panel~(b) shows the corresponding jet/floor tracer fluxes,
$\dot{M}_{\rm jet}^{\rm in}$ and $\dot{M}_{\rm jet}^{\rm out}$; and
panel~(c) shows the net fluxes (in--out) for both tracers. At each radius in the $20$--$80\,r_{\rm g}$ band, we identify the cells adjacent to the $\sigma=1$ surface, compute the tracer-weighted mass flux $\rho\,\theta_{\rm tr}\,u_n$ normal to it, where $u_n$ is the component of the fluid four-velocity along $\nabla\sigma$ (positive into the jet), and sum the inward (`in') and outward (`out') contributions over the surface in both hemispheres (see Appendix~\ref{sec:funnel_flux} for details).

In the leadup to the eruption ($t\approx11.1$--$11.3\times10^{3}t_{\rm g}$) the disk inflow is comparable to or smaller than the outflow, and the net disk flux in panel~(c) is close to zero or slightly negative. Across the eruption ($t\approx11.4$--$11.6\times10^{3}t_{\rm g}$) the balance reverses: $\dot{M}_{\rm disk}^{\rm in}$ rises to $\sim(9$--$12)\times10^{1}$ while $\dot{M}_{\rm disk}^{\rm out}$ drops to $\sim(4.3$--$4.9)\times10^{1}$, producing a strongly positive net disk flux in panel~(c).  This shows that, during the eruption, more disk material crosses the jet boundary into the high-$\sigma$ region than leaves it in the $20$--$80\,r_{\rm g}$ band.

The jet/floor tracer in panel~(b) exhibits a parallel enhancement:
$\dot{M}_{\rm jet}^{\rm in}$ increases by an order of magnitude over the eruption window while $\dot{M}_{\rm jet}^{\rm out}$ stays nearly
constant.  The net jet/floor flux in panel~(c) therefore remains small but positive, indicating a weak net mass loading of the jet with former jet material. Combined with the $dM/d\ln\sigma$ behaviour, this suggests that the flux tubes ejected during the eruption move through the $20$--$80\,r_{\rm g}$ band, entrain disk plasma, and bring both former jet material (loaded into the tubes through reconnection) and disk material to the jet boundary.  The increase in disk tracer mass in the sheath and jet bands, together with the positive net disk flux across the jet boundary, indicates that flux eruptions enhance the loading of the jet boundary and sheath with disk material. Flux eruptions are therefore mass-transfer events: former jet material that has passed through the flux tubes, mixed with disk material, flows into the jet sheath and enters the high-$\sigma$ region at $20$--$80\,r_{\rm g}$, but does not reach the jet spine.

\subsection{Flux tubes as mass transfer highways}\label{mixing}

We now investigate whether the flux tubes ejected during the eruption mediate the loading of the jet, i.e. whether they can explain the injection of disk material into the jet that we find at 20--80\,$r_{\rm g}$, as well as the increased content of former jet material in the disk and sheath at larger radii. We note that this radial range likely reflects how far the tubes have travelled during the analysed period rather than a preferred location: at later times, and with higher resolution at large radii, the same process may operate further out.

\begin{figure*}
    \centering
    \includegraphics[width=0.85\textwidth]{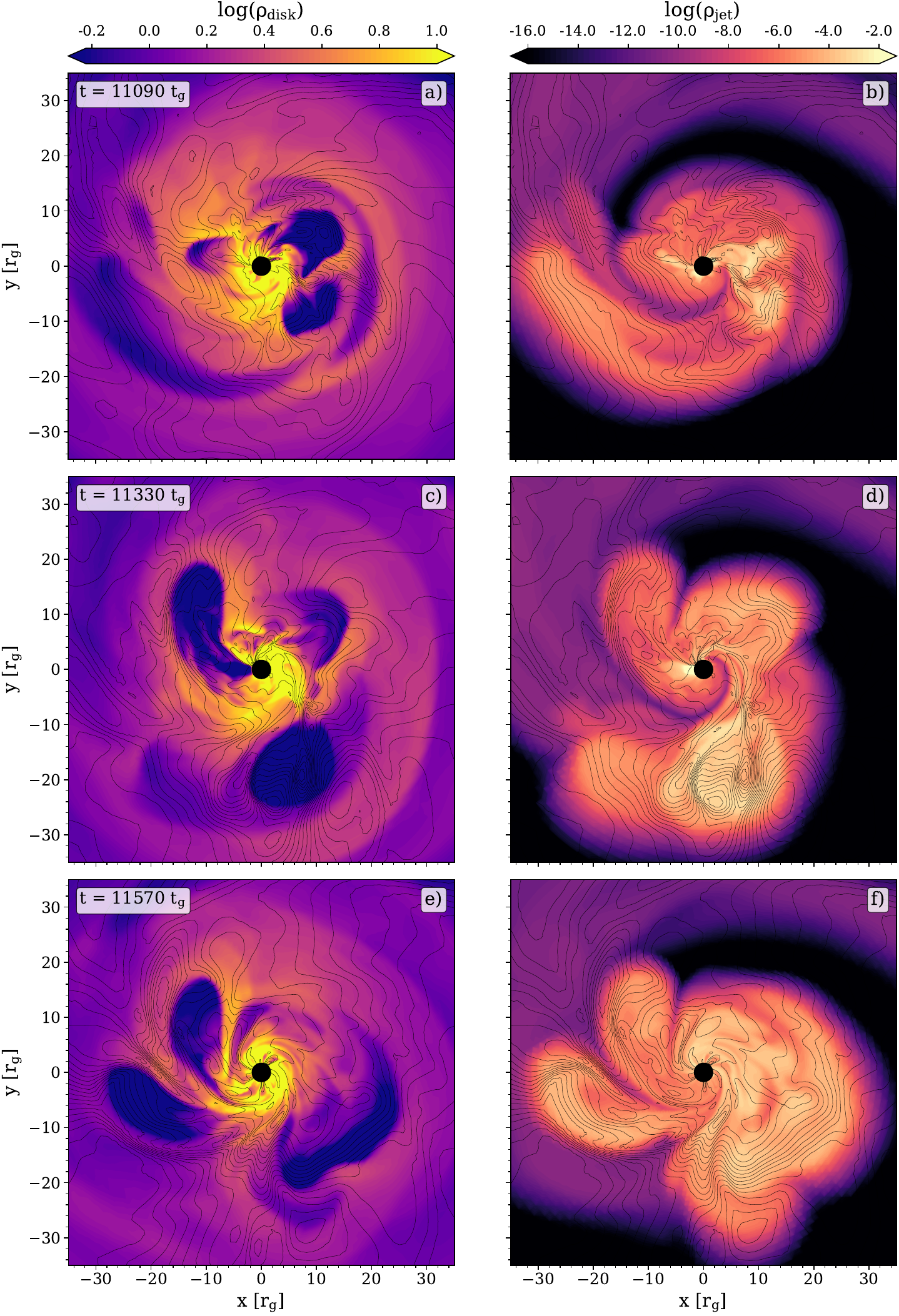}
    \caption{
    Equatorial ($z=0$) $x$–$y$ slices of disk- and jet-tracer density across the eruption, at $t=11090$, $11340$ and $11590\,t_g$ (top to bottom). Panels~(a), (c) and (e) show $\log_{10}\rho_{\rm disk}$ and panels~(b), (d) and (f) show $\log_{10}\rho_{\rm jet}$; black lines are magnetic potential contours.
    }
    \label{fig:xy_midplane}
\end{figure*}





\begin{figure*}
    \centering
    \includegraphics[width=\textwidth]{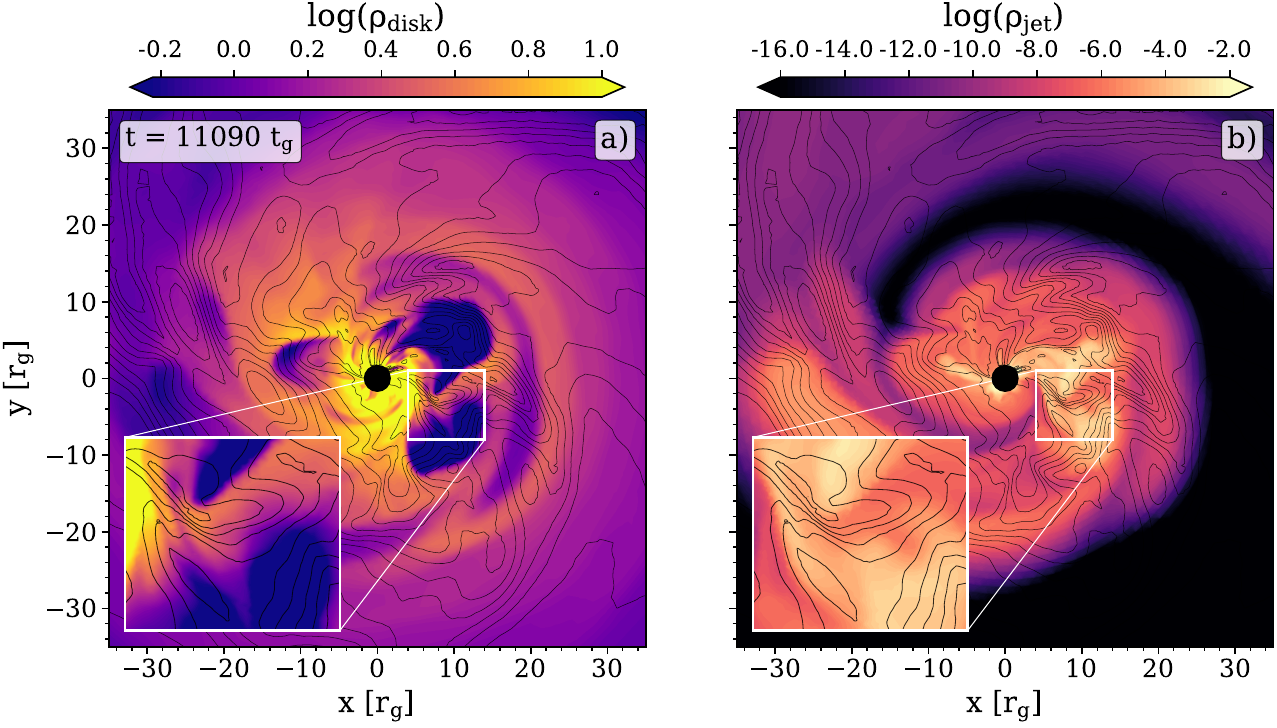}
    \caption{
    Zoomed view of the equatorial ($z=0$) slice at $t=11090\,t_g$ (top row of Fig.~\ref{fig:xy_midplane}). Panel~(a) shows $\log_{10}\rho_{\rm disk}$ and panel~(b) $\log_{10}\rho_{\rm jet}$; black lines are magnetic potential contours. The white box marks the region magnified in the inset, at the boundary between the low-density flux tubes and the surrounding disk. A tongue of disk material extends between two flux tubes, and the potential contours bend along it, pinching disk material into the flux tubes.
    }
    \label{fig:xy_zoom_335}
\end{figure*}

\begin{figure*}
    \centering
    \includegraphics[width=\textwidth]{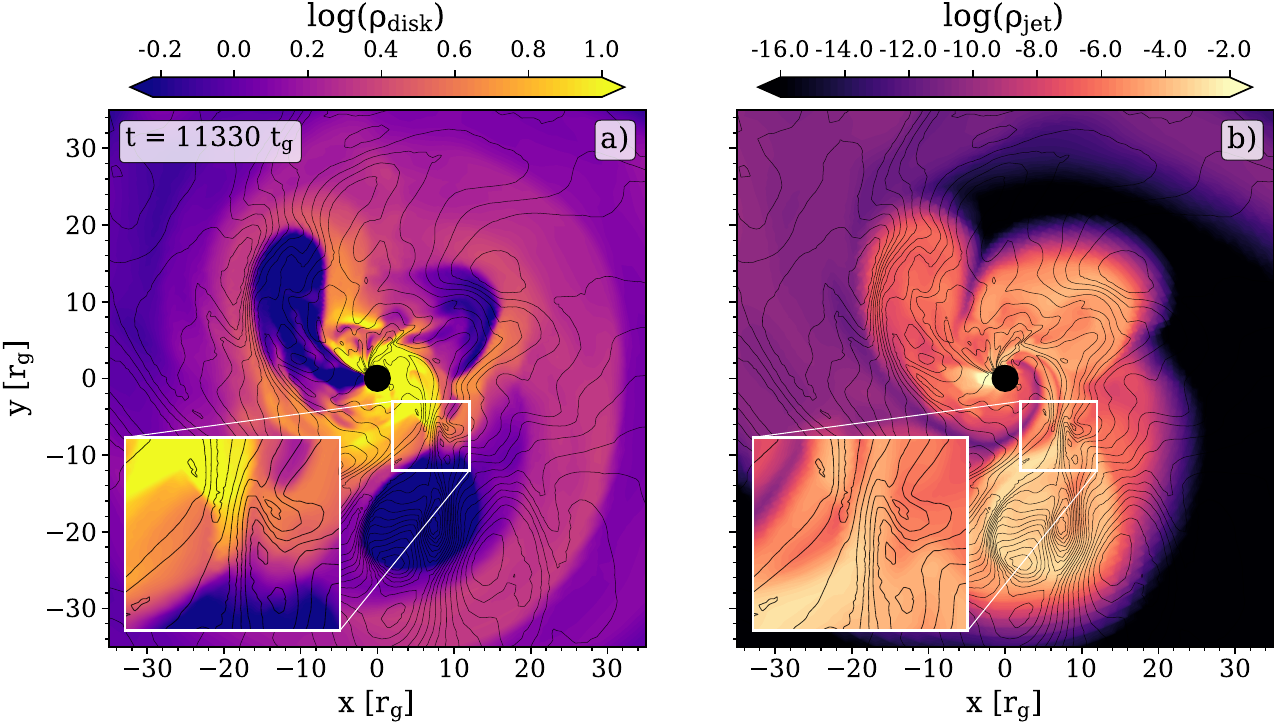}
    \caption{
    As Fig.~\ref{fig:xy_zoom_335}, at $t=11340\,t_g$ (middle row of Fig.~\ref{fig:xy_midplane}). The inset magnifies the inner edge of the large flux tube below the black hole, where disk material borders the flux tube and the magnetic potential contours are strongly bunched and distorted, with the signature of disk material loading onto a forming Rayleigh–Taylor-like plume.
    }
    \label{fig:xy_zoom_345}
\end{figure*}

\begin{figure*}
    \centering
    \includegraphics[width=\textwidth]{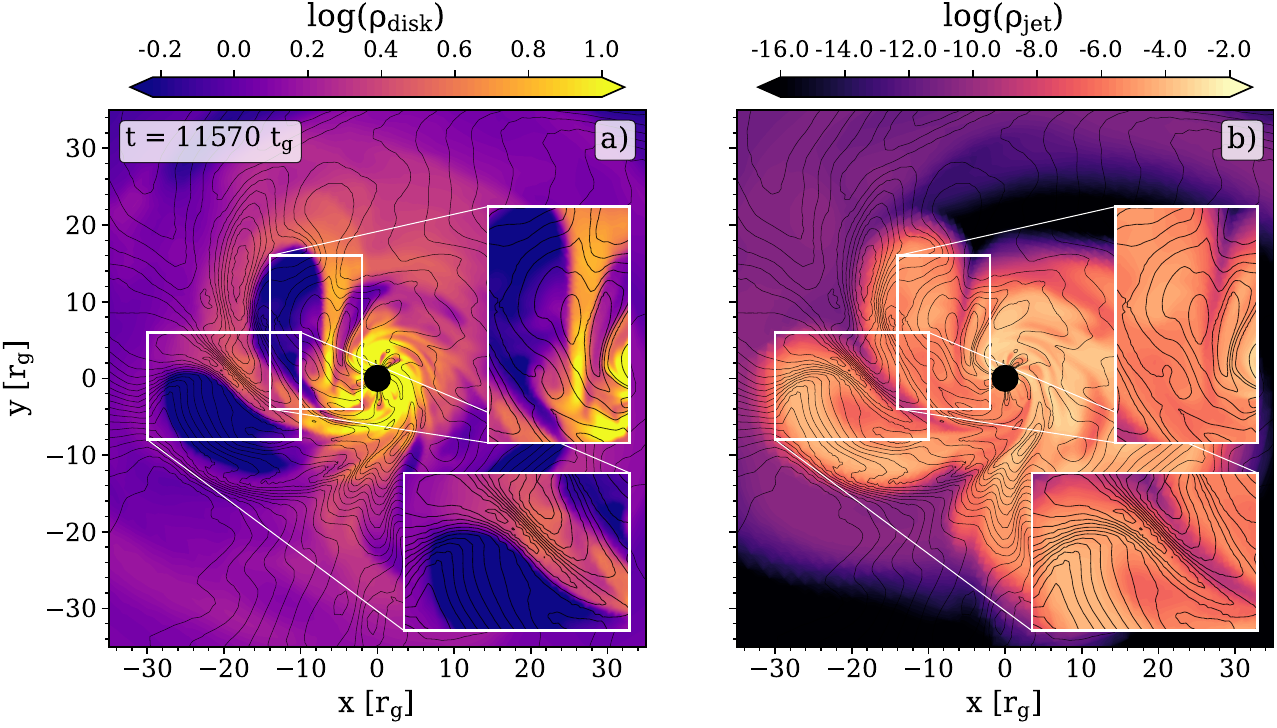}
    \caption{
    As Fig.~\ref{fig:xy_zoom_335}, at $t=11590\,t_g$ (bottom row of Fig.~\ref{fig:xy_midplane}). Two regions are magnified. Upper inset: the black-hole-facing side of a flux tube to the upper left of the black hole, where elongated fingers of lower-density gas alternate with denser disk material pinched through the flux-tube edge. Lower inset: the elongated flux tube to the left of the black hole, whose upper edge is lined by compressed magnetic potential contours with loaded disk material.
    }
    \label{fig:xy_zoom_355}
\end{figure*}

The equatorial ($z=0$) $x$–$y$ slices in Fig.~\ref{fig:xy_midplane} show where material is exchanged between the flux tubes and the disk. At early times, the tubes appear as narrow, low-density channels that are almost completely devoid of disk-origin plasma and are instead filled with material injected in the jet (our jet/floor tracer): the panels tracking the material originating in the jet show that the flux tubes are filled with former jet material, which subsequently mixes with disk material. As the eruption progresses, the tubes are pushed into the disk, where they orbit the black hole. At the boundaries between the tubes and the disk, Rayleigh–Taylor-like fingers of high $\rho_{\rm disk}$ develop within the inner $\sim 40\,r_g$, mixing disk gas into the tubes and former jet material into the disk; in the jet-tracer panels, these regions coincide with a reduction in $\rho_{\rm jet}$. The zoomed views in Figs.~\ref{fig:xy_zoom_335}--\ref{fig:xy_zoom_355} show these boundaries in detail: a tongue of disk material extends between two flux tubes and is pinched into them (Fig.~\ref{fig:xy_zoom_335}); along the edge of a flux tube the magnetic potential contours are strongly bunched and distorted, with the signature of disk material loading onto a forming Rayleigh–Taylor-like plume (Fig.~\ref{fig:xy_zoom_345}); and at $t=11590\,t_g$ elongated fingers of lower-density gas alternate with denser disk material pinched through the black-hole-facing edge of a flux tube (Fig.~\ref{fig:xy_zoom_355}). The disk panels show the low-density cavities that the tubes carve out of the midplane, while the jet panels show the pockets of jet-origin material that they leave behind in the disk. In these cuts, magnetically dominated ($\sigma>1$) gas is present only close to the black hole, where the jet base is narrowest, indicating that in the midplane the mixing occurs at the boundary between the flux tubes and the disk, and not at the jet boundary. This motivates a complementary view in the meridional $x$–$z$ plane, where the vertical structure of the tubes and their connection to the jet can be followed.

\begin{figure*}
    \centering
    \includegraphics[width=\textwidth]{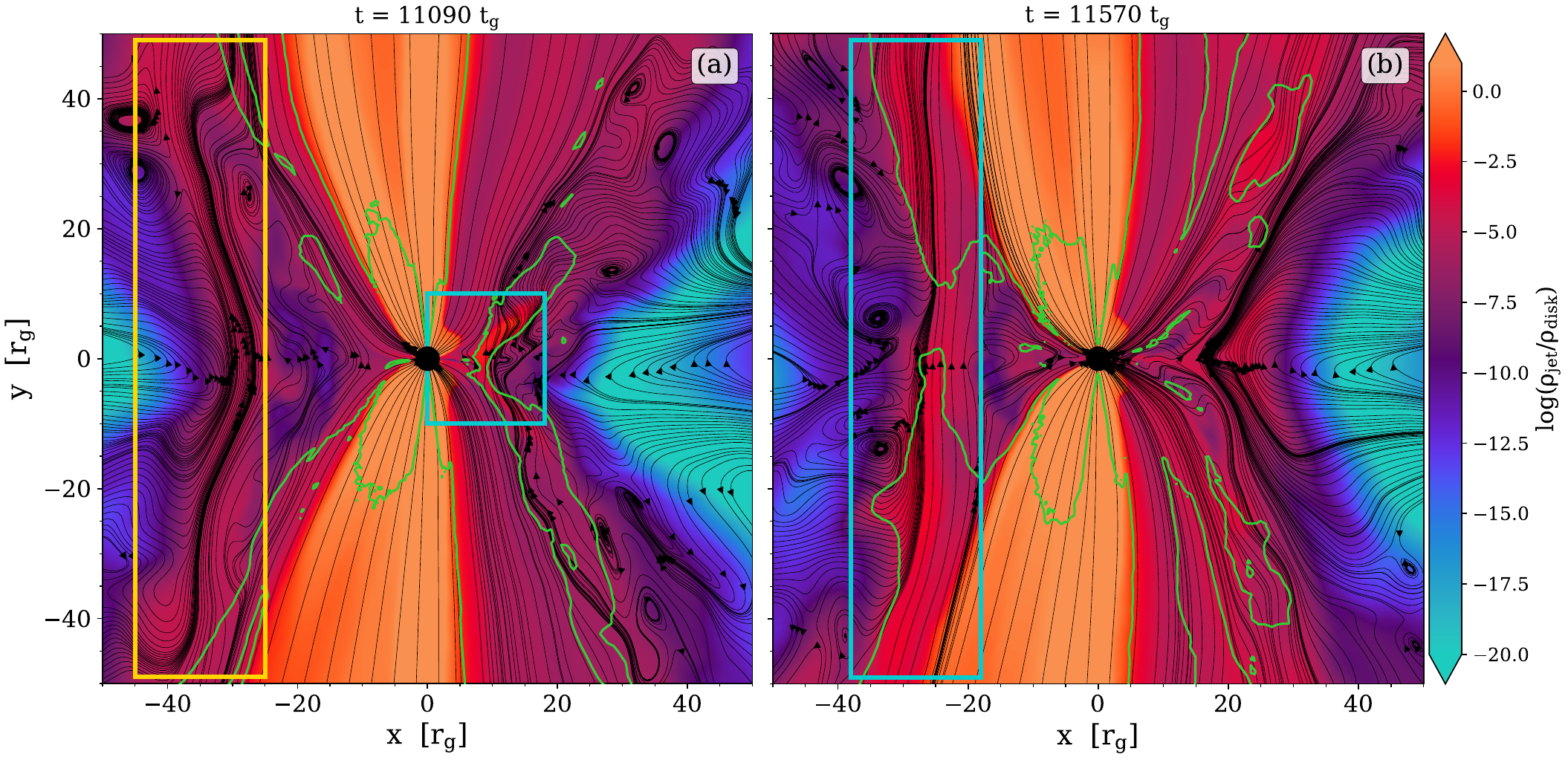}
    \caption{
    Meridional $x$–$z$ slices of the jet–to–disk composition ratio across the eruption. The colour scale shows $\log_{10}(\rho_{\rm jet}/\rho_{\rm disk})$, so that red–orange regions are jet dominated and blue-purple regions are disk dominated. Black streamlines trace the in-plane component of the magnetic field; the green contour marks the $\mathrm{Be}=1.02$ surface. \emph{Left:} At the beginning of the eruption. The yellow box encloses a remnant flux tube from the previous event. The turquoise box captures a portion of a forming flux tube. \emph{Right:} end of the eruption. The turquoise box shows that the nascent tube from the left panel has completed roughly half an orbit around the black hole, and is now more coherent and predominantly poloidal. Notably, it has vortical features now appearing along the length of the disk-facing edge of the tube.
    }
    \label{fig:xz_flux_tubes}
\end{figure*}

\begin{figure}
    \centering
    \includegraphics[width=\columnwidth]{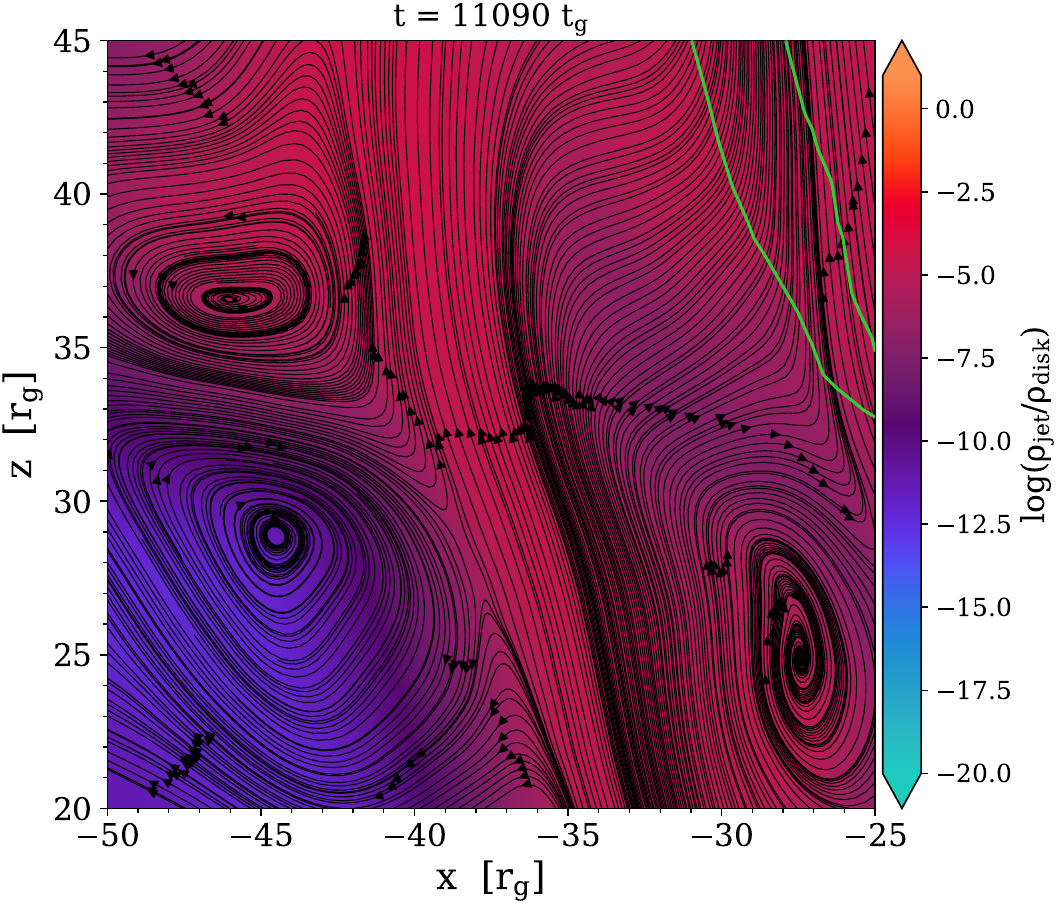}
    \caption{
    Zoom into the upper part of the remnant flux tube (yellow box in the left panel of Fig.~\ref{fig:xz_flux_tubes}) at $t=11090\,t_g$, covering $-50\leq x\leq -25\,r_g$ and $20\leq z\leq 45\,r_g$. As in Fig.~\ref{fig:xz_flux_tubes}, the colour scale shows $\log_{10}(\rho_{\rm jet}/\rho_{\rm disk})$, black streamlines trace the in-plane magnetic field and the green line is the $\mathrm{Be}=1.02$ contour. A bundle of nearly vertical field lines (the remnant tube) runs through the centre of the panel; on either side of it the in-plane field forms closed loops (at $x\approx -45\,r_g$, $z\approx 29$ and $36\,r_g$, and at $x\approx -28\,r_g$, $z\approx 25\,r_g$). The $\mathrm{Be}=1.02$ contour enters only the upper-right corner.
    }
    \label{fig:xz_zoom_remnant}
\end{figure}

\begin{figure}
    \centering
    \includegraphics[width=\columnwidth]{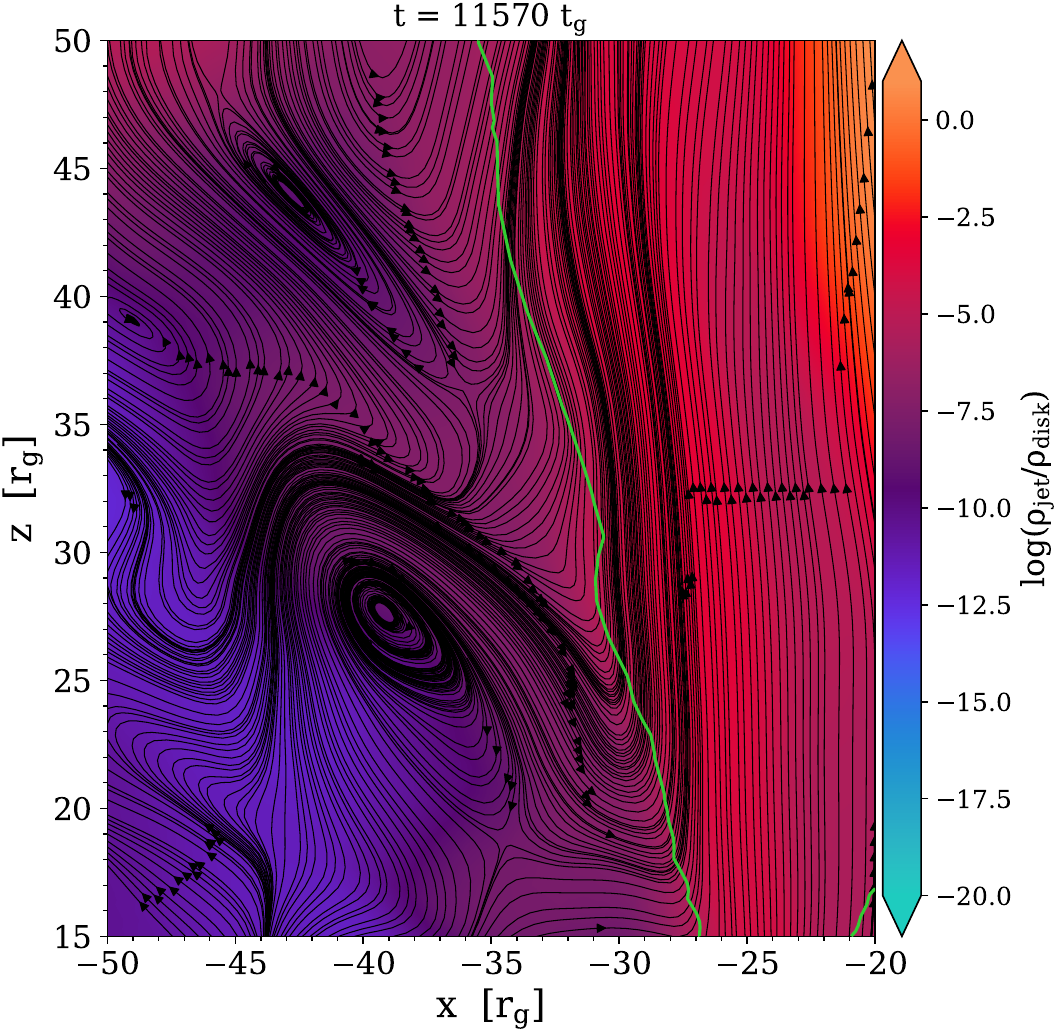}
    \caption{
    Zoom into the disk-facing edge of the flux tube highlighted by the turquoise box in the right panel of Fig.~\ref{fig:xz_flux_tubes}, at $t=11590\,t_g$, covering $-50\leq x\leq -20\,r_g$ and $15\leq z\leq 50\,r_g$ (extending $\sim 10\,r_g$ further from the axis than the box); quantities as in Fig.~\ref{fig:xz_zoom_remnant}. The jet-dominated tube (right) is bounded on its disk-facing side by the $\mathrm{Be}=1.02$ contour (green). On the disk side of this contour the in-plane field curls into closed loops, a large one at $x\approx -39\,r_g$, $z\approx 28\,r_g$ and a smaller one at $x\approx -42\,r_g$, $z\approx 44\,r_g$; these are the vortical features noted in Fig.~\ref{fig:xz_flux_tubes}.
    }
    \label{fig:xz_zoom_355}
\end{figure}

\begin{figure}
    \centering
    \includegraphics[width=\columnwidth]{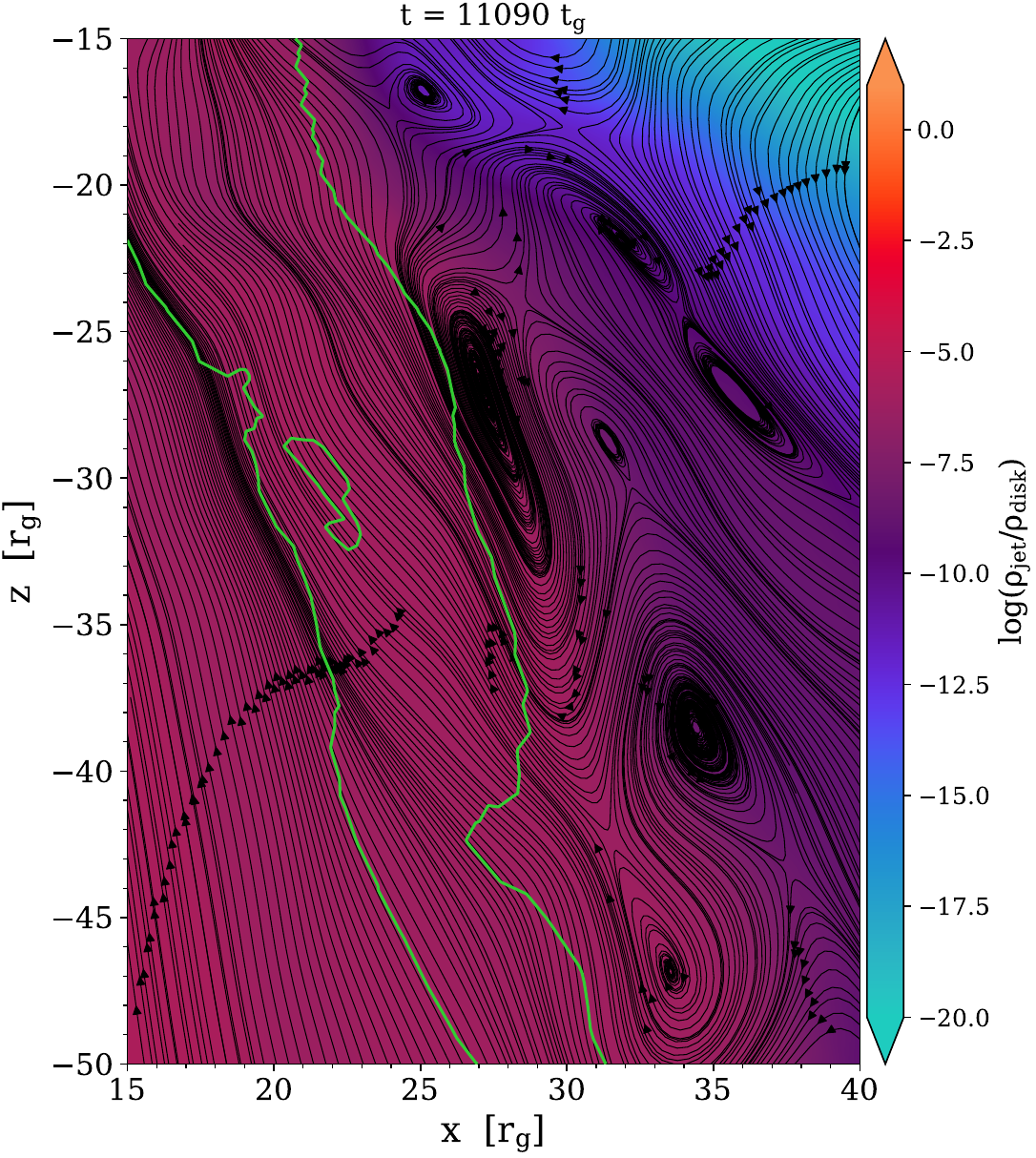}
    \caption{
    Zoom into the region below the midplane to the right of the jet at $t=11090\,t_g$ (left panel of Fig.~\ref{fig:xz_flux_tubes}), covering $15\leq x\leq 40\,r_g$ and $-50\leq z\leq -15\,r_g$; quantities as in Fig.~\ref{fig:xz_zoom_remnant}. Gas with a higher jet fraction ($\log_{10}(\rho_{\rm jet}/\rho_{\rm disk})\approx -5$, left) is separated from more disk-dominated gas (right) at the outer $\mathrm{Be}=1.02$ contour (green). Outside this contour the in-plane field forms several closed loops; the largest, elongated one lies directly against the contour at $z\approx -24$ to $-32\,r_g$.
    }
    \label{fig:xz_zoom_jetboundary}
\end{figure}

The meridional $x$–$z$ slices in Fig.~\ref{fig:xz_flux_tubes} show how these flux tubes evolve in height during the eruption. We plot $\log_{10}(\rho_{\rm jet}/\rho_{\rm disk})$ so that regions dominated by jet-origin material appear red–orange and disk-dominated regions appear blue–purple. The green contour marks $\mathrm{Be}=1.02$, where $\mathrm{Be}=-hu^t$ is the Bernoulli parameter, with $h$ the specific enthalpy and $u^t$ the contravariant time component of the four-velocity; it encloses unbound, outflowing gas \citep{Tchekhovskoy2011, McKinney2012}. We adopt this threshold following \citet{Moscibrodzka2014}, who used $-hu_t=1.02$ to separate the jet from the disk and its slower wind. A different threshold would change the width of the sheath and would therefore affect our results quantitatively; however, the mean magnetisation of the gas with $\mathrm{Be}>1.02$ is close to $\sigma_{\rm sh}=10^{-1.8}$, the value at which the flux-tube field lines settle in the sheath (Section~\ref{sec:dmdsigma}). We note that the inner $\sim 30\,r_g$ is dominated by former floor material in these slices, so that the composition there is largely set by the density floors. In the left panel, taken near the beginning of the eruption, the yellow box marks a remnant flux tube from a previous event. This tube has already been distorted by the disk pushing on it, and disk material is already mixed into it along much of its length. Near $z \sim 40\,r_g$, where the remnant tube connects to the jet sheath, it develops localised knots and spiral distortions, suggestive of tube material mixing into the jet; a zoom just below this region (Fig.~\ref{fig:xz_zoom_remnant}) shows closed loops of the in-plane field on either side of the tube. By contrast, the turquoise box in the right panel highlights a much more coherent, nearly poloidal flux tube from the end-of-eruption snapshot. This tube is just forming at smaller radii in the left panel, as a result of a reconnection event in the midplane within the inner few $r_g$, and becomes fully developed by the time of the right panel. A zoom into the disk-facing edge of this tube (Fig.~\ref{fig:xz_zoom_355}) shows the in-plane field curling into closed loops on the disk side of the $\mathrm{Be}=1.02$ contour, the largest at $x\approx -39\,r_g$, $z\approx 28\,r_g$. Such vortices could arise from turbulence or from kink-like distortions of the tube; we defer an investigation of their origin to a follow-up paper. In the left panel, closed loops of the in-plane field also appear just outside the outer $\mathrm{Be}=1.02$ contour in the lower hemisphere (Fig.~\ref{fig:xz_zoom_jetboundary}); these are instabilities around a surface wave running down the jet–disk boundary through the sheath. Whether the disk material mixed into the tubes is carried along them into the jet is examined with field-line tracing below.

\begin{figure}
    \centering
    \includegraphics[width=\columnwidth]{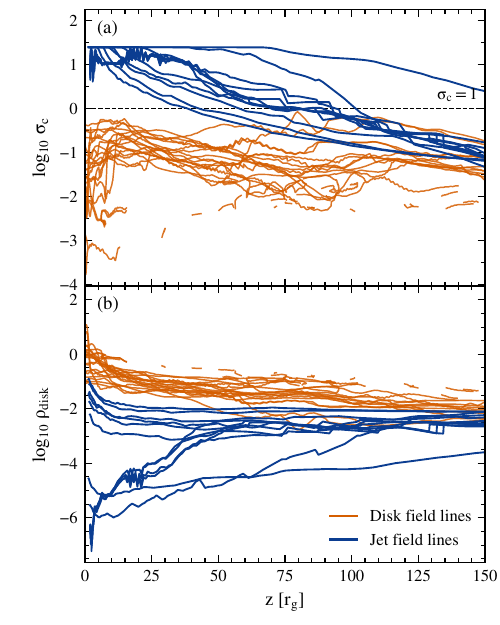}
    \caption{
    Magnetisation and disk-tracer density along representative magnetic field lines. Top: $\log_{10}\sigma$ as a function of height $z$; bottom: $\log_{10}\rho_{\rm disk}$ along the same lines. The blue curve shows a jet line, seeded near the horizon, that starts at high magnetisation and negligible disk content. Orange curves show disk lines seeded in the equatorial midplane at $10<r<40\,r_g$. Disk lines begin at low $\sigma$ and high $\rho_{\rm disk}$, rise through the flux tube to the jet sheath, and coil around the jet line, gradually losing disk mass as they mix with jet-origin plasma. Correspondingly, the jet line exhibits a clear increase in $\rho_{\rm disk}$ as it passes through $z\simeq 40$–$60\,r_g$, where the tubes interact most strongly. Above this region the combined bundle of disk and jet lines bends away from the axis into the sheath. The horizontal line in the top panel marks $\sigma=1$, separating matter-dominated ($\sigma<1$) from magnetically dominated ($\sigma>1$) segments along each field line.
    }
    \label{fig:streamline_profiles}
\end{figure}

\begin{figure*}
    \centering
    \includegraphics[width=\textwidth]{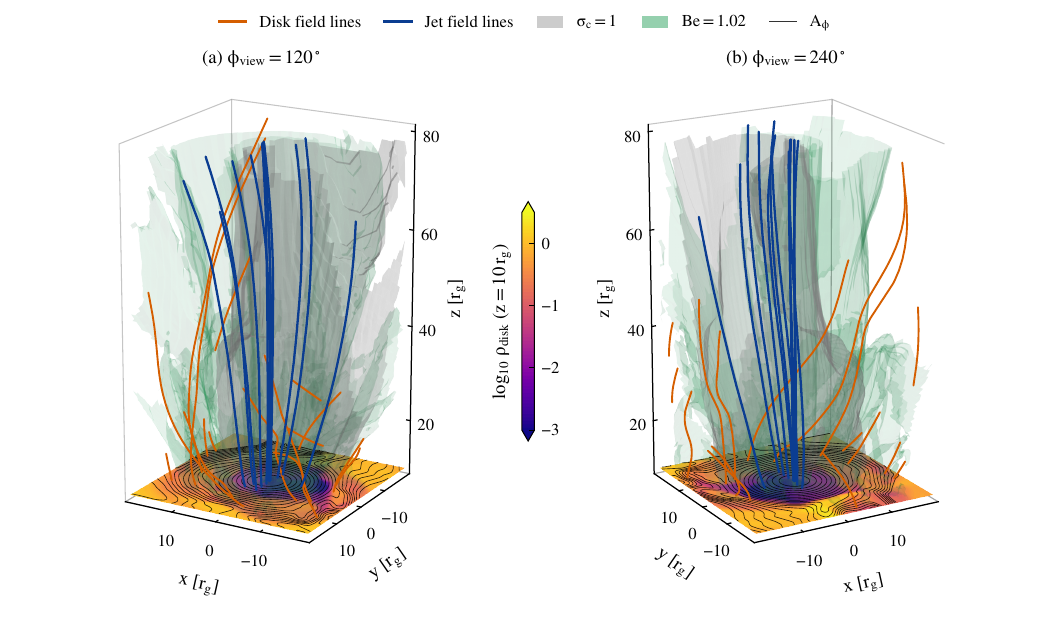}
    \caption{
    Zoomed 3D view of a flux tube resulting from an eruption. Both panels show magnetic field lines, clipped to $|x|,|y|\leq 20\,r_g$ and $10\leq z \leq 80\,r_g$. Orange lines are seeded in the equatorial midplane (disk lines), blue lines are seeded near the horizon in the polar cone (jet lines). Semi-transparent yellow and cyan surfaces mark, respectively, the $\sigma=1$ iso-surface (the jet boundary) and the $\mathrm{Be}=1.02$ iso-surface, which encloses all unbound, outflowing gas (the jet sheath lies between the two surfaces), restricted to the same box. The left and right panels correspond to viewing azimuths $\phi_{\rm view}=120^\circ$ and $240^\circ$. The disk lines start out predominantly toroidal in the disk and run predominantly vertically through the flux tube; they coil among themselves and then wrap around the jet lines where the tube connects to the jet sheath, before both sets of lines bend away from the axis into the sheath. The full height of the tube, down to the equatorial plane, is shown in Fig.~\ref{fig:flux_tube_3d}.}
    \label{fig:flux_tube_zoom}
\end{figure*}

\begin{figure*}
    \centering
    \includegraphics[width=\textwidth]{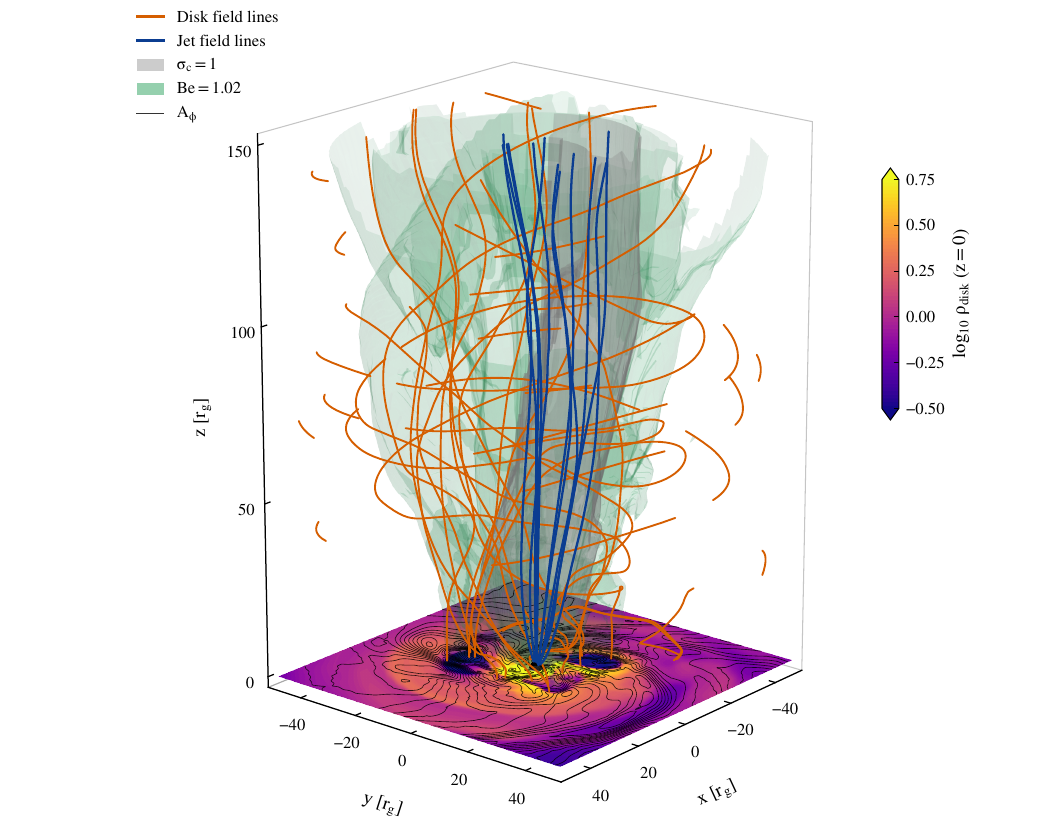}
    \caption{
    Three-dimensional view of disk (orange) and jet (blue) magnetic field lines over the full height of the flux tube, from the equatorial plane up to $z\approx 150\,r_g$, complementing the clipped view in Fig.~\ref{fig:flux_tube_zoom}. The equatorial slice shows $\log_{10}\rho_{\rm disk}$ at $z=0$ with magnetic potential ($A_\phi$) contours as black lines, so that the footpoints of the disk lines are visible. Semi-transparent grey and green surfaces mark the $\sigma=1$ and $\mathrm{Be}=1.02$ iso-surfaces, respectively.
    }
    \label{fig:flux_tube_3d}
\end{figure*}

To connect the midplane and meridional views with the actual flow of material along magnetic structures, we integrate representative magnetic field lines in full 3D with a fourth-order Runge–Kutta integrator along the local magnetic field (Figs.~\ref{fig:streamline_profiles} and \ref{fig:flux_tube_zoom}). We distinguish `jet lines', seeded just outside the horizon in a narrow polar cone, from `disk lines', seeded in the equatorial midplane at $10<r<40\,r_g$; Fig.~\ref{fig:flux_tube_3d} shows disk and jet lines down to the footpoints of the disk lines in the equatorial plane. For each line we record the magnetisation and the loading with disk material as a function of height, $\log_{10}\sigma(z)$ and $\log_{10}\rho_{\rm disk}(z)$. Figure~\ref{fig:streamline_profiles} plots these quantities for a single representative jet line (blue) and several disk lines (orange). The jet line starts near the horizon at high magnetisation and negligible disk-tracer content, remaining loaded almost purely with jet-origin plasma until it reaches $z\simeq 40$–$60\,r_g$. In this range the disk lines, which begin at the midplane with low $\sigma$ and high $\rho_{\rm disk}$, have already risen through the flux tube to the jet sheath. They first coil among themselves and then around the jet line (as seen in Fig.~\ref{fig:flux_tube_zoom}), in agreement with \citet{Nalewajko2026}, gradually transferring mass into the jet line. This exchange appears in Fig.~\ref{fig:streamline_profiles} as a modest but clear rise in $\rho_{\rm disk}$ along the jet line while its magnetisation declines from $\sigma\gtrsim 10$ toward intermediate values. Above $z\simeq 60\,r_g$ the bundle of disk and jet lines bends away from the axis, consistent with the helical distortion of the jet that follows a flux eruption \citep{Nalewajko2026}, and the lines settle into the sheath, characterised by intermediate magnetisation and mixed (but still disk-dominated) composition, consistent with the $dM/d\ln\sigma$ diagnostic and $x$–$z$ plots discussed earlier.

\section{Discussion}\label{sec:discussion}

The redistribution of disk and jet material that we find during flux eruptions can be understood within the framework of flux eruption dynamics \citep{McKinney2007, Tchekhovskoy2011, Ripperda2022}. In the pre-eruption state, magnetic flux accumulates on the horizon and powers the jet through the Blandford–Znajek mechanism. Once the flux saturates, the magnetic pressure balances the ram pressure of the accreting gas, and the accretion flow near the horizon is squeezed into a thin equatorial current sheet that becomes tearing unstable and reconnects. This reconnection ejects a flux tube into the disk, which then orbits the black hole while moving outwards in radius \citep{Ripperda2022}. Both the reconnection and the ejected flux tubes can accelerate particles \citep{Hakobyan2023, Zhdankin_2023}. Our tracer diagnostics reveal the compositional signature of this process for the first time in 3D.

In the mass distributions over magnetisation, this process appears as disk material being shifted from disk-like to sheath-like magnetisations as the eruption proceeds. We interpret this as follows: the ejected flux tubes mix with the disk at their outer boundaries through Rayleigh–Taylor-like instabilities, and the disk material mixed into a tube can then move vertically along it to where the tube connects to the jet sheath, while the tube itself moves outwards.

As shown in Section~\ref{sec:dmdsigma} (Fig.~\ref{fig:funnel_flux}), the net flux of disk material across the jet boundary changes from close to zero to strongly positive during the eruption.

In terms of the mass distribution over magnetisation ($dM/d\ln\sigma$), the mass of disk material at sheath-like ($\sigma_{\rm sh}<\sigma<1$) and jet-like ($\sigma>1$) magnetisations increases during the eruption, while over the same period (between $11090$ and $11590\,t_g$) the mass of former jet material at disk-like and sheath-like magnetisations ($\sigma<1$) also increases, indicating that both disk and former jet material are carried by the ejected flux tubes. The jet boundary is thus not compositionally static: it is repeatedly and intermittently loaded with disk-origin plasma.

From an observational perspective, this redistribution of material may have important multi-wavelength and multi-messenger consequences. In real systems, the floored material is expected to be dominated by pair plasma. We stress, however, that the floor tracer only records where this material resides: it carries no information on the actual pair content, which is set by pair-production physics not included in our GRMHD simulation. With this caveat, such material could supply a leptonic population that, if a fraction is accelerated to relativistic energies, can produce synchrotron emission and thereby increase the local synchrotron photon density. These photons can subsequently act as seed photons for synchrotron self-Compton (SSC) scattering, potentially enhancing the inverse-Compton emission from the inner jet \citep{Boettcher2013}. The disk-origin component could instead primarily increase the baryonic content and inertia of the outflow. If a fraction of the ions from the disk were subsequently accelerated, they could provide an energetic proton population potentially capable of producing secondary emission through hadronic interactions. In proton–proton ($pp$) interactions, the surrounding disk-origin baryons may provide the target material for the accelerated protons, whereas in proton–photon ($p\gamma$) interactions the target is the local photon field, which may include synchrotron photons produced by the accelerated leptonic component; both channels produce pions, whose decays yield neutrinos and $\gamma$-rays. Additionally, if disk material is accelerated in the jet boundary it can potentially be a source of cosmic rays. Thus, if the two tracer populations are taken as proxies for pairs and baryons, the material they track at the jet boundary could influence the high-energy emission through distinct but coupled channels. During flux eruptions, the order-unity enhancement of disk-origin material at the jet boundary and in the inner sheath found in our simulations could therefore transiently modify the composition, baryon density, and available non-thermal particle reservoir there, while concurrent changes in the floor-origin component (a proxy for, but not a measure of, the pair content) could alter the synchrotron photon field relevant for SSC and $p\gamma$ interactions. Although our simulations do not directly model particle acceleration or radiative transfer, these changes in the disk- and floor-origin content of the jet boundary provide a physically motivated pathway by which variations in mass loading could modulate VHE $\gamma$-ray and neutrino emission without requiring the jet as a whole to be hadron dominated at all times.

Black hole X-ray binaries (BHXRBs) in luminous states are in a different regime from the radiatively inefficient flow simulated here: their accretion flows are radiatively cooled and need to be modelled with radiative GRMHD. When such flows reach the MAD regime, the disk truncates and magnetic flux is expelled over much longer timescales, rather than in the recurrent eruption cycle seen in our simulation \citep{Liska2022trunc}. The decay of magnetic flux in such truncated disks also produces flux tubes (\citealt{Liska2022trunc}; Musoke et al., in prep.), which can potentially connect disk material to the jet in the same way as the flux tubes in our simulation. We therefore do not relate the timescales of our eruption cycle to the rapid, correlated X-ray and optical/infrared variability observed in BHXRBs \citep[e.g.,][]{Gandhi_2008,Gandhi_2010,Casella2010,Gandhi_2017}, but we note that flux tubes may act as highways for disk material into the jet in these systems as well.

Semi-analytical multi-zone jet models \citep[e.g.,][]{Lucchini_2022}, which enable broadband modelling of black hole X-ray binaries, place the onset of non-thermal particle acceleration on compact jet scales, although the inferred distance is source-dependent; for example, modelling of GX~339--4 places it at $z_{\rm acc}\sim10^{3.5}R_g$ \citep{Connors_2019}. More importantly for the present work, the semi-analytical mass-loaded jet model of \citet{Lucchini_2023} identifies $z_{\rm diss}\simeq100R_g$ as the fiducial location at which mass entrainment begins, explicitly linking the onset of mass loading to the inner acceleration/dissipation region. These models demonstrate that substantial dynamical processing and the injection of baryonic material can begin at $\mathcal{O}(10^2R_g)$, well inside the region subsequently probed by large-scale jet--environment interactions \citep{Lucchini_2022,Lucchini_2023}. This provides a natural connection to the recent detection of $>100$~TeV $\gamma$-ray emission from Galactic jet sources by LHAASO, including V4641~Sgr and SS~433, which demonstrates that particle acceleration can persist or be re-established on macroscopic scales \citep{Cao_2025}. In particular, the extended emission associated with these systems indicates that the highest-energy particles need not be accelerated exclusively in the compact inner jet, but can instead arise from interactions between the evolved outflow and its surrounding interstellar medium. Within this multi-scale picture, the efficient disk mass loading found in our GRMHD simulations, as traced in the jet boundary and sheath, provides the physical connection between the two regimes. Material entrained and accelerated near the launching region establishes the baryonic content, inertia, and momentum flux of the outflow, while the associated magnetic-to-kinetic energy conversion determines the kinetic power ultimately transported to larger distances.

Similarly, the IceCube neutrino event IC\,170922A was associated with a $\gamma$-ray flare of the blazar TXS\,0506+056 \citep{Aartsen2018multimessenger}. As discussed in Section~\ref{sec:intro}, this association points to hadronic acceleration that likely requires significant baryon loading close to the black hole. TXS\,0506+056 is, however, a luminous source, whose accretion flow is likely radiatively efficient and therefore not directly described by our simulation. Moreover, the duration of flux eruptions depends on the reconnection rate, which is underestimated in ideal GRMHD simulations such as ours, so we do not compare timescales directly. Luminous, truncated disks can nevertheless also expel magnetic flux and produce flux tubes that connect the disk to the jet (\citealt{Liska2022trunc}; Musoke et al., in prep.). During flux eruptions in our simulation (Section~\ref{sec:dmdsigma}), the exchange of disk material across the jet boundary at $20$--$80\,r_g$ changes from nearly balanced ($\dot{M}_{\rm disk}^{\rm in}\lesssim\dot{M}_{\rm disk}^{\rm out}$) to an inflow that exceeds the outflow by a factor of $\sim2$--$3$. If luminous, truncated disks produce flux tubes that behave similarly, such episodes of enhanced sheath loading could provide a trigger for flaring of this kind.

Using the mass of M87* ($M_{\rm BH}=6.5\times10^9\,M_\odot$), the
gravitational time is $t_g = GM/c^3 \approx 3.2\times10^4$\,s
$\approx 0.37$\,days. Hence an eruption cycle period of
$1000$–$2000\,t_g$ translates to $1$–$2$\,years, the MAD phase duration
($\sim430\,t_g$) to $\sim160$\,days, and the fast eruption drop
($\sim340\,t_g$) to $\sim125$\,days. These numbers are in rough agreement with the
long‑term X‑ray/radio variability patterns reported for the M87* jet
\citep[e.g.,][]{Hada2016, Kim2018}. In particular, the $\sim10$\,day rise time of individual eruptions matches the fastest flares from the nucleus of M87 seen by
\textit{Chandra} and \textit{Swift} \citep[see also][]{Ripperda2022}. The intraday X‑ray variability of
the M87 core, which traces emission from the base of the jet on scales of $\sim r_g$, has been observed with a timescale of $\lesssim 20$\,days (e.g., with \textit{Chandra}), consistent with the rapid fluctuations seen in our simulation during the buildup and eruption phases. Our phase classification therefore provides a physical framework for interpreting the multi‑wavelength variability of M87* and other low-luminosity systems that may be in a MAD state.

An important implication of this flux-tube picture is that the mass loading of the jet boundary is intermittent and tied to the eruption cycle, in agreement with \citet{chow2026baryoncontentmagneticallyarrested}. The Rayleigh–Taylor-like plumes at the tube–disk interface \citep{Ripperda2022}, together with the helical, kink-like distortions of the jet where the tubes connect to it \citep{Nalewajko2026}, provide natural sites where dense gas from the disk is injected into regions of intermediate magnetisation ($\sigma \gtrsim 1$) and relativistic bulk motion at $z < 100\,r_g$. In these zones, magnetic energy may be converted into non-thermal particle energy via reconnection and current-driven or shear-driven instabilities, which have been shown to accelerate particles \citep{Davelaar2020, Sironi_2021}, while the enhanced baryon fraction may supply the hadronic targets required for $pp$ and $p\gamma$ interactions. This suggests that flux eruptions may seed localised, baryon-rich patches in the jet sheath and jet boundary that may be capable of powering the VHE flares and neutrino production discussed in Section ~\ref{sec:intro}, without requiring the jet as a whole to be hadron-dominated. Our results therefore support a scenario in which repeated flux eruptions in MAD systems act as a `highway' that episodically injects disk material into the magnetised jet sheath, potentially linking accretion dynamics to the transient VHE and neutrino activity of accreting black holes.

\section{Conclusions}\label{sec:conclusions}

In this work we have used a high–spin MAD GRMHD simulation with rotated
coordinates, adaptive mesh refinement, and strictly conserved eulerian
tracers to investigate how accretion disks interact with, and mass-load, their relativistic jets.
Our main results can be summarised as follows.

\begin{enumerate}
    \item \textbf{Flux eruptions load the jet sheath through flux tubes.}
    During a flux eruption, reconnection in the equatorial current sheet ejects a coherent flux tube into the disk \citep{Ripperda2022}, and the reconnection fills the tube with former jet material. At the outer boundary of the tube, Rayleigh–Taylor-like instabilities mix this former jet material into the disk, leaving behind low-density cavities in the midplane, while the same instabilities mix disk material into the tube. As the flux tube moves outwards, its connection point to the jet moves upwards \citep{Nalewajko2026}. Disk material moves along the flux tube and is mixed into the jet at the tube–jet boundary, mass-loading the jet sheath.

    \item \textbf{Streamline view of disk–jet exchange.}
    By integrating 3D field lines from disk and jet seeds, we follow
    individual flux tubes up to the jet. Jet lines start near
    the horizon with high $\sigma$ and negligible disk content, while
    disk lines originate at the midplane with low $\sigma$ and
    high $\rho_{\rm disk}$. Magnetised field lines originating at the jet base and weakly magnetised field lines originating in the disk connect where the flux tube connects to the jet, and mix there, indicating mass loading: the jet line acquires a modest but distinct increase in $\rho_{\rm disk}$ as it passes through the point where the flux tube connects to the jet. Above this region the jet shows signs of a helical, kink-like distortion \citep{Nalewajko2026}.

    \item \textbf{Implications for hadronic loading and emission.}
    Although the absolute amount of jet-origin (floor) material at low
    $\sigma$ remains small, the eruption-driven exchange between disk and
    jet tracer components enhances the loading of the jet sheath and jet boundary with disk material by order-unity factors. The ejected flux tubes therefore offer a plausible mechanism for intermittently
    supplying the hadronic content required by high-energy $\gamma$-ray
    and neutrino observations, without demanding that the jet as a whole be
    hadron dominated.
\end{enumerate}




Future work will build on these results constraining where and when baryons enter the jet, to quantifying how turbulence in the baryon‑loaded sheath shapes particle acceleration and the multi–messenger output of accreting black holes.

\section*{Data Availability}
Plotting scripts and relevant data objects can be found on 10.5281/zenodo.24139886. The H-AMR code itself, along with the the dumps can be provided under reasonable request by S.P., or, M.T.P.L..

\section*{Acknowledgements}
S.P. thanks Oliver Porth for his useful discussions and suggestions. S.P., A.L.S. and S.M. are supported by
a European Research Council Synergy Grant ‘Blackholistic’ (grant agreement no. 101071643). This work was supported by a grant from the Simons Foundation (MP-SCMPS-00001470). L.S. acknowledges support from DoE Early Career Award DE-SC0023015, NASA ATP 80NSSC24K1238, NASA ATP 80NSSC24K1826, NSF AST-2307202, and the Multimessenger Plasma Physics Center (MPPC), grant NSF PHY-2206609. B.R. acknowledges the support of the Natural Sciences and Engineering Research Council of Canada (NSERC) funding reference numbers RGPIN-2023-04844 and ALLRP 592744. B.R. is supported by a grant from the Simons Foundation (MP-SCMPS-00001470). B.R. is supported by the Canadian Space Agency (23JWGO2A01). The computational resources and services used in this work were partially provided by SciNet, Compute Ontario, Calcul Quebec and the
Digital Research Alliance of Canada (alliancecan.ca), compute allocation rrg-ripperda and rrg-dhaggard. ML was supported by the NASA Hubble Fellowship Program fellowships, NSF grants AST-2407809 and AST-2606780, and NASA grant 80NSSC22K0817. S.P. also credits use of A.I., specifically claude for help with data visualisation and grammar/word checks.

\bibliographystyle{mnras}
\bibliography{main} 

@ARTICLE{Kantzas2021,
       author = {{Kantzas}, D. and {Markoff}, S. and {Beuchert}, T. and {Lucchini}, M. and {Chhotray}, A. and others},
        title = "{A new lepto-hadronic model applied to the first simultaneous multiwavelength data set for Cygnus X-1}",
      journal = {\mnras},
         year = 2021,
       volume = {500},
       number = {2},
        pages = {2112-2126},
          doi = {10.1093/mnras/staa3349}
}

@ARTICLE{Hakobyan2025,
       author = {{Hakobyan}, H. and {Levinson}, A. and {Sironi}, L. and {Philippov}, A. and {Ripperda}, B.},
        title = "{Reconnection-driven Flares in M87*: Proton--Synchrotron-powered GeV Emission}",
      journal = {\apjl},
         year = 2025,
       volume = {995},
       number = {2},
          eid = {L73},
        pages = {L73},
          doi = {10.3847/2041-8213/ae286a},
archivePrefix = {arXiv},
       eprint = {2507.14002},
 primaryClass = {astro-ph.HE}
}

@ARTICLE{Nalewajko2026,
       author = {{Nalewajko}, K. and {Kapusta}, M. and {Ripperda}, B. and {Philippov}, A.~A.},
        title = "{Distortion of a Relativistic Jet Echoing a Magnetic Flux Eruption}",
      journal = {\apj},
         year = 2026,
       volume = {1003},
       number = {2},
          eid = {174},
        pages = {174},
          doi = {10.3847/1538-4357/ae6653},
archivePrefix = {arXiv},
       eprint = {2604.25996},
 primaryClass = {astro-ph.HE}
}

@ARTICLE{Chernoglazov2021,
       author = {{Chernoglazov}, A. and {Ripperda}, B. and {Philippov}, A.~A.},
        title = "{Dynamic Alignment and Plasmoid Formation in Relativistic Magnetohydrodynamic Turbulence}",
      journal = {\apjl},
         year = 2021,
       volume = {923},
       number = {1},
          eid = {L13},
        pages = {L13},
          doi = {10.3847/2041-8213/ac3afa},
archivePrefix = {arXiv},
       eprint = {2111.08188},
 primaryClass = {astro-ph.HE}
}

@ARTICLE{Davelaar2020,
       author = {{Davelaar}, J. and {Philippov}, A.~A. and {Bromberg}, O. and {Singh}, C.~B.},
        title = "{Particle Acceleration in Kink-unstable Jets}",
      journal = {\apjl},
         year = 2020,
       volume = {896},
       number = {2},
          eid = {L31},
        pages = {L31},
          doi = {10.3847/2041-8213/ab95a2},
archivePrefix = {arXiv},
       eprint = {1910.13370},
 primaryClass = {astro-ph.HE}
}

@ARTICLE{Liska2022trunc,
       author = {{Liska}, M.~T.~P. and {Musoke}, G. and {Tchekhovskoy}, A. and {Porth}, O. and {Beloborodov}, A.~M.},
        title = "{Formation of Magnetically Truncated Accretion Disks in 3D Radiation-transport Two-temperature GRMHD Simulations}",
      journal = {\apjl},
         year = 2022,
       volume = {935},
       number = {1},
          eid = {L1},
        pages = {L1},
          doi = {10.3847/2041-8213/ac84db},
archivePrefix = {arXiv},
       eprint = {2201.03526},
 primaryClass = {astro-ph.HE}
}

@ARTICLE{EHTMWL2024,
       author = {{Algaba}, J.~C. and {Balokovi{\'c}}, M. and {Chandra}, S. and {Cheong}, W.-Y. and {Cui}, Y.-Z. and others},
        title = "{Broadband multi-wavelength properties of M87 during the 2018 EHT campaign including a very high energy flaring episode}",
      journal = {\aap},
         year = 2024,
       volume = {692},
          eid = {A140},
        pages = {A140},
          doi = {10.1051/0004-6361/202450497},
archivePrefix = {arXiv},
       eprint = {2404.17623},
 primaryClass = {astro-ph.HE}
}

@ARTICLE{Gandhi_2008,
       author = {{Gandhi}, P. and {Makishima}, K. and {Durant}, M. and {Fabian}, A.~C. and {Dhillon}, V.~S. and {Marsh}, T.~R. and {Miller}, J.~M. and {Shahbaz}, T. and {Spruit}, H.~C.},
        title = "{Rapid optical and X-ray timing observations of GX 339-4: flux correlations at the onset of a low/hard state}",
      journal = {\mnras},
         year = 2008,
       volume = {390},
       number = {1},
        pages = {L29-L33},
          doi = {10.1111/j.1745-3933.2008.00529.x},
archivePrefix = {arXiv},
       eprint = {0807.1529},
 primaryClass = {astro-ph}
}

@ARTICLE{Gandhi_2010,
       author = {{Gandhi}, P. and {Dhillon}, V.~S. and {Durant}, M. and {Fabian}, A.~C. and {Kubota}, A. and {Makishima}, K. and {Malzac}, J. and {Marsh}, T.~R. and {Miller}, J.~M. and {Shahbaz}, T. and {Spruit}, H.~C. and {Casella}, P.},
        title = "{Rapid optical and X-ray timing observations of GX 339-4: multicomponent optical variability in the low/hard state}",
      journal = {\mnras},
         year = 2010,
       volume = {407},
       number = {4},
        pages = {2166-2192},
          doi = {10.1111/j.1365-2966.2010.17083.x},
archivePrefix = {arXiv},
       eprint = {1005.4685},
 primaryClass = {astro-ph.HE}
}

@ARTICLE{Casella2010,
       author = {{Casella}, P. and {Maccarone}, T.~J. and {O'Brien}, K. and {Fender}, R.~P. and {Russell}, D.~M. and others},
        title = "{Fast infrared variability from a relativistic jet in GX 339-4}",
      journal = {\mnras},
         year = 2010,
       volume = {404},
       number = {1},
        pages = {L21-L25},
          doi = {10.1111/j.1745-3933.2010.00826.x},
archivePrefix = {arXiv},
       eprint = {1002.1233},
 primaryClass = {astro-ph.HE}
}

@ARTICLE{Gandhi_2017,
       author = {{Gandhi}, P. and {Bachetti}, M. and {Dhillon}, V.~S. and {Fender}, R.~P. and {Hardy}, L.~K. and others},
        title = "{An elevation of 0.1 light-seconds for the optical jet base in an accreting Galactic black hole system}",
      journal = {Nature Astronomy},
         year = 2017,
       volume = {1},
       number = {12},
        pages = {859-864},
          doi = {10.1038/s41550-017-0273-3},
archivePrefix = {arXiv},
       eprint = {1710.09838},
 primaryClass = {astro-ph.HE}
}

@article{IceCube2018neutrino,
  title={Neutrino emission from the direction of the blazar TXS 0506+056 prior to the IceCube-170922A alert},
  author={{IceCube Collaboration}},
  journal={Science},
  volume={361},
  number={6398},
  pages={147--151},
  year={2018},
  doi={10.1126/science.aat2890}
}

@article{Shukla2020,
  title={Short-timescale $\gamma$-ray variability in CTA 102},
  author={Shukla, Arti and Mannheim, Karl and Patel, Sanjay and others},
  journal={Nature Communications},
  volume={11},
  number={1},
  pages={4176},
  year={2020},
  doi={10.1038/s41467-020-17912-z}
}

@article{BlandfordZnajek1977,
  title={Electromagnetic extraction of energy from Kerr black holes},
  author={Blandford, Roger D and Znajek, Roman L},
  journal={Monthly Notices of the Royal Astronomical Society},
  volume={179},
  number={3},
  pages={433--456},
  year={1977},
  doi={10.1093/mnras/179.3.433}
}

@article{McKinney2006,
  title={General relativistic magnetohydrodynamic simulations of the jet formation and large-scale propagation from black hole accretion systems},
  author={McKinney, Jonathan C},
  journal={Monthly Notices of the Royal Astronomical Society},
  volume={368},
  number={4},
  pages={1561--1582},
  year={2006},
  doi={10.1111/j.1365-2966.2006.10256.x}
}

@article{Tchekhovskoy2011,
  title={Efficient generation of jets from magnetically arrested accretion on a rapidly spinning black hole},
  author={Tchekhovskoy, Alexander and Narayan, Ramesh and McKinney, Jonathan C},
  journal={Monthly Notices of the Royal Astronomical Society},
  volume={418},
  number={1},
  pages={L79--L83},
  year={2011},
  doi={10.1111/j.1745-3933.2011.01147.x}
}

@article{Hada2016,
  title={High-sensitivity 86 GHz (3.5 mm) VLBI observations of M87: Deep imaging of the jet base at a resolution of 10 Schwarzschild radii},
  author={Hada, Kazuhiro and Kino, Motoki and Doi, Akihiro and others},
  journal={The Astrophysical Journal},
  volume={817},
  number={2},
  pages={131},
  year={2016},
  doi={10.3847/0004-637X/817/2/131}
}

@article{Kim2018,
  title={The limb-brightened jet of M87 down to the 7 Schwarzschild radii scale},
  author={Kim, Jae-Young and Krichbaum, Thomas P and Roy, Alan L and others},
  journal={Astronomy \& Astrophysics},
  volume={616},
  pages={A188},
  year={2018},
  doi={10.1051/0004-6361/201832921}
}

@article{Curtis1918,
  title={Descriptions of 762 nebulae and clusters photographed with the Crossley reflector},
  author={Curtis, Heber Doust},
  journal={Publications of Lick Observatory},
  volume={13},
  pages={9--42},
  year={1918}
}

@ARTICLE{Baade1956,
       author = {{Baade}, W.},
        title = "{Polarization in the Jet of Messier 87.}",
      journal = {\apj},
         year = 1956,
        month = may,
       volume = {123},
        pages = {550-551},
          doi = {10.1086/146194},
       adsurl = {https://ui.adsabs.harvard.edu/abs/1956ApJ...123..550B}
}

@article{Schmidt1963,
  title={3C 273: a star-like object with large red-shift},
  author={Schmidt, Maarten},
  journal={Nature},
  volume={197},
  number={4872},
  pages={1040},
  year={1963},
  doi={10.1038/1971040a0}
}

@article{Lynden1971,
  title={Violent phenomena in the nuclei of galaxies},
  author={Lynden-Bell, Donald},
  journal={Monthly Notices of the Royal Astronomical Society},
  volume={155},
  number={1},
  pages={95--118},
  year={1971},
  doi={10.1093/mnras/155.1.95}
}

@article{Mirabel1994,
  title={A superluminal source in the Galaxy},
  author={Mirabel, I F{\'e}lix and Rodr{\'\i}guez, Luis F},
  journal={Nature},
  volume={371},
  number={6492},
  pages={46--48},
  year={1994},
  doi={10.1038/371046a0}
}

@misc{Fender2004,
      title={Jets from X-ray binaries}, 
      author={Rob Fender},
      year={2003},
      eprint={astro-ph/0303339},
      archivePrefix={arXiv},
      primaryClass={astro-ph},
      url={https://arxiv.org/abs/astro-ph/0303339}, 
}

@article{Bridle1984,
  title={Extragalactic radio jets},
  author={Bridle, Alan H and Perley, Richard A},
  journal={Annual Review of Astronomy and Astrophysics},
  volume={22},
  number={1},
  pages={319--358},
  year={1984},
  doi={10.1146/annurev.aa.22.090184.001535}
}

@article{Begelman1984,
  title={Theory of extragalactic radio sources},
  author={Begelman, Mitchell C and Blandford, Roger D and Rees, Martin J},
  journal={Reviews of Modern Physics},
  volume={56},
  number={2},
  pages={255--351},
  year={1984},
  doi={10.1103/RevModPhys.56.255}
}

@article{Mooley2018,
  title={Superluminal motion of a relativistic jet in the neutron-star merger GW170817},
  author={Mooley, Kunal P and Deller, Adam T and Gottlieb, Ore and others},
  journal={Nature},
  volume={561},
  number={7723},
  pages={355--359},
  year={2018},
  doi={10.1038/s41586-018-0486-3}
}

@article{Bloom2011,
  title={A possible relativistic jetted outburst from a massive black hole fed by a tidally disrupted star},
  author={Bloom, Joshua S and Giannios, Dimitrios and Metzger, Brian D and others},
  journal={Science},
  volume={333},
  number={6039},
  pages={203--206},
  year={2011},
  doi={10.1126/science.1207150}
}

@article{Mizuno2012,
  title={Recollimation shocks in magnetized relativistic jets},
  author={Mizuno, Yosuke and Pohl, Martin and Niemiec, Jacek and others},
  journal={The Astrophysical Journal},
  volume={757},
  number={1},
  pages={16},
  year={2012},
  doi={10.1088/0004-637X/757/1/16}
}

@article{Mbarek2019,
  title={Reconnection-driven flares in 3D black hole magnetospheres},
  author={Mbarek, Razieh and Sironi, Lorenzo},
  journal={The Astrophysical Journal},
  volume={886},
  number={1},
  pages={8},
  year={2019},
  doi={10.3847/1538-4357/ab4a08}
}

@article{Liska2018,
  title={Formation of precessing jets by tilted black hole discs in 3D general relativistic MHD simulations},
  author={Liska, Matthew and Hesp, Chris and Tchekhovskoy, Alexander and others},
  journal={Monthly Notices of the Royal Astronomical Society},
  volume={474},
  number={1},
  pages={L81--L85},
  year={2018},
  doi={10.1093/mnrasl/slx174}
}

@article{Mbarek2022,
   title={Relativistic Asymmetric Magnetic Reconnection},
   volume={128},
   ISSN={1079-7114},
   url={http://dx.doi.org/10.1103/PhysRevLett.128.145101},
   DOI={10.1103/physrevlett.128.145101},
   number={14},
   journal={Physical Review Letters},
   publisher={American Physical Society (APS)},
   author={Mbarek, Rostom and Haggerty, Colby and Sironi, Lorenzo and Shay, Michael and Caprioli, Damiano},
   year={2022},
   month=apr }

@article{Chow2023,
   title={The Kelvin–Helmholtz Instability at the Boundary of Relativistic Magnetized Jets},
   volume={951},
   ISSN={2041-8213},
   url={http://dx.doi.org/10.3847/2041-8213/acdfcf},
   DOI={10.3847/2041-8213/acdfcf},
   number={2},
   journal={The Astrophysical Journal Letters},
   publisher={American Astronomical Society},
   author={Chow, Anthony and Davelaar, Jordy and Rowan, Michael E. and Sironi, Lorenzo},
   year={2023},
   month=jul, pages={L23} }

@ARTICLE{Ripperda2020,
       author = {{Ripperda}, B. and {Bacchini}, F. and {Philippov}, A.~A.},
        title = "{Magnetic Reconnection and Hot Spot Formation in Black Hole Accretion Disks}",
      journal = {\apj},
         year = 2020,
       volume = {900},
       number = {2},
          eid = {100},
        pages = {100},
          doi = {10.3847/1538-4357/ababab},
archivePrefix = {arXiv},
       eprint = {2003.04330},
 primaryClass = {astro-ph.HE}
}

@ARTICLE{Hakobyan2023,
       author = {{Hakobyan}, H. and {Ripperda}, B. and {Philippov}, A.~A.},
        title = "{Radiative Reconnection-powered TeV Flares from the Black Hole Magnetosphere in M87}",
      journal = {\apjl},
         year = 2023,
       volume = {943},
       number = {2},
          eid = {L29},
        pages = {L29},
          doi = {10.3847/2041-8213/acb264},
archivePrefix = {arXiv},
       eprint = {2209.02105},
 primaryClass = {astro-ph.HE}
}

@ARTICLE{McKinney2007,
       author = {{McKinney}, Jonathan C. and {Narayan}, Ramesh},
        title = "{Disc-jet coupling in black hole accretion systems - I. General relativistic magnetohydrodynamical models}",
      journal = {\mnras},
         year = 2007,
        month = feb,
       volume = {375},
       number = {2},
        pages = {513-530},
          doi = {10.1111/j.1365-2966.2006.11301.x},
archivePrefix = {arXiv},
       eprint = {astro-ph/0607575},
 primaryClass = {astro-ph},
       adsurl = {https://ui.adsabs.harvard.edu/abs/2007MNRAS.375..513M}
}

@article{EHT2021magnetic,
  title={First M87 Event Horizon Telescope results. VIII. Magnetic field structure near the event horizon},
  author={{Event Horizon Telescope Collaboration}},
  journal={The Astrophysical Journal Letters},
  volume={910},
  number={1},
  pages={L13},
  year={2021},
  doi={10.3847/2041-8213/abe4de}
}

@article{Gammie2003,
  title={HARM: a numerical scheme for general relativistic magnetohydrodynamics},
  author={Gammie, Charles F and McKinney, Jonathan C and T{\'o}th, G{\'a}bor},
  journal={The Astrophysical Journal},
  volume={589},
  number={1},
  pages={444--457},
  year={2003},
  doi={10.1086/374594}
}

@article{Ryan2017,
  title={The structure and evolution of a trans-Alfvénic jet driven by a black hole},
  author={Ryan, Benjamin R and Ressler, Sean M and Dolence, Joshua C and others},
  journal={The Astrophysical Journal Letters},
  volume={844},
  number={2},
  pages={L24},
  year={2017},
  doi={10.3847/2041-8213/aa8034}
}

@article{Wong2021tracers,
  title={Tracing jet--disk mass exchange in three-dimensional GRMHD simulations with passive scalars},
  author={Wong, George N and Prather, Ben S and Dhruv, Vikram and others},
  journal={The Astrophysical Journal},
  volume={914},
  number={1},
  pages={55},
  year={2021},
  doi={10.3847/1538-4357/abf8b8}
}

@article{Boettcher2013,
  title={Leptonic and hadronic modeling of Fermi-detected blazars},
  author={B{\"o}ttcher, Markus and Reimer, Anita and Sweeney, Kyle and others},
  journal={The Astrophysical Journal},
  volume={768},
  number={1},
  pages={54},
  year={2013},
  doi={10.1088/0004-637X/768/1/54}
}

@article{Cerruti2015,
  title={A hadronic origin for ultra-high-frequency-peaked BL Lac objects},
  author={Cerruti, Matteo and Zech, Andreas and Boisson, Catherine and others},
  journal={Monthly Notices of the Royal Astronomical Society},
  volume={448},
  number={1},
  pages={910--927},
  year={2015},
  doi={10.1093/mnras/stu2691}
}

@article{Ansoldi2018magic,
  title={The Blazar TXS 0506+056 Associated with a High-energy Neutrino: Insights into Extragalactic Jets and Cosmic-Ray Acceleration},
  author={Ansoldi, S and Antonelli, LA and Arcaro, C and others},
  journal={The Astrophysical Journal Letters},
  volume={863},
  number={1},
  pages={L10},
  year={2018},
  doi={10.3847/2041-8213/aad083}
}

@article{Petropoulou2015hadronic,
  title={Hadronic supercriticality as a trigger for $\gamma$-ray emission from blazars},
  author={Petropoulou, Maria and Dimitrakoudis, Stavros and Padovani, Paolo and others},
  journal={Monthly Notices of the Royal Astronomical Society},
  volume={448},
  number={3},
  pages={2412--2429},
  year={2015},
  doi={10.1093/mnras/stv179}
}

@article{Ackermann2016cosmic,
  title={Resolving the extragalactic $\gamma$-ray background above 50 GeV with the Fermi Large Area Telescope},
  author={Ackermann, M and Ajello, M and Albert, A and others},
  journal={Physical Review Letters},
  volume={116},
  number={15},
  pages={151105},
  year={2016},
  doi={10.1103/PhysRevLett.116.151105}
}

@article{Aartsen2017lowenergy,
  title={The contribution of Fermi-2LAC blazars to the diffuse TeV-PeV neutrino flux},
  author={Aartsen, MG and Abraham, K and Ackermann, M and others},
  journal={The Astrophysical Journal},
  volume={835},
  number={1},
  pages={45},
  year={2017},
  doi={10.3847/1538-4357/835/1/45}
}

@article{Stecker2013cosmic,
  title={Cosmic ray proton-initiated cascade models for the extragalactic gamma-ray background},
  author={Stecker, Floyd W and Venters, Tonia M},
  journal={The Astrophysical Journal},
  volume={736},
  number={1},
  pages={40},
  year={2013},
  doi={10.1088/0004-637X/736/1/40}
}

@article{Sahakyan2018,
   title={Lepto-hadronic γ-Ray and Neutrino Emission from the Jet of TXS 0506+056},
   volume={866},
   ISSN={1538-4357},
   url={http://dx.doi.org/10.3847/1538-4357/aadade},
   DOI={10.3847/1538-4357/aadade},
   number={2},
   journal={The Astrophysical Journal},
   publisher={American Astronomical Society},
   author={Sahakyan, Narek},
   year={2018},
   month=oct, pages={109} }

@article{Porth2019event,
  title={The event horizon general relativistic magnetohydrodynamic code comparison project},
  author={Porth, Oliver and Chatterjee, Koushik and Narayan, Ramesh and others},
  journal={The Astrophysical Journal Supplement Series},
  volume={243},
  number={2},
  pages={26},
  year={2019},
  doi={10.3847/1538-4365/ab29fd}
}

@article{Abbasi2022time,
  title={Time-integrated neutrino source searches with 10 years of IceCube data},
  author={Abbasi, R and Ackermann, M and Adams, J and others},
  journal={Physical Review Letters},
  volume={129},
  number={5},
  pages={051103},
  year={2022},
  doi={10.1103/PhysRevLett.129.051103}
}

@article{Eichmann_2022,
   title={Solving the Multimessenger Puzzle of the AGN-starburst Composite Galaxy NGC 1068},
   volume={939},
   ISSN={1538-4357},
   url={http://dx.doi.org/10.3847/1538-4357/ac9588},
   DOI={10.3847/1538-4357/ac9588},
   number={1},
   journal={The Astrophysical Journal},
   publisher={American Astronomical Society},
   author={Eichmann, Björn and Oikonomou, Foteini and Salvatore, Silvia and Dettmar, Ralf-Jürgen and Tjus, Julia Becker},
   year={2022},
   month=nov, pages={43} }

@article{Chatterjee_2019J,
   title={Accelerating AGN jets to parsec scales using general relativistic MHD simulations},
   volume={490},
   ISSN={1365-2966},
   url={http://dx.doi.org/10.1093/mnras/stz2626},
   DOI={10.1093/mnras/stz2626},
   number={2},
   journal={Monthly Notices of the Royal Astronomical Society},
   publisher={Oxford University Press (OUP)},
   author={Chatterjee, K and Liska, M and Tchekhovskoy, A and Markoff, S B},
   year={2019},
   month=sep, pages={2200–2218} }

@article{Zhdankin_2023,
   title={Particle acceleration by magnetic Rayleigh-Taylor instability: Mechanism for flares in black hole accretion flows},
   volume={5},
   ISSN={2643-1564},
   url={http://dx.doi.org/10.1103/PhysRevResearch.5.043023},
   DOI={10.1103/physrevresearch.5.043023},
   number={4},
   journal={Physical Review Research},
   publisher={American Physical Society (APS)},
   author={Zhdankin, Vladimir and Ripperda, Bart and Philippov, Alexander A.},
   year={2023},
   month=oct }

@article{Kerr1963,
  author = {Kerr, Roy P.},
  title = {Gravitational field of a spinning mass as an example of algebraically special metrics},
  journal = {Physical Review Letters},
  volume = {11},
  number = {5},
  pages = {237--238},
  year = {1963},
  publisher = {American Physical Society},
  doi = {10.1103/PhysRevLett.11.237}
}

@article{Schild1967,
  author = {Schild, Alfred},
  title = {Electromagnetic field of a particle moving in a Kerr metric},
  journal = {Physical Review},
  volume = {160},  
  number = {5},
  pages = {1177--1179},
  year = {1967},
  publisher = {American Physical Society},
  doi = {10.1103/PhysRev.160.1177}
}

@ARTICLE{Liska2022,
       author = {{Liska}, M.~T.~P. and {Chatterjee}, K. and {Issa}, D. and {Yoon}, D. and {Kaaz}, N. and {Tchekhovskoy}, A. and {van Eijnatten}, D. and {Musoke}, G. and {Hesp}, C. and {Rohoza}, V. and {Markoff}, S. and {Ingram}, A. and {van der Klis}, M.},
        title = "{H-AMR: A New GPU-accelerated GRMHD Code for Exascale Computing with 3D Adaptive Mesh Refinement and Local Adaptive Time Stepping}",
      journal = {\apjs},
         year = 2022,
       volume = {263},
       number = {2},
          eid = {26},
        pages = {26},
          doi = {10.3847/1538-4365/ac9966},
archivePrefix = {arXiv},
       eprint = {1912.10192},
 primaryClass = {astro-ph.HE}
}

@article{Fishbone1976,
  title={Relativistic fluid disks in orbit around Kerr black holes},
  author={Fishbone, L. G. and Moncrief, V.},
  journal={The Astrophysical Journal},
  volume={207},
  pages={962--976},
  year={1976},
  doi={10.1086/154565}
}

@article{Beskin1992,
  title={Filling the magnetosphere of a supermassive black-hole with plasma},
  author={Beskin, V. S. and Istomin, Ya N. and Par'ev, V. I.},
  journal={Soviet Astronomy},
  volume={36},
  pages={642},
  year={1992}
}

@article{Levinson2000,
  title={Particle Acceleration and Curvature TeV Emission by Rotating, Supermassive Black Holes},
  author={Levinson, A.},
  journal={Physical Review Letters},
  volume={85},
  pages={912--915},
  year={2000},
  doi={10.1103/PhysRevLett.85.912}
}

@article{Hirotani2016,
  title={Lepton Acceleration in the Vicinity of the Event Horizon: Very High Energy Emissions from Supermassive Black Holes},
  author={Hirotani, K. and Pu, H.-Y.},
  journal={The Astrophysical Journal},
  volume={818},
  pages={50},
  year={2016},
  doi={10.3847/0004-637X/818/1/50}
}

@article{Parfrey2019,
  title={First-Principles Plasma Simulations of Black-Hole Jet Launching},
  author={Parfrey, K. and Philippov, A. and Cerutti, B.},
  journal={Physical Review Letters},
  volume={122},
  pages={035101},
  year={2019},
  doi={10.1103/PhysRevLett.122.035101}
}

@article{Crinquand2020,
  title={Multidimensional Simulations of Ergospheric Pair Discharges around Black Holes},
  author={Crinquand, B. and Cerutti, B. and Philippov, A. and Parfrey, K. and Dubus, G.},
  journal={Physical Review Letters},
  volume={124},
  pages={145101},
  year={2020},
  doi={10.1103/PhysRevLett.124.145101}
}

@article{Crinquand2021,
  title={Synthetic gamma-ray lightcurves from pair-producing gap discharges in black hole magnetospheres},
  author={Crinquand, B. and Cerutti, B. and Dubus, G. and Parfrey, K. and Philippov, A.},
  journal={Astronomy \& Astrophysics},
  volume={650},
  pages={A163},
  year={2021},
  doi={10.1051/0004-6361/202040158}
}

@article{McKinney2012,
  title={General relativistic magnetohydrodynamic simulations of magnetically choked accretion flows around black holes},
  author={McKinney, J. C. and Tchekhovskoy, A. and Blandford, R. D.},
  journal={Monthly Notices of the Royal Astronomical Society},
  volume={423},
  pages={3083--3117},
  year={2012},
  doi={10.1111/j.1365-2966.2012.21074.x}
}

@article{Keppens2014,
  title={MPI-AMRVAC for Solar and Astrophysics},
  author={Keppens, R. and Porth, O. and Xia, C. and Teunissen, J. and Monceau-Baroux, R. and et al.},
  journal={The Astrophysical Journal Supplement Series},
  volume={214},
  pages={4},
  year={2014},
  doi={10.1088/0067-0049/214/1/4}
}

@article{BALSARA2001614,
title = {Divergence-Free Adaptive Mesh Refinement for Magnetohydrodynamics},
journal = {Journal of Computational Physics},
volume = {174},
number = {2},
pages = {614-648},
year = {2001},
issn = {0021-9991},
doi = {https://doi.org/10.1006/jcph.2001.6917},
url = {https://www.sciencedirect.com/science/article/pii/S0021999101969177},
author = {Dinshaw S. Balsara}
}

@ARTICLE{Courant,
       author = {{Courant}, R. and {Friedrichs}, K. and {Lewy}, H.},
        title = "{{\"U}ber die partiellen Differenzengleichungen der mathematischen Physik}",
      journal = {Mathematische Annalen},
         year = 1928,
        month = jan,
       volume = {100},
        pages = {32-74},
          doi = {10.1007/BF01448839},
       adsurl = {https://ui.adsabs.harvard.edu/abs/1928MatAn.100...32C}
}

@article{Berger1984,
  title={Adaptive mesh refinement for hyperbolic partial differential equations},
  author={Berger, Marsha J. and Oliger, Joseph},
  journal={Journal of Computational Physics},
  volume={53},
  pages={484--512},
  year={1984},
  doi={10.1016/0021-9991(84)90073-1}
}

@article{Berger1989,
  title={A conservative adaptive projection method for the variable density incompressible Navier-Stokes equations},
  author={Berger, Marsha J. and Colella, Phillip},
  journal={Journal of Computational Physics},
  volume={82},
  pages={64--84},
  year={1989},
  doi={10.1016/0021-9991(89)90035-1}
}

@article{Moscibrodzka2013,
  title={General relativistic magnetohydrodynamical simulations of the jet in M 87},
  author={Mościbrodzka, Monika and Falcke, Heino and Shiokawa, Hotaka},
  journal={Astronomy \& Astrophysics},
  volume={586},
  pages={A38},
  year={2016},
  doi={10.1051/0004-6361/201526630}
}

@article{Nalewajko2014,
  title={Energetic constraints on a rapid gamma-ray flare in PKS 1222+216},
  author={Nalewajko, Krzysztof and Begelman, Mitchell C. and Cerutti, Benoît and Uzdensky, Dmitri A. and Sikora, Marek},
  journal={The Astrophysical Journal},
  volume={789},
  pages={161},
  year={2014},
  doi={10.1088/0004-637X/789/2/161}
}

@article{Ripperda2021,
  title={Current Driven Instability and Particle Acceleration in Relativistic Jets},
  author={Ripperda, Bart and Sironi, Lorenzo and Spitkovsky, Anatoly},
  journal={The Astrophysical Journal Letters},
  volume={918},
  pages={L34},
  year={2021},
  doi={10.3847/2041-8213/ac22af}
}

@article{Harten1983,
  title={On upstream differencing and Godunov-type schemes for hyperbolic conservation laws},
  author={Harten, A. and Lax, P. D. and van Leer, B.},
  journal={SIAM Review},
  volume={25},
  pages={35--61},
  year={1983},
  doi={10.1137/1025002}
}

@article{Gardiner2005,
  title={An unsplit Godunov method for ideal MHD via constrained transport},
  author={Gardiner, Thomas A. and Stone, James M.},
  journal={Journal of Computational Physics},
  volume={205},
  pages={509--539},
  year={2005},
  doi={10.1016/j.jcp.2004.11.016}
}

@article{Colella1984,
  title={The piecewise parabolic method (PPM) for gas-dynamical simulations},
  author={Colella, Phillip and Woodward, Paul R.},
  journal={Journal of Computational Physics},
  volume={54},
  pages={174--201},
  year={1984},
  doi={10.1016/0021-9991(84)90143-8}
}

@article{Narayan2003,
  title={Magnetically Arrested Disk: an Energetically Efficient Accretion Flow},
  author={Narayan, Ramesh and Igumenshchev, Igor V. and Abramowicz, Marek A.},
  journal={Publications of the Astronomical Society of Japan},
  volume={55},
  pages={L69--L72},
  year={2003},
  doi={10.1093/pasj/55.6.L69}
}

@article{Igumenshchev2003,
  title={Three-dimensional Magnetohydrodynamic Simulations of Radiatively Inefficient Accretion Flows},
  author={Igumenshchev, Igor V. and Narayan, Ramesh and Abramowicz, Marek A.},
  journal={The Astrophysical Journal},
  volume={592},
  pages={1042--1059},
  year={2003},
  doi={10.1086/375769}
}

@article{Chatterjee_2022,
   title={Flux Eruption Events Drive Angular Momentum Transport in Magnetically Arrested Accretion Flows},
   volume={941},
   ISSN={1538-4357},
   url={http://dx.doi.org/10.3847/1538-4357/ac9d97},
   DOI={10.3847/1538-4357/ac9d97},
   number={1},
   journal={The Astrophysical Journal},
   publisher={American Astronomical Society},
   author={Chatterjee, K. and Narayan, R.},
   year={2022},
   month=dec, pages={30} }

@article{eht2021,
  author={{Event Horizon Telescope Collaboration}},
  title={First M87 Event Horizon Telescope Results. VII. Polarization of the Ring},
  journal={The Astrophysical Journal Letters},
  volume={910},
  number={1},
  pages={L12},
  year={2021},
  doi={10.3847/2041-8213/abe71d}
}

@ARTICLE{Ripperda2022,
       author = {{Ripperda}, B. and {Liska}, M. and {Chatterjee}, K. and {Musoke}, G. and {Philippov}, A.~A. and {Markoff}, S.~B. and {Tchekhovskoy}, A. and {Younsi}, Z.},
        title = "{Black Hole Flares: Ejection of Accreted Magnetic Flux through 3D Plasmoid-mediated Reconnection}",
      journal = {\apjl},
         year = 2022,
       volume = {924},
       number = {2},
          eid = {L32},
        pages = {L32},
          doi = {10.3847/2041-8213/ac46a1},
archivePrefix = {arXiv},
       eprint = {2109.15115},
 primaryClass = {astro-ph.HE}
}

@misc{salas2024resolutionanalysismagneticallyarrested,
      title={Resolution analysis of magnetically arrested disk simulations}, 
      author={León Salas and Gibwa Musoke and Koushik Chatterjee and Sera Markoff and Oliver Porth and Matthew Liska and Bart Ripperda},
      year={2024},
      eprint={2405.00564},
      archivePrefix={arXiv},
      primaryClass={astro-ph.HE},
      url={https://arxiv.org/abs/2405.00564}, 
}

@misc{cendes2024ubiquitouslateradioemission,
      title={Ubiquitous Late Radio Emission from Tidal Disruption Events}, 
      author={Yvette Cendes and Edo Berger and Kate D. Alexander and Ryan Chornock and Raffaella Margutti and Brian Metzger and Mark H. Wieringa and Michael F. Bietenholz and Aprajita Hajela and Tanmoy Laskar and Michael C. Stroh and Giacomo Terreran},
      year={2024},
      eprint={2308.13595},
      archivePrefix={arXiv},
      primaryClass={astro-ph.HE},
      url={https://arxiv.org/abs/2308.13595}, 
}

@misc{abbasi2024searchneutrinoemissionhard,
      title={Search for neutrino emission from hard X-ray AGN with IceCube}, 
      author={R. Abbasi and M. Ackermann and J. Adams and S. K. Agarwalla and J. A. Aguilar and M. Ahlers and J. M. Alameddine and N. M. Amin and K. Andeen and C. Argüelles and Y. Ashida and S. Athanasiadou and L. Ausborm and S. N. Axani and X. Bai and A. Balagopal V. and M. Baricevic and S. W. Barwick and S. Bash and V. Basu and R. Bay and J. J. Beatty and J. Becker Tjus and J. Beise and C. Bellenghi and C. Benning and S. BenZvi and D. Berley and E. Bernardini and D. Z. Besson and E. Blaufuss and L. Bloom and S. Blot and F. Bontempo and J. Y. Book Motzkin and C. Boscolo Meneguolo and S. Böser and O. Botner and J. Böttcher and J. Braun and B. Brinson and J. Brostean-Kaiser and L. Brusa and R. T. Burley and D. Butterfield and M. A. Campana and I. Caracas and K. Carloni and J. Carpio and S. Chattopadhyay and N. Chau and Z. Chen and D. Chirkin and S. Choi and B. A. Clark and A. Coleman and G. H. Collin and A. Connolly and J. M. Conrad and P. Coppin and R. Corley and P. Correa and D. F. Cowen and P. Dave and C. De Clercq and J. J. DeLaunay and D. Delgado and S. Deng and A. Desai and P. Desiati and K. D. de Vries and G. de Wasseige and T. DeYoung and A. Diaz and J. C. Díaz-Vélez and P. Dierichs and M. Dittmer and A. Domi and L. Draper and H. Dujmovic and K. Dutta and M. A. DuVernois and T. Ehrhardt and L. Eidenschink and A. Eimer and P. Eller and E. Ellinger and S. El Mentawi and D. Elsässer and R. Engel and H. Erpenbeck and J. Evans and P. A. Evenson and K. L. Fan and K. Fang and K. Farrag and A. R. Fazely and A. Fedynitch and N. Feigl and S. Fiedlschuster and C. Finley and L. Fischer and D. Fox and A. Franckowiak and S. Fukami and P. Fürst and J. Gallagher and E. Ganster and A. Garcia and M. Garcia and G. Garg and E. Genton and L. Gerhardt and A. Ghadimi and C. Girard-Carillo and C. Glaser and T. Glüsenkamp and J. G. Gonzalez and S. Goswami and A. Granados and D. Grant and S. J. Gray and O. Gries and S. Griffin and S. Griswold and K. M. Groth and C. Günther and P. Gutjahr and C. Ha and C. Haack and A. Hallgren and L. Halve and F. Halzen and H. Hamdaoui and M. Ha Minh and M. Handt and K. Hanson and J. Hardin and A. A. Harnisch and P. Hatch and A. Haungs and J. Häußler and K. Helbing and J. Hellrung and J. Hermannsgabner and L. Heuermann and N. Heyer and S. Hickford and A. Hidvegi and C. Hill and G. C. Hill and K. D. Hoffman and S. Hori and K. Hoshina and M. Hostert and W. Hou and T. Huber and K. Hultqvist and M. Hünnefeld and R. Hussain and K. Hymon and A. Ishihara and W. Iwakiri and M. Jacquart and S. Jain and O. Janik and M. Jansson and G. S. Japaridze and M. Jeong and M. Jin and B. J. P. Jones and N. Kamp and D. Kang and W. Kang and X. Kang and A. Kappes and D. Kappesser and L. Kardum and T. Karg and M. Karl and A. Karle and A. Katil and U. Katz and M. Kauer and J. L. Kelley and M. Khanal and A. Khatee Zathul and A. Kheirandish and J. Kiryluk and S. R. Klein and A. Kochocki and R. Koirala and H. Kolanoski and T. Kontrimas and L. Köpke and C. Kopper and D. J. Koskinen and P. Koundal and M. Kovacevich and M. Kowalski and T. Kozynets and J. Krishnamoorthi and K. Kruiswijk and E. Krupczak and A. Kumar and E. Kun and N. Kurahashi and N. Lad and C. Lagunas Gualda and M. Lamoureux and M. J. Larson and S. Latseva and F. Lauber and J. P. Lazar and J. W. Lee and K. Leonard DeHolton and A. Leszczyńska and J. Liao and M. Lincetto and Y. T. Liu and M. Liubarska and C. Love and C. J. Lozano Mariscal and L. Lu and F. Lucarelli and W. Luszczak and Y. Lyu and J. Madsen and E. Magnus and K. B. M. Mahn and Y. Makino and E. Manao and S. Mancina and W. Marie Sainte and I. C. Mariş and S. Marka and Z. Marka and M. Marsee and I. Martinez-Soler and R. Maruyama and F. Mayhew and F. McNally and J. V. Mead and K. Meagher and S. Mechbal and A. Medina and M. Meier and Y. Merckx and L. Merten and J. Micallef and J. Mitchell and T. Montaruli and R. W. Moore and Y. Morii and R. Morse and M. Moulai and T. Mukherjee and R. Naab and R. Nagai and M. Nakos and U. Naumann and J. Necker and A. Negi and L. Neste and M. Neumann and H. Niederhausen and M. U. Nisa and K. Noda and A. Noell and A. Novikov and A. Obertacke Pollmann and V. O'Dell and B. Oeyen and A. Olivas and R. Orsoe and J. Osborn and E. O'Sullivan and V. Palusova and H. Pandya and N. Park and G. K. Parker and E. N. Paudel and L. Paul and C. Pérez de los Heros and T. Pernice and J. Peterson and S. Philippen and A. Pizzuto and M. Plum and A. Pontén and Y. Popovych and M. Prado Rodriguez and B. Pries and G. C. Privon and R. Procter-Murphy and G. T. Przybylski and C. Raab and J. Rack-Helleis and M. Ravn and K. Rawlins and Z. Rechav and A. Rehman and P. Reichherzer and E. Resconi and S. Reusch and W. Rhode and B. Riedel and A. Rifaie and E. J. Roberts and S. Robertson and S. Rodan and G. Roellinghoff and M. Rongen and A. Rosted and C. Rott and T. Ruhe and L. Ruohan and D. Ryckbosch and I. Safa and J. Saffer and D. Salazar-Gallegos and P. Sampathkumar and A. Sandrock and M. Santander and S. Sarkar and S. Sarkar and J. Savelberg and P. Savina and P. Schaile and M. Schaufel and H. Schieler and S. Schindler and L. Schlickmann and B. Schlüter and F. Schlüter and N. Schmeisser and T. Schmidt and J. Schneider and F. G. Schröder and L. Schumacher and S. Sclafani and D. Seckel and M. Seikh and M. Seo and S. Seunarine and P. Sevle Myhr and R. Shah and S. Shefali and N. Shimizu and M. Silva and B. Skrzypek and B. Smithers and R. Snihur and J. Soedingrekso and A. Søgaard and D. Soldin and P. Soldin and G. Sommani and C. Spannfellner and G. M. Spiczak and C. Spiering and M. Stamatikos and T. Stanev and T. Stezelberger and T. Stürwald and T. Stuttard and G. W. Sullivan and I. Taboada and S. Ter-Antonyan and A. Terliuk and M. Thiesmeyer and W. G. Thompson and J. Thwaites and S. Tilav and K. Tollefson and C. Tönnis and S. Toscano and D. Tosi and A. Trettin and R. Turcotte and J. P. Twagirayezu and M. A. Unland Elorrieta and A. K. Upadhyay and K. Upshaw and A. Vaidyanathan and N. Valtonen-Mattila and J. Vandenbroucke and N. van Eijndhoven and D. Vannerom and J. van Santen and J. Vara and F. Varsi and J. Veitch-Michaelis and M. Venugopal and M. Vereecken and S. Verpoest and D. Veske and A. Vijai and C. Walck and A. Wang and C. Weaver and P. Weigel and A. Weindl and J. Weldert and A. Y. Wen and C. Wendt and J. Werthebach and M. Weyrauch and N. Whitehorn and C. H. Wiebusch and D. R. Williams and L. Witthaus and A. Wolf and M. Wolf and G. Wrede and X. W. Xu and J. P. Yanez and E. Yildizci and S. Yoshida and R. Young and S. Yu and T. Yuan and Z. Zhang and P. Zhelnin and P. Zilberman and M. Zimmerman},
      year={2024},
      eprint={2406.06684},
      archivePrefix={arXiv},
      primaryClass={astro-ph.HE},
      url={https://arxiv.org/abs/2406.06684}, 
}

@ARTICLE{Biretta,
       author = {{Biretta}, J.~A. and {Sparks}, W.~B. and {Macchetto}, F.},
        title = "{Hubble Space Telescope Observations of Superluminal Motion in the M87 Jet}",
      journal = {\apj},
         year = 1999,
        month = aug,
       volume = {520},
       number = {2},
        pages = {621-626},
          doi = {10.1086/307499},
       adsurl = {https://ui.adsabs.harvard.edu/abs/1999ApJ...520..621B}
}

@ARTICLE{Reid,
       author = {{Reid}, M.~J. and {Biretta}, J.~A. and {Junor}, W. and {Muxlow}, T.~W.~B. and {Spencer}, R.~E.},
        title = "{Subluminal Motion and Limb Brightening in the Nuclear Jet of M87}",
      journal = {\apj},
         year = 1989,
        month = jan,
       volume = {336},
        pages = {112},
          doi = {10.1086/166998},
       adsurl = {https://ui.adsabs.harvard.edu/abs/1989ApJ...336..112R}
}

@ARTICLE{1989ApJ...342..128B,
       author = {{Biretta}, J.~A. and {Owen}, F.~N. and {Cornwell}, T.~J.},
        title = "{A Search for Motion and Flux Variations in the M87 Jet}",
      journal = {\apj},
         year = 1989,
        month = jul,
       volume = {342},
        pages = {128},
          doi = {10.1086/167581},
       adsurl = {https://ui.adsabs.harvard.edu/abs/1989ApJ...342..128B}
}

@article{Sironi_2021,
   title={Reconnection-driven Particle Acceleration in Relativistic Shear Flows},
   volume={907},
   ISSN={2041-8213},
   url={http://dx.doi.org/10.3847/2041-8213/abd9bc},
   DOI={10.3847/2041-8213/abd9bc},
   number={2},
   journal={The Astrophysical Journal Letters},
   publisher={American Astronomical Society},
   author={Sironi, Lorenzo and Rowan, Michael E. and Narayan, Ramesh},
   year={2021},
   month=feb, pages={L44} }

@misc{figueiredo2024compositionasymmetricshearedrelativisticmagnetic,
      title={Composition-asymmetric and sheared relativistic magnetic reconnection}, 
      author={Enzo Figueiredo and Benoît Cerutti and John Mehlhaff and Nicolas Scepi},
      year={2024},
      eprint={2409.13495},
      archivePrefix={arXiv},
      primaryClass={astro-ph.HE},
      url={https://arxiv.org/abs/2409.13495}, 
}

@ARTICLE{Jiang2014,
       author = {{Jiang}, Yan-Fei and {Stone}, James M. and {Davis}, Shane W.},
        title = "{A Global Three-dimensional Radiation Magneto-hydrodynamic Simulation of Super-Eddington Accretion Disks}",
      journal = {\apj},
         year = 2014,
        month = dec,
       volume = {796},
       number = {2},
          eid = {106},
        pages = {106},
          doi = {10.1088/0004-637X/796/2/106},
archivePrefix = {arXiv},
       eprint = {1410.0678},
 primaryClass = {astro-ph.HE},
       adsurl = {https://ui.adsabs.harvard.edu/abs/2014ApJ...796..106J}
}

@article{Trigo_2013,
   title={Baryons in the relativistic jets of the stellar-mass black-hole candidate 4U 1630-47},
   volume={504},
   ISSN={1476-4687},
   url={http://dx.doi.org/10.1038/nature12672},
   DOI={10.1038/nature12672},
   number={7479},
   journal={Nature},
   publisher={Springer Science and Business Media LLC},
   author={Trigo, María Díaz and Miller-Jones, James C. A. and Migliari, Simone and Broderick, Jess W. and Tzioumis, Tasso},
   year={2013},
   month=nov, pages={260–262} }

@article{Trueba_2019,
   title={A Comprehensive Chandra Study of the Disk Wind in the Black Hole Candidate 4U 1630-472},
   volume={886},
   ISSN={1538-4357},
   url={http://dx.doi.org/10.3847/1538-4357/ab4f70},
   DOI={10.3847/1538-4357/ab4f70},
   number={2},
   journal={The Astrophysical Journal},
   publisher={American Astronomical Society},
   author={Trueba, N. and Miller, J. M. and Kaastra, J. and Zoghbi, A. and Fabian, A. C. and Kallman, T. and Proga, D. and Raymond, J.},
   year={2019},
   month=nov, pages={104} }

@ARTICLE{Madejski2016,
       author = {{Madejski}, Grzegorz (Greg) and {Sikora}, Marek},
        title = "{Gamma-Ray Observations of Active Galactic Nuclei}",
      journal = {\araa},
         year = 2016,
        month = sep,
       volume = {54},
        pages = {725-760},
          doi = {10.1146/annurev-astro-081913-040044},
       adsurl = {https://ui.adsabs.harvard.edu/abs/2016ARA&A..54..725M}
}

@article{Blandford_2019,
   title={Relativistic Jets from Active Galactic Nuclei},
   volume={57},
   ISSN={1545-4282},
   url={http://dx.doi.org/10.1146/annurev-astro-081817-051948},
   DOI={10.1146/annurev-astro-081817-051948},
   number={1},
   journal={Annual Review of Astronomy and Astrophysics},
   publisher={Annual Reviews},
   author={Blandford, Roger and Meier, David and Readhead, Anthony},
   year={2019},
   month=Aug, pages={467–509} }

@article{Homan_2015,
   title={MOJAVE. XII. ACCELERATION AND COLLIMATION OF BLAZAR JETS ON PARSEC SCALES},
   volume={798},
   ISSN={1538-4357},
   url={http://dx.doi.org/10.1088/0004-637X/798/2/134},
   DOI={10.1088/0004-637x/798/2/134},
   number={2},
   journal={The Astrophysical Journal},
   publisher={American Astronomical Society},
   author={Homan, D. C. and Lister, M. L. and Kovalev, Y. Y. and Pushkarev, A. B. and Savolainen, T. and Kellermann, K. I. and Richards, J. L. and Ros, E.},
   year={2015},
   month=Jan, pages={134} }

@article{Zech_2021,
   title={Electron-proton co-acceleration on relativistic shocks in extreme-TeV blazars},
   volume={654},
   ISSN={1432-0746},
   url={http://dx.doi.org/10.1051/0004-6361/202141062},
   DOI={10.1051/0004-6361/202141062},
   journal={Astronomy &amp; Astrophysics},
   publisher={EDP Sciences},
   author={Zech, Andreas and Lemoine, Martin},
   year={2021},
   month=Oct, pages={A96} }

@article{Nalewajko_2014,
   title={CONSTRAINING THE LOCATION OF GAMMA-RAY FLARES IN LUMINOUS BLAZARS},
   volume={789},
   ISSN={1538-4357},
   url={http://dx.doi.org/10.1088/0004-637X/789/2/161},
   DOI={10.1088/0004-637x/789/2/161},
   number={2},
   journal={The Astrophysical Journal},
   publisher={American Astronomical Society},
   author={Nalewajko, Krzysztof and Begelman, Mitchell C. and Sikora, Marek},
   year={2014},
   month=June, pages={161} }

@article{Marshall_2002,
   title={The High‐Resolution X‐Ray Spectrum of SS 433 Using the<i>Chandra</i>HETGS},
   volume={564},
   ISSN={1538-4357},
   url={http://dx.doi.org/10.1086/324398},
   DOI={10.1086/324398},
   number={2},
   journal={The Astrophysical Journal},
   publisher={American Astronomical Society},
   author={Marshall, Herman L. and Canizares, Claude R. and Schulz, Norbert S.},
   year={2002},
   month=Jan, pages={941–952} }

@ARTICLE{Margon_79,
       author = {{Margon}, B. and {Ford}, H.~C. and {Katz}, J.~I. and {Kwitter}, K.~B. and {Ulrich}, R.~K. and {Stone}, R.~P.~S. and {Klemola}, A.},
        title = "{The bizarre spectrum of SS 433.}",
      journal = {\apjl},
         year = 1979,
        month = may,
       volume = {230},
        pages = {L41-L45},
          doi = {10.1086/182958},
       adsurl = {https://ui.adsabs.harvard.edu/abs/1979ApJ...230L..41M}
}

@ARTICLE{Abell_79,
       author = {{Abell}, G.~O. and {Margon}, B.},
        title = "{A kinematic model for SS433}",
      journal = {\nat},
         year = 1979,
        month = jun,
       volume = {279},
       number = {5715},
        pages = {701-703},
          doi = {10.1038/279701a0},
       adsurl = {https://ui.adsabs.harvard.edu/abs/1979Natur.279..701A}
}

@misc{bruni20263c403candidateneutrinoemittingradio,
      title={3C403: a candidate neutrino-emitting radio galaxy}, 
      author={Gabriele Bruni and Loredana Bassani and Sergio Alves Garre and Manuela Molina and Angela Malizia and Mariateresa Fiocchi and James Rodi and Antoine Kouchner and Alexis Coleiro and Julien Aublin and Giulia Illuminati and Francesca Panessa and Angela Bazzano and Lorenzo Natalucci and Pietro Ubertini},
      year={2026},
      eprint={2506.21403},
      archivePrefix={arXiv},
      primaryClass={astro-ph.HE},
      url={https://arxiv.org/abs/2506.21403}, 
}

@ARTICLE{Aartsen2018multimessenger,
       author = {{IceCube Collaboration} and {Aartsen}, M.~G. and {Ackermann}, M. and {Adams}, J. and {Aguilar}, J.~A. and {Ahlers}, M. and {Ahrens}, M. and {Al Samarai}, I. and {Altmann}, D. and {Andeen}, K. and {Anderson}, T. and {Ansseau}, I. and {Anton}, G. and {Arg{\"u}elles}, C. and {Auffenberg}, J. and {Axani}, S. and {Bagherpour}, H. and {Bai}, X. and {Barron}, J.~P. and {Barwick}, S.~W. and {Baum}, V. and {Bay}, R. and {Beatty}, J.~J. and {Becker Tjus}, J. and {Becker}, K.-H. and {BenZvi}, S. and {Berley}, D. and {Bernardini}, E. and {Besson}, D.~Z. and {Binder}, G. and {Bindig}, D. and {Blaufuss}, E. and {Blot}, S. and {Bohm}, C. and {B{\"o}rner}, M. and {Bos}, F. and {B{\"o}ser}, S. and {Botner}, O. and {Bourbeau}, E. and {Bourbeau}, J. and {Bradascio}, F. and {Braun}, J. and {Brenzke}, M. and {Bretz}, H.-P. and {Bron}, S. and {Brostean-Kaiser}, J. and {Burgman}, A. and {Busse}, R.~S. and {Carver}, T. and {Cheung}, E. and {Chirkin}, D. and {Christov}, A. and {Clark}, K. and {Classen}, L. and {Coenders}, S. and {Collin}, G.~H. and {Conrad}, J.~M. and {Coppin}, P. and {Correa}, P. and {Cowen}, D.~F. and {Cross}, R. and {Dave}, P. and {Day}, M. and {de Andr{\'e}}, J.~P.~A.~M. and {De Clercq}, C. and {DeLaunay}, J.~J. and {Dembinski}, H. and {De Ridder}, S. and {Desiati}, P. and {de Vries}, K.~D. and {de Wasseige}, G. and {de With}, M. and {DeYoung}, T. and {D{\'\i}az-V{\'e}lez}, J.~C. and {di Lorenzo}, V. and {Dujmovic}, H. and {Dumm}, J.~P. and {Dunkman}, M. and {Dvorak}, E. and {Eberhardt}, B. and {Ehrhardt}, T. and {Eichmann}, B. and {Eller}, P. and {Evenson}, P.~A. and {Fahey}, S. and {Fazely}, A.~R. and {Felde}, J. and {Filimonov}, K. and {Finley}, C. and {Flis}, S. and {Franckowiak}, A. and {Friedman}, E. and {Fritz}, A. and {Gaisser}, T.~K. and {Gallagher}, J. and {Gerhardt}, L. and {Ghorbani}, K. and {Glauch}, T. and {Gl{\"u}senkamp}, T. and {Goldschmidt}, A. and {Gonzalez}, J.~G. and {Grant}, D. and {Griffith}, Z. and {Haack}, C. and {Hallgren}, A. and {Halzen}, F. and {Hanson}, K. and {Hebecker}, D. and {Heereman}, D. and {Helbing}, K. and {Hellauer}, R. and {Hickford}, S. and {Hignight}, J. and {Hill}, G.~C. and {Hoffman}, K.~D. and {Hoffmann}, R. and {Hoinka}, T. and {Hokanson-Fasig}, B. and {Hoshina}, K. and {Huang}, F. and {Huber}, M. and {Hultqvist}, K. and {H{\"u}nnefeld}, M. and {Hussain}, R. and {In}, S. and {Iovine}, N. and {Ishihara}, A. and {Jacobi}, E. and {Japaridze}, G.~S. and {Jeong}, M. and {Jero}, K. and {Jones}, B.~J.~P. and {Kalaczynski}, P. and {Kang}, W. and {Kappes}, A. and {Kappesser}, D. and {Karg}, T. and {Karle}, A. and {Katz}, U. and {Kauer}, M. and {Keivani}, A. and {Kelley}, J.~L. and {Kheirandish}, A. and {Kim}, J. and {Kim}, M. and {Kintscher}, T. and {Kiryluk}, J. and {Kittler}, T. and {Klein}, S.~R. and {Koirala}, R. and {Kolanoski}, H. and {K{\"o}pke}, L. and {Kopper}, C. and {Kopper}, S. and {Koschinsky}, J.~P. and {Koskinen}, D.~J. and {Kowalski}, M. and {Krings}, K. and {Kroll}, M. and {Kr{\"u}ckl}, G. and {Kunwar}, S. and {Kurahashi}, N. and {Kuwabara}, T. and {Kyriacou}, A. and {Labare}, M. and {Lanfranchi}, J.~L. and {Larson}, M.~J. and {Lauber}, F. and {Leonard}, K. and {Lesiak-Bzdak}, M. and {Leuermann}, M. and {Liu}, Q.~R. and {Lozano Mariscal}, C.~J. and {Lu}, L. and {L{\"u}nemann}, J. and {Luszczak}, W. and {Madsen}, J. and {Maggi}, G. and {Mahn}, K.~B.~M. and {Mancina}, S. and {Maruyama}, R. and {Mase}, K. and {Maunu}, R. and {Meagher}, K. and {Medici}, M. and {Meier}, M. and {Menne}, T. and {Merino}, G. and {Meures}, T. and {Miarecki}, S. and {Micallef}, J. and {Moment{\'e}}, G. and {Montaruli}, T. and {Moore}, R.~W. and {Morse}, R. and {Moulai}, M. and {Nahnhauer}, R. and {Nakarmi}, P. and {Naumann}, U. and {Neer}, G.},
        title = "{Multimessenger observations of a flaring blazar coincident with high-energy neutrino IceCube-170922A}",
      journal = {Science},
         year = 2018,
        month = jul,
       volume = {361},
       number = {6398},
          eid = {eaat1378},
        pages = {eaat1378},
          doi = {10.1126/science.aat1378},
archivePrefix = {arXiv},
       eprint = {1807.08816},
 primaryClass = {astro-ph.HE},
       adsurl = {https://ui.adsabs.harvard.edu/abs/2018Sci...361.1378I}
}

@article{Murase2020AGNcores,
   title={Hidden Cores of Active Galactic Nuclei as the Origin of Medium-Energy Neutrinos: Critical Tests with the MeV Gamma-Ray Connection},
   volume={125},
   ISSN={1079-7114},
   url={http://dx.doi.org/10.1103/PhysRevLett.125.011101},
   DOI={10.1103/physrevlett.125.011101},
   number={1},
   journal={Physical Review Letters},
   publisher={American Physical Society (APS)},
   author={Murase, Kohta and Kimura, Shigeo S. and Mészáros, Peter},
   year={2020},
   month=June }

@article{Murase2016hidden,
   title={Hidden Cosmic-Ray Accelerators as an Origin of TeV-PeV Cosmic Neutrinos},
   volume={116},
   ISSN={1079-7114},
   url={http://dx.doi.org/10.1103/PhysRevLett.116.071101},
   DOI={10.1103/physrevlett.116.071101},
   number={7},
   journal={Physical Review Letters},
   publisher={American Physical Society (APS)},
   author={Murase, Kohta and Guetta, Dafne and Ahlers, Markus},
   year={2016},
   month=Feb }

@article{Inoue2019coronae,
   title={On the Origin of High-energy Neutrinos from NGC 1068: The Role of Nonthermal Coronal Activity},
   volume={891},
   ISSN={2041-8213},
   url={http://dx.doi.org/10.3847/2041-8213/ab7661},
   DOI={10.3847/2041-8213/ab7661},
   number={2},
   journal={The Astrophysical Journal Letters},
   publisher={American Astronomical Society},
   author={Inoue, Yoshiyuki and Khangulyan, Dmitry and Doi, Akihiro},
   year={2020},
   month=Mar, pages={L33} }

@misc{chow2026baryoncontentmagneticallyarrested,
      title={The baryon content of magnetically arrested black hole disks and jets}, 
      author={Anthony Chow and Lorenzo Sironi and Bart Ripperda and Amir Levinson},
      year={2026},
      eprint={2603.10122},
      archivePrefix={arXiv},
      primaryClass={astro-ph.HE},
      url={https://arxiv.org/abs/2603.10122}, 
}

@article{Romero_2017,
   title={Relativistic Jets in Active Galactic Nuclei and Microquasars},
   volume={207},
   ISSN={1572-9672},
   url={http://dx.doi.org/10.1007/s11214-016-0328-2},
   DOI={10.1007/s11214-016-0328-2},
   number={1-4},
   journal={Space Science Reviews},
   publisher={Springer Science and Business Media LLC},
   author={Romero, Gustavo E. and Boettcher, M. and Markoff, S. and Tavecchio, F.},
   year={2017},
   month=Jan, pages={5–61} }

@article{Connors_2019,
  author  = {Connors, R. M. T. and van Eijnatten, D. and Markoff, S. and
             Ceccobello, C. and Grinberg, V. and Heil, L. and Kantzas, D. and
             Lucchini, M. and Crumley, P.},
  title   = {Combining timing characteristics with physical broad-band spectral
             modelling of black hole X-ray binary GX 339--4},
  journal = {Monthly Notices of the Royal Astronomical Society},
  year    = {2019},
  volume  = {485},
  number  = {3},
  pages   = {3696--3714},
  doi     = {10.1093/mnras/stz604}
}

@article{Lucchini_2022,
  author  = {Lucchini, M. and Ceccobello, C. and Markoff, S. B. and
             Kini, Y. and Chhotray, A. and Connors, R. M. T. and
             Crumley, P. K. and Falcke, H. and Kantzas, D. and Maitra, D.},
  title   = {BHJet: a public multizone, steady state jet + thermal corona
             spectral model},
  journal = {Monthly Notices of the Royal Astronomical Society},
  year    = {2022},
  volume  = {517},
  number  = {4},
  pages   = {5853--5881},
  doi     = {10.1093/mnras/stac2904}
}

@article{Lucchini_2023,
  author  = {Kantzas, D. and Markoff, S. and Lucchini, M. and Ceccobello, C.
             and Chatterjee, K.},
  title   = {Exploring the role of composition and mass loading on the
             properties of hadronic jets},
  journal = {Monthly Notices of the Royal Astronomical Society},
  year    = {2023},
  volume  = {520},
  number  = {4},
  pages   = {6017--6039},
  doi     = {10.1093/mnras/stad521}
}

@article{Cao_2025,
  author  = {Cao, Z. and Aharonian, F. and Bai, Y.-X. and Bao, Y.-W. and
             Bastieri, D. and Bi, X.-J. and Bi, Y.-J. and Bian, W.-Y. and
             Bukevich, A. V. and Cai, C. and Cao, W.-Y. and Cao, Z. and
             Chang, J. and Chang, J.-F. and Chen, A. and Chen, E.-S. and
             Chen, G. and Chen, H.-X. and Chen, L. and Chen, M.-J. and
             Chen, M.-L. and Chen, Q.-H. and Chen, S. and Chen, S.-H. and
             Chen, S.-Z. and Chen, T.-L. and Chen, X.-B. and Chen, X.-J. and
             Chen, Y. and Cheng, N. and Cheng, Y.-D. and Chu, M. C. and
             Cui, M.-Y. and Cui, S.-W. and Cui, X.-H. and Cui, Y.-D. and
             Dai, B.-Z. and Dai, H.-L. and Dai, Z. and Dong, X.-Q. and
             Duan, K.-K. and Fan, J.-H. and Fan, Y.-Z. and Fang, J. and
             Fang, K. and Feng, C.-F. and Feng, H. and Feng, L. and
             Feng, S. and Feng, W. and Feng, Y. and Feng, Y.-L. and
             Gabici, S. and Gao, B. and Gao, C.-D. and Gao, Q. and
             Gao, W. and Gao, W.-K. and Ge, M. and Ge, T.-T. and
             Geng, L. and Giacinti, G. and Gong, G. and Gou, Q. and
             Gu, F.-L. and Gu, J. and Guo, F.-L. and Guo, X.-L. and
             Guo, Y.-Q. and Han, Y.-A. and He, H.-H. and He, H.-N. and
             He, X. and He, Y. and He, Y. and Huang, D.-H. and Huang, S.-C.
             and Huang, X.-Y. and Huang, Y. and Ji, X.-L. and Jia, H.-Y. and
             Jiang, H.-B. and Jiang, K. and Jiang, X.-W. and Jia, Y.-D. and
             Khangulyan, D. and Kuleshov, D. and Li, B.-B. and Li, C. and
             Li, D. and Li, F. and Li, H. and Li, H. and Li, J. and
             Li, K. and Li, L. and Liu, C. and Liu, D. and Liu, H.-D. and
             Liu, J. and Liu, J.-L. and Liu, M.-Y. and Liu, R.-Y. and
             Liu, S.-M. and Liu, W. and Luo, Y. and Lv, H.-K. and
             Ma, B.-Q. and Ma, L.-L. and Mao, J.-R. and Mitthumsiri, W. and
             Mou, G.-B. and Neronov, A. and Ng, K. C. Y. and Ni, M.-Y. and
             Nie, L. and Ou, L.-J. and Pattarakijwanich, P. and Pei, Z.-Y. and
             Qi, J.-C. and Qi, M.-Y. and Qin, J.-J. and Raza, A. and
             Ruffolo, D. and S{\'a}iz, A. and Semikoz, D. and Shao, L. and
             Shchegolev, O. and Shen, Y.-Z. and Shi, Z. and Shu, F.-W. and
             Song, H.-C. and Su, Y. and Sun, H. and Sun, Q. and
             Tang, Q. and Tang, R. and Tam, P. H. T. and Wang, J. and
             Wang, K. and Wang, L. and Wang, W. and Wang, X.-Y. and
             Wang, Y. and Wang, Z.-H. and Wei, D.-M. and Wei, J.-J. and
             Wei, Y.-J. and Wu, X.-F. and Wu, Y.-S. and Xi, S.-Q. and
             Xia, J. and Xing, Y. and Xiong, D.-R. and Xu, D.-L. and
             Xu, R.-X. and Yang, C.-W. and Yang, F.-F. and Yang, M.-J. and
             Yao, Z.-G. and Ye, X.-A. and Yin, L.-Q. and You, X.-H. and
             You, Z.-Y. and Yuan, Q. and Yue, H. and Zeng, H.-D. and
             Zeng, T.-X. and Zhang, B.-T. and Zhang, C. and Zhang, F. and
             Zhang, H.-F. and Zhang, L. and Zhang, P.-F. and Zhang, R. and
             Zhang, X. and Zhang, X.-P. and Zhang, Y. and Zhang, Z.-P. and
             Zhao, J. and Zhao, L. and Zhao, L.-Z. and Zhao, S.-P. and
             Zhou, H. and Zhou, J.-N. and Zhu, B.-Y. and Zhu, C.-G. and
             Zhu, F.-R. and Zhu, H. and Zou, Y.-C. and Zuo, X.},
  title   = {Ultrahigh-energy gamma-ray emission associated with black
             hole--jet systems},
  journal = {National Science Review},
  year    = {2025},
  volume  = {12},
  number  = {12},
  pages   = {nwaf496},
  doi     = {10.1093/nsr/nwaf496}
}

@article{Acciari_2009,
  author  = {Acciari, V. A. and others},
  title   = {A Science With M87: Detection of Very High Energy Gamma Rays
             from the Edge of a Supermassive Black Hole},
  journal = {Science},
  year    = {2009},
  volume  = {325},
  number  = {5939},
  pages   = {444--448},
  doi     = {10.1126/science.1175408},
  eprint  = {0908.0511},
  archivePrefix = {arXiv},
  primaryClass  = {astro-ph.HE}
}

@ARTICLE{najafi_2024,
       author = {{Najafi-Ziyazi}, Mahdi and {Davelaar}, Jordy and {Mizuno}, Yosuke and {Porth}, Oliver},
        title = "{Flares in the Galactic centre - II. Polarization signatures of flares at mm-wavelengths}",
      journal = {\mnras},
         year = 2024,
        month = jul,
       volume = {531},
       number = {4},
        pages = {3961-3972},
          doi = {10.1093/mnras/stae1343},
archivePrefix = {arXiv},
       eprint = {2308.16740},
 primaryClass = {astro-ph.HE},
       adsurl = {https://ui.adsabs.harvard.edu/abs/2024MNRAS.531.3961N}
}

@article{Stawarz_2006,
   title={Dynamics and high-energy emission of the flaring HST-1 knot in the M 87 jet},
   volume={370},
   ISSN={1365-2966},
   url={http://dx.doi.org/10.1111/j.1365-2966.2006.10525.x},
   DOI={10.1111/j.1365-2966.2006.10525.x},
   number={2},
   journal={Monthly Notices of the Royal Astronomical Society},
   publisher={Oxford University Press (OUP)},
   author={Stawarz, L. and Aharonian, F. and Kataoka, J. and Ostrowski, M. and Siemiginowska, A. and Sikora, M.},
   year={2006},
   month=Aug, pages={981–992} }

@ARTICLE{Harris2009,
       author = {{Harris}, D.~E. and {Cheung}, C.~C. and {Stawarz}, {\L}ukasz and {Biretta}, J.~A. and {Perlman}, E.~S.},
        title = "{Variability Timescales in the M87 Jet: Signatures of E $^{2}$ Losses, Discovery of a Quasi Period in HST-1, and the Site of TeV Flaring}",
      journal = {\apj},
         year = 2009,
        month = jul,
       volume = {699},
       number = {1},
        pages = {305-314},
          doi = {10.1088/0004-637X/699/1/305},
archivePrefix = {arXiv},
       eprint = {0904.3925},
 primaryClass = {astro-ph.CO},
       adsurl = {https://ui.adsabs.harvard.edu/abs/2009ApJ...699..305H}
}

@article{Stehle2001,
    author = {Stehle, R. and Spruit, H. C.},
    title = {Stability of accretion discs threaded by a strong magnetic field},
    journal = {Monthly Notices of the Royal Astronomical Society},
    volume = {323},
    number = {3},
    pages = {587-600},
    year = {2001},
    month = {05},
    issn = {0035-8711},
    doi = {10.1046/j.1365-8711.2001.04186.x},
    url = {https://doi.org/10.1046/j.1365-8711.2001.04186.x},
    eprint = {https://academic.oup.com/mnras/article-pdf/323/3/587/3961940/323-3-587.pdf},
}

@misc{medvedev2026electronpositroncascademagnetospheressupermassive,
      title={Electron-positron cascade in magnetospheres of supermassive Kerr black holes and the origin of relativistic AGN jets}, 
      author={Mikhail V. Medvedev and Andrzej A. Zdziarski},
      year={2026},
      eprint={2602.00313},
      archivePrefix={arXiv},
      primaryClass={astro-ph.HE},
      url={https://arxiv.org/abs/2602.00313}, 
}

@ARTICLE{Fiorillo24b,
       author = {{Fiorillo}, Damiano F.~G. and {Comisso}, Luca and {Peretti}, Enrico and {Petropoulou}, Maria and {Sironi}, Lorenzo},
        title = "{A Magnetized Strongly Turbulent Corona as the Source of Neutrinos from NGC 1068}",
      journal = {\apj},
         year = 2024,
        month = oct,
       volume = {974},
       number = {1},
          eid = {75},
        pages = {75},
          doi = {10.3847/1538-4357/ad7021},
archivePrefix = {arXiv},
       eprint = {2407.01678},
 primaryClass = {astro-ph.HE},
       adsurl = {https://ui.adsabs.harvard.edu/abs/2024ApJ...974...75F}
}

@ARTICLE{Fiorillo24a,
       author = {{Fiorillo}, Damiano F.~G. and {Petropoulou}, Maria and {Comisso}, Luca and {Peretti}, Enrico and {Sironi}, Lorenzo},
        title = "{TeV Neutrinos and Hard X-Rays from Relativistic Reconnection in the Corona of NGC 1068}",
      journal = {\apjl},
         year = 2024,
        month = jan,
       volume = {961},
       number = {1},
          eid = {L14},
        pages = {L14},
          doi = {10.3847/2041-8213/ad192b},
archivePrefix = {arXiv},
       eprint = {2310.18254},
 primaryClass = {astro-ph.HE},
       adsurl = {https://ui.adsabs.harvard.edu/abs/2024ApJ...961L..14F}
}

@article{Moscibrodzka2014,
  title={Observational appearance of inefficient accretion flows and jets in {3D GRMHD} simulations: Application to {Sgr A*}},
  author={Mościbrodzka, Monika and Falcke, Heino and Shiokawa, Hotaka and Gammie, Charles F.},
  journal={Astronomy \& Astrophysics},
  volume={570},
  pages={A7},
  year={2014},
  doi={10.1051/0004-6361/201424358}
}



\appendix

\section{Flipped metric derivation for coordinate flip}\label{flipGR}

We begin with the standard spherical coordinates $(t, r, \theta, \phi)$ where $\theta = 0$ corresponds to the positive $z$-axis (jet direction) and $\theta = \pi/2$ corresponds to the equatorial plane (disk midplane). We transform to new coordinates $(t', r', \theta', \phi')$ via a 90-degree rotation about the $y$-axis:
\begin{equation}
\begin{pmatrix}
x' \\
y' \\
z'
\end{pmatrix}
=
\begin{pmatrix}
0 & 0 & 1 \\
0 & 1 & 0 \\
-1 & 0 & 0
\end{pmatrix}
\begin{pmatrix}
x \\
y \\
z
\end{pmatrix}
\end{equation}
In spherical coordinates, this transformation becomes:
\begin{align}
t' &= t \\
r' &= r \\
\sin\theta' \cos\phi' &= \cos\theta \\
\sin\theta' \sin\phi' &= \sin\theta \sin\phi \\
\cos\theta' &= -\sin\theta \cos\phi
\end{align}
From these relations, we derive:
\begin{align}
\theta' &= \arccos(-\sin\theta \cos\phi) \\
\phi' &= \arctan\left(\frac{\sin\phi}{\cot\theta}\right)
\end{align}
In spherical Kerr-Schild coordinates $(t, r, \theta, \phi)$, the metric components are \citep{Kerr1963,Schild1967}:
\begin{align}
g_{tt} &= -1 + \frac{2Mr}{r^2 + a^2\cos^2\theta} \\
g_{tr} &= \frac{2Mr}{r^2 + a^2\cos^2\theta} \\
g_{t\phi} &= -\frac{2Mar\sin^2\theta}{r^2 + a^2\cos^2\theta} \\
g_{rr} &= 1 + \frac{2Mr}{r^2 + a^2\cos^2\theta} \\
g_{r\phi} &= -\frac{2Ma\sin^2\theta}{r^2 + a^2\cos^2\theta} \\
g_{\theta\theta} &= r^2 + a^2\cos^2\theta \\
g_{\phi\phi} &= (r^2 + a^2)\sin^2\theta + \frac{2Ma^2r\sin^4\theta}{r^2 + a^2\cos^2\theta}
\end{align}
For the coordinate transformation, the transformed metric components are given by:
\begin{equation}
g'_{\mu'\nu'} = \frac{\partial x^\mu}{\partial x'^{\mu'}} \frac{\partial x^\nu}{\partial x'^{\nu'}} g_{\mu\nu}
\end{equation}
The Jacobian matrix elements for our 90-degree rotation are:
\begin{align}
\frac{\partial t}{\partial t'} &= 1, \quad \frac{\partial r}{\partial r'} = 1 \\
\frac{\partial \theta}{\partial \theta'} &= \frac{-\cos\theta' \cos\phi'}{\sin\theta}, \quad 
\frac{\partial \theta}{\partial \phi'} = \frac{\sin\theta' \sin\phi'}{\sin\theta} \\
\frac{\partial \phi}{\partial \theta'} &= -\frac{\sin\phi'}{\sin^2\theta'(1 + \cot^2\theta'/\sin^2\phi')}, \\
\frac{\partial \phi}{\partial \phi'} &= \frac{\cos\phi'}{\sin^2\theta' + \cos^2\theta'/\sin^2\phi'}
\end{align}
where the inverse relations are:
\begin{align}
\cos\theta &= \sin\theta' \cos\phi' \\
\sin\theta \sin\phi &= \sin\theta' \sin\phi' \\
\sin\theta \cos\phi &= -\cos\theta'
\end{align}
After applying the rotation transformation, the metric in the new coordinates $(t', r', \theta', \phi')$ becomes:
\begin{align}
g'_{t't'} &= -1 + \frac{2Mr}{r^2 + a^2\sin^2\theta'\cos^2\phi'} \\
g'_{t'r'} &= \frac{2Mr}{r^2 + a^2\sin^2\theta'\cos^2\phi'} \\
g'_{t'\phi'} &= -\frac{2Mar\cos^2\theta'}{r^2 + a^2\sin^2\theta'\cos^2\phi'} \\
g'_{r'r'} &= 1 + \frac{2Mr}{r^2 + a^2\sin^2\theta'\cos^2\phi'} \\
g'_{r'\phi'} &= -\frac{2Ma\cos^2\theta'}{r^2 + a^2\sin^2\theta'\cos^2\phi'} \\
g'_{\theta'\theta'} &= r^2 + a^2\sin^2\theta'\cos^2\phi' \\
g'_{\phi'\phi'} &= (r^2 + a^2)\cos^2\theta' + \frac{2Ma^2r\cos^4\theta'}{r^2 + a^2\sin^2\theta'\cos^2\phi'}
\end{align}
Note that in the rotated coordinates, the jet axis (originally at $\theta = 0$) now corresponds to $\phi' = 0, \pi$, eliminating the coordinate singularity.

\section{Initial tracer distribution} \label{sec:inittracers}
\begin{figure}
    \centering
    \includegraphics[width=\columnwidth]{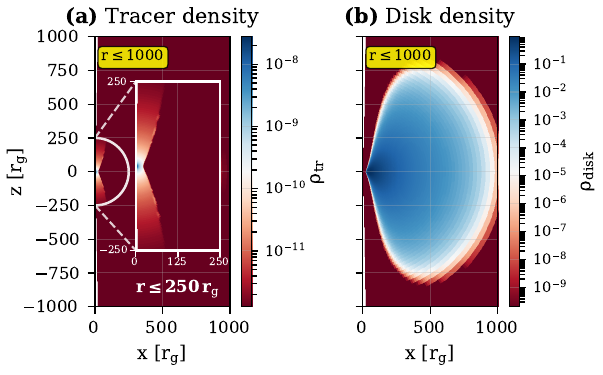}
    \caption{\textbf{Initial density configuration at $t = 0$} showing the decomposition into floor and disk components in the meridional plane ($\phi = 0$). Panel (a) displays the floor density $\rho_{\mathrm{floor}}$ and panel (b) shows the disk density $\rho_{\mathrm{disk}}$, both in logarithmic scale.}
    \label{fig2}
\end{figure}

At $t=0$, we initialise the tracer fractions based on the initial plasma distribution. All plasma within the Fishbone-Moncrief torus \citep{Fishbone1976} is assigned disk tracer fraction $\theta_{\mathrm{disk}} = 1$ and floor tracer fraction $\theta_{\mathrm{floor}} = 0$. Conversely, all plasma outside the torus is initialised with $\theta_{\mathrm{disk}} = 0$ and $\theta_{\mathrm{floor}} = 1$. These fractions satisfy the conservation constraint $\theta_{\mathrm{disk}} + \theta_{\mathrm{floor}} = 1$ at all times and locations. Fig.~\ref{fig2} shows the initial tracer distribution at $t = 0$.

\section{Numerical mass conservation with tracers}

\begin{figure}
    \centering
    \includegraphics[width=\columnwidth]{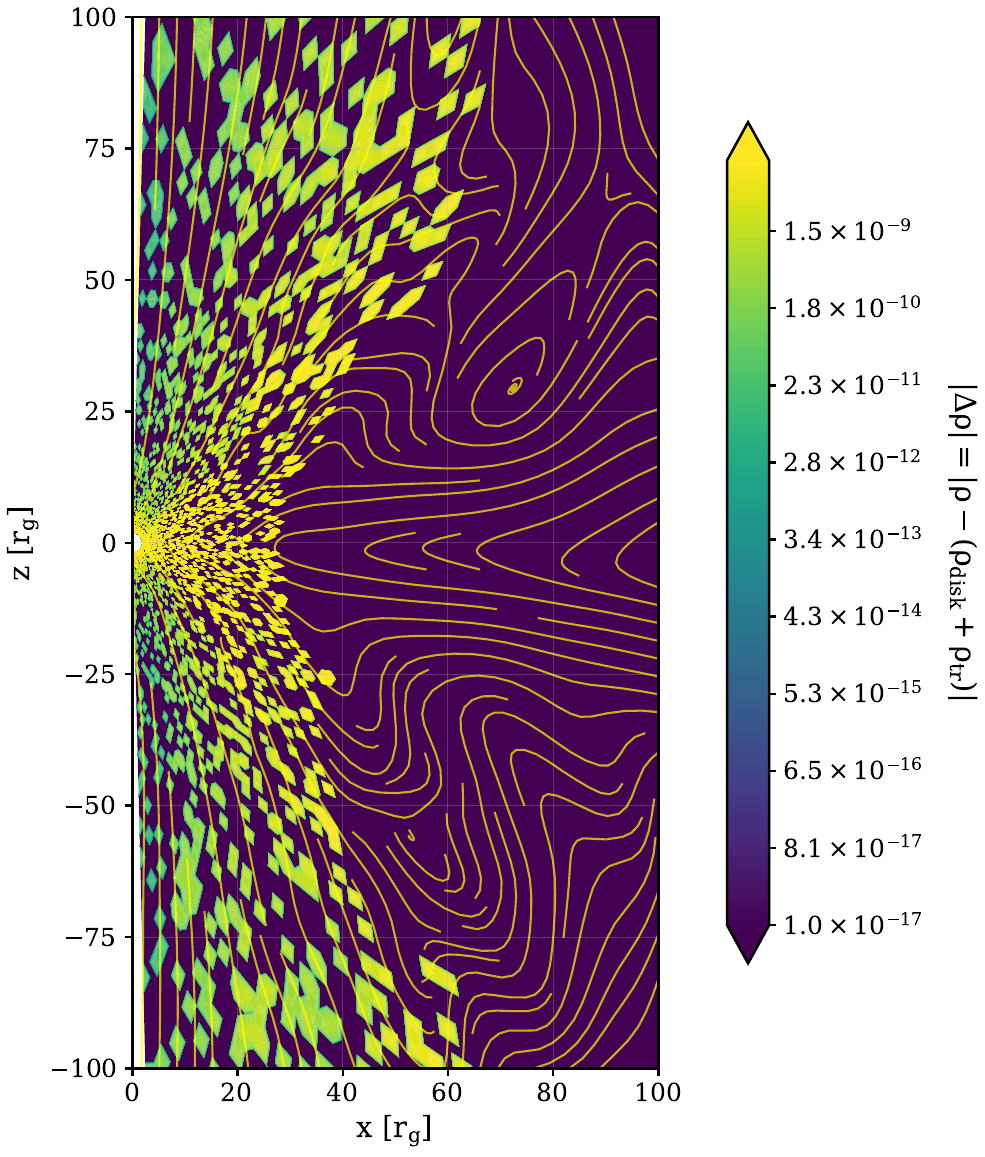}
    \caption{\textbf{Contour plot of the tracer numerical error} $\Delta\rho = \rho - (\rho_\mathrm{disk} + \rho_\mathrm{floor})$ in the meridional plane ($\phi = 0$) of the GRMHD simulation.}
    \label{fig3}
\end{figure}

+We study the magnitude of the numerical error (mass divergence) using logarithmic normalization to capture deviations from mass continuity across multiple orders of magnitude in Fig.~\ref{fig3}. The yellow contour indicates the field lines projected to the x-z slice, showing correspondence of the white noise region with poloidal field dominated regions in the jet. The numerical errors are well within machine precision and arise from the limiting contribution of the grid and the numerical schemes employed to evolve the fluid equations in the strong-field regime.

\section{Adaptive mesh refinement}\label{sec:amrapp}
To efficiently resolve the multi-scale physics governing jet-disk interactions while maintaining computational tractability, we employ a block-based AMR approach implemented within the H-AMR framework \citep{Liska2022}. H-AMR utilises different prolongation and restriction operators for cell-centered \citep{Berger1984, Berger1989} and face-centered variables \citep{BALSARA2001614}. This offers stable dynamic spatial refinement based on the following discussion, coupled with local adaptive time-stepping based on the Courant–Friedrichs–Lewy (CFL) condition \citep{Courant}, offering an efficient test bed to study our physics goals.

This strategy allows us to reach an effective resolution of $6224 \times 2208 \times 2496$ cells in the most refined simulation state. We apply uniform AMR, i.e. all refinement criteria trigger the same maximum refinement level. The refinement criteria are as follows:

\textbf{Jet refinement}: The jet is identified using the magnetization parameter $\sigma = b^2/\rho c^2$ and the Bernoulli parameter $\mathrm{Be} = -hu^t$, where $h$ is the specific enthalpy and $u^t$ the contravariant time component of the four-velocity \citep{Tchekhovskoy2011, McKinney2012}:
\begin{equation}
\mathcal{R}_{\mathrm{jet}} = \begin{cases}
1 & \text{if } \sigma > 1 \text{ and } Be > 1.02 \\
0 & \text{otherwise}
\end{cases}
\end{equation}

This criterion targets the highly magnetised ($\sigma > 1$) and unbound ($Be > 1.02$) regions characteristic of relativistic jets \citep{Moscibrodzka2013, Moscibrodzka2014}.

\textbf{Jet sheath refinement}: The jet sheath is identified by moderate magnetization and outflowing Bernoulli parameter values \citep{Porth2019event, Chatterjee_2019J}:
\begin{equation}
\mathcal{R}_{\mathrm{sheath}} = \begin{cases}
1 & \text{if } 0.1 < \sigma < 1 \text{ and } Be > 1.02 \\
0 & \text{otherwise}
\end{cases}
\end{equation}
This range captures the transition region between the magnetically-dominated jet core and the thermally-dominated disk \citep{Nalewajko2014, Wong2021tracers}.

\textbf{Turbulence and gradient refinement}: To capture magnetic reconnection and turbulent structures, we employ a gradient-based criterion using the magnetic field strength \citep{Ripperda2021, Chernoglazov2021}:
\begin{equation}
\mathcal{R}_{\mathrm{turb}} = \min\left(1, \frac{|\nabla b|}{b/\Delta x}\right)
\end{equation}
where $\Delta x$ is the local cell size ($\Delta x = \min\left(\Delta r, r \Delta\theta, r \sin\theta \Delta\phi\right)$).

\begin{figure*}
\centering
\includegraphics[width=0.95\textwidth]{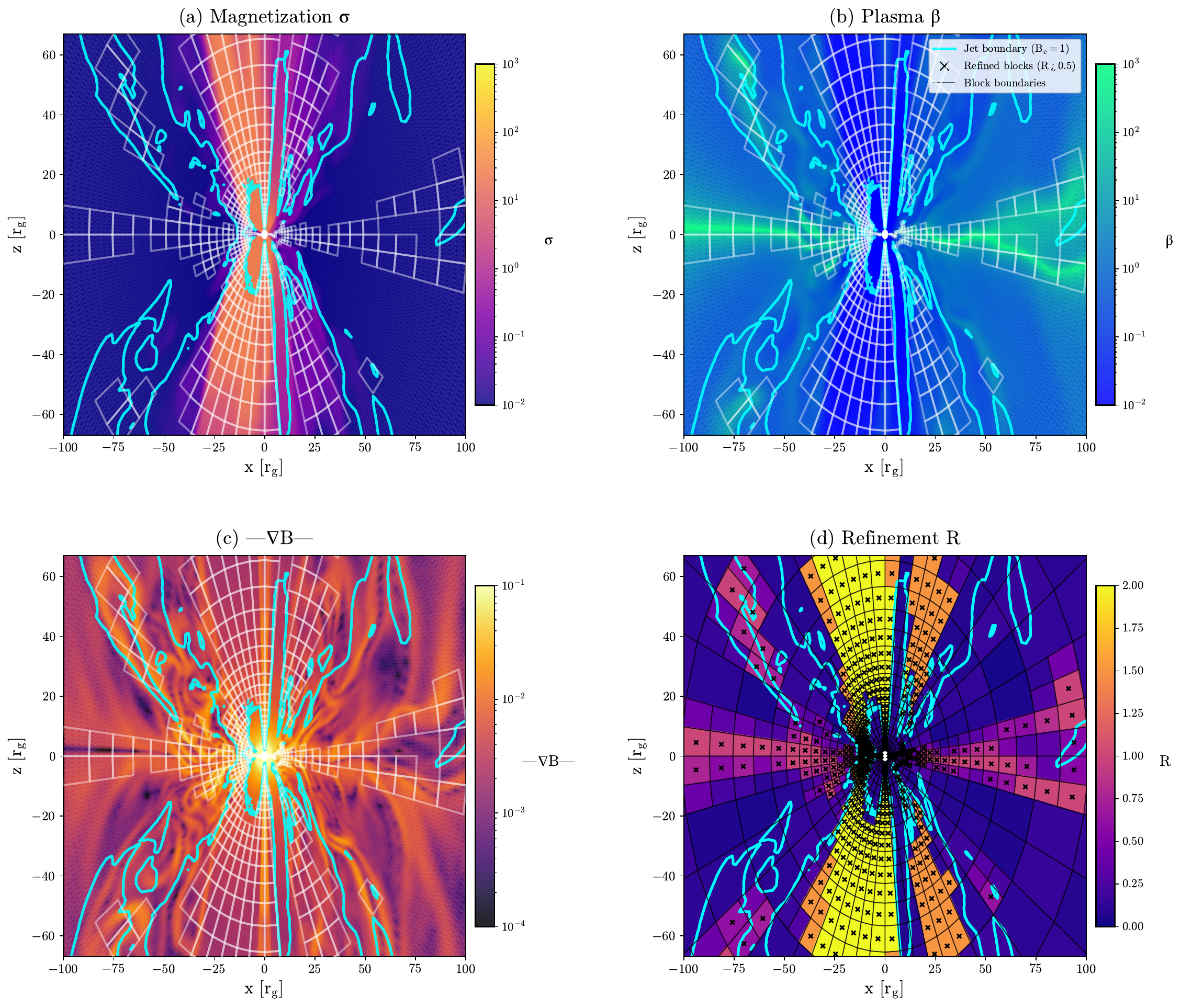}
\caption{
\textbf{Adaptive mesh refinement structure in GRMHD black hole jet-disk simulations.} 
Meridional slices showing AMR block structure and key plasma diagnostics for a high-spin black hole simulation ($a = 0.9375$) at $t \simeq 10^4 t_g$. Cyan contours at $\mathrm{Be} = 1.02$ mark the boundary of the outflow, and black lines mark the block boundaries.
\textbf{(a)} Magnetization parameter $\sigma = b^2/\rho$ revealing magnetically dominated regions in the jet ($\sigma \gg 1$) versus magnetically sub-dominant disk regions ($\sigma \ll 1$).
\textbf{(b)} Plasma $\beta = p_{\rm gas}/p_{\rm mag}$ delineating the jet-disk interface where $\beta \sim 1$, with regions of $\beta \ll 1$ indicating magnetically-dominated plasma in the jet and $\beta \gg 1$ showing pressure-dominated gas in the accretion disk.
\textbf{(c)} Magnetic field gradient magnitude $|\nabla B|$ highlighting sites of strong field structure variation.
\textbf{(d)} Combined refinement metric $\mathcal{R}$ incorporating three criteria: jet core refinement ($\sigma > 1$ and $Be > 1.02$, colored dark green), jet sheath refinement ($0.1 < \sigma < 1$ and $Be > 1.02$, intermediate green), and turbulence-based refinement (sensitive to $|\nabla B|$, yellow/orange).
Black crosses mark the center of each block which meets the refinement criteria in this snapshot ($\mathcal{R} > 0.5$). Blocks the same size but not having these crosses were part of the same parent block that met the criteria before but not post refinement}
\label{fig:amr_structure}
\end{figure*}

The refinement decision is based on a composite criterion that combines several plasma parameters:
\begin{equation}
\mathcal{R} = \max(\mathcal{R}_{\mathrm{jet}}, \mathcal{R}_{\mathrm{sheath}}, \mathcal{R}_{\mathrm{turb}})
\end{equation}
where refinement is triggered when $\mathcal{R} > 0.5$. Fig.~\ref{fig:amr_structure} shows the plasma parameters and the refinement criterion over the grid.

\section{\texorpdfstring{$dM/d\ln\sigma$}{dM/dlnsigma} diagnostics: Calculations and practical considerations} \label{appendixdM}
To quantify the composition of jet and disk material across the simulation domain we employ a diagnostic that measures the mass distribution per logarithmic interval of magnetisation, $\sigma\equiv b^2/\rho$, defined as $dM/d\ln\sigma$.  This diagnostic, computed at multiple radial cuts $r_{\rm cut}\in\{20,\,40,\,60,\,80,\,100\}\,r_g$, provides a scale-dependent view of how mass is distributed between weakly magnetised (disk-like, $\sigma<\sigma_{\rm sh}$), moderately magnetised (sheath-like, $\sigma_{\rm sh}<\sigma<1$), and strongly magnetised (jet-like, $\sigma>1$) plasma states, where the disk--sheath boundary is defined as $\sigma_{\rm sh}=10^{-1.8}\approx0.016$ (the mean sigma measured along the magnetic field streamlines along the sheath).  Following the volume-element integration scheme of \citet{McKinney2006} and
\citet{Porth2019event}, we compute
\begin{equation}
\frac{dM}{d\ln\sigma}
  = \frac{1}{d\ln\sigma}
    \int_{r<r_{\rm cut}}
    \rho\,\theta_{\rm tr}\,\sqrt{-g}\;d\phi\,dr\,d\theta,
\label{eq:dMdlnsigma}
\end{equation}
where $\theta_{\rm tr}$ is the tracer fraction for either disk-origin
material ($\theta_{\rm disk}$, tracer~1) or floor/jet-origin material
($\theta_{\rm floor}$, tracer~2), and $\sqrt{-g}$ is the metric
determinant serving as the covariant volume element.  In practice, the
integral over $\phi$ is replaced by a discrete sum over the $n_\phi$
azimuthal cells with uniform spacing $d\phi=2\pi/n_\phi$, while the
integrals over $r$ and $\theta$ are replaced by a sum over all grid cells
within $r<r_{\rm cut}$.  The per-cell mass contribution is therefore
\begin{equation}
dm = \rho\,\theta_{\rm tr}\,\sqrt{-g}\;d\phi\,dr\,d\theta,
\end{equation}
where $\sqrt{-g}$ is evaluated at each $(r,\theta)$ grid point and
broadcast to all $n_\phi$ azimuthal cells.

We bin $\sigma$ logarithmically over $n_{\rm bins}=50$ bins spanning the
range $[\sigma_{\rm min},\sigma_{\rm max}]$, where $\sigma_{\rm min}$ is
the minimum positive $\sigma$ in the domain (floored at $10^{-20}$ to
avoid numerical zeros) and $\sigma_{\rm max}$ is the maximum.  The bin
edges are
\begin{equation}
\sigma_i = \sigma_{\rm min}
  \left(\frac{\sigma_{\rm max}}{\sigma_{\rm min}}\right)^{i/(n_{\rm bins}-1)},
  \quad i=0,\ldots,n_{\rm bins}-1,
\end{equation}
and the bin centre is taken as the geometric mean
$\bar\sigma_i=\sqrt{\sigma_i\,\sigma_{i+1}}$.  Within each bin the
distribution is approximated as
\begin{equation}
\left.\frac{dM}{d\ln\sigma}\right|_{\bar\sigma_i}
  = \frac{\displaystyle\sum_{j\in\mathrm{bin}\,i} dm_j}
         {\ln(\sigma_{i+1}/\sigma_i)}.
\end{equation}
This ensures adequate sampling across the $\sim5$ orders of magnitude
dynamic range from $\sigma\sim10^{-4}$ (disk dominated) to $\sigma\sim25$
(the maximum $\sigma$ in the simulation, set by the density floor and reached in the jet core).

\subsubsection{Mass integrals and composition diagnostics}

By integrating the $dM/d\ln\sigma$ curves over $\ln\sigma$ we obtain the
total tracer-weighted masses
\begin{equation}
M_{\rm disk} = \int d\ln\sigma\;\frac{dM_{\rm disk}}{d\ln\sigma},\qquad
M_{\rm jet}  = \int d\ln\sigma\;\frac{dM_{\rm jet}}{d\ln\sigma}.
\label{eq:Mtot}
\end{equation}
The integration employs Simpson's rule when the number of non-empty bins
exceeds two, with a trapezoidal fallback for under-sampled curves and a
single-bin approximation ($M\approx (dM/d\ln\sigma)\cdot\Delta\ln\sigma$)
otherwise.

To characterise the composition in specific magnetisation bands we
additionally integrate over two restricted $\sigma$ ranges.  In the
high-$\sigma$ band $\sigma>\sigma_{\rm sh}$ we compute
\begin{equation}
M_{\sigma>\sigma_{\rm sh}}^{\rm disk}
  =\int_{\sigma>\sigma_{\rm sh}}d\ln\sigma\;
   \frac{dM_{\rm disk}}{d\ln\sigma}, 
M_{\sigma>\sigma_{\rm sh}}^{\rm jet}
  =\int_{\sigma>\sigma_{\rm sh}}d\ln\sigma\;
   \frac{dM_{\rm jet}}{d\ln\sigma},
\end{equation}
and in the low-$\sigma$ band $\sigma<1$ we compute $M_{\sigma<1}^{\rm
disk}$ and $M_{\sigma<1}^{\rm jet}$ via the same procedure.  These
quantities are used later when we discuss the relative contributions of
disk and jet/floor material in the disk, sheath, and jet bands.

\subsubsection{Jet-shell mass budget}

To isolate the composition of the jet ($\sigma>1$) we
additionally compute the disk and jet/floor masses directly inside the
magnetisation-defined jet shell, defined as the set of cells satisfying
$\sigma>\sigma_{\rm jet}=1$ and $r_{\rm min}<r<r_{\rm max}$ with
$r_{\rm min}=20\,r_g$ and $r_{\rm max}=80\,r_g$.  For each such cell the
tracer-weighted mass is
\begin{equation}
dM_{\rm tr}^{\rm shell}
  = \rho\,\theta_{\rm tr}\,\sqrt{-g}\,d\phi\,dr\,d\theta,
\end{equation}
and the total masses in the jet shell are
\begin{equation}
M_{\rm disk}^{\rm shell}
  = \sum_{\rm shell\;cells} dM_{\rm disk}^{\rm shell},\qquad
M_{\rm jet}^{\rm shell}
  = \sum_{\rm shell\;cells} dM_{\rm jet}^{\rm shell}.
\end{equation}
Unlike the $dM/d\ln\sigma$ integrals, which are computed via logarithmic
binning, the jet-shell masses are computed by direct cell summation
without binning, making them sensitive to the instantaneous composition of the jet.

\subsubsection{Jet-boundary mass flux}
\label{sec:funnel_flux}

To track the time-resolved exchange of disk and jet/floor material across
the jet boundary, we compute the tracer-weighted mass flux through the
$\sigma=\sigma_{\rm jet}=1$ surface.  We define the jet boundary at each
radius $r$ as the set of cells satisfying $\sigma>\sigma_{\rm jet}$ and
having at least one meridional ($\theta$) neighbour with
$\sigma\le\sigma_{\rm jet}$.  This identifies the outermost layer of
jet cells, adjacent to the lower-magnetisation sheath and disk.

The normal to the jet boundary is approximated by the gradient of $\sigma$
in the $(r,\theta)$ plane,
\begin{equation}
\hat{n}_i \propto \partial_i\sigma,\qquad i\in\{r,\theta\},
\end{equation}
computed via second-order centred finite differences,
\begin{align}
\partial_r\sigma\big|_{i_r}
  &\approx \frac{\sigma_{i_r+1,\,i_\theta}-\sigma_{i_r-1,\,i_\theta}}
               {r_{i_r+1}-r_{i_r-1}},
\label{eq:grad_r}\\
\partial_\theta\sigma\big|_{i_\theta}
  &\approx \frac{\sigma_{i_r,\,i_\theta+1}-\sigma_{i_r,\,i_\theta-1}}
               {\theta_{i_\theta+1}-\theta_{i_\theta-1}}.
\label{eq:grad_th}
\end{align}
The normal component of the four-velocity at each wall cell is then
\begin{equation}
u_n = u^r\,\partial_r\sigma + u^\theta\,\partial_\theta\sigma,
\label{eq:un}
\end{equation}
where $u^r$ and $u^\theta$ are the contravariant radial and meridional
components of the fluid four-velocity.  The tracer-weighted mass flux
through the area element $dA = \sqrt{-g}\,d\phi\,dr$ is
\begin{equation}
d\dot{M}_{\rm tr}
  = \rho\,\theta_{\rm tr}\,|u_n|\,\sqrt{-g}\,d\phi\,dr,
\label{eq:dMdot}
\end{equation}
where $|u_n|$ gives the magnitude of the normal flux.  The direction
(into or out of the jet) of mass transport across the jet boundary is determined from the velocity component normal to the boundary, rather than from $u^\theta$ alone. Since the jet is defined by $\sigma>\sigma_{\rm jet}$, the gradient $\nabla\sigma$ points toward increasing magnetization and therefore toward the interior of the jet. We define the normal component of the velocity as
\begin{equation}
u_n =
u^r \frac{\partial \sigma}{\partial r}
+
u^\theta \frac{\partial \sigma}{\partial \theta}.
\end{equation}
The magnitude of the normal velocity is therefore $|u_n|$. The sign of $u_n$ determines the direction of transport across the jet boundary: $u_n>0$ corresponds to flow into the jet, while $u_n<0$ corresponds to flow out of the jet. This definition accounts for both radial and polar motion and does not assume that the sign of $u^\theta$ alone determines whether gas enters or leaves the funnel.

The total inward and outward fluxes for each tracer are consequently
\begin{equation}
\dot{M}_{\rm tr}^{\rm in}
  = \sum_{\rm wall,\;into} d\dot{M}_{\rm tr},\qquad
\dot{M}_{\rm tr}^{\rm out}
  = \sum_{\rm wall,\;out} d\dot{M}_{\rm tr},
\label{eq:Mdot_inout}
\end{equation}
, integrated over the radial band $r_{\rm min}<r<r_{\rm max}$ and summed
over both hemispheres.  We report these quantities separately for the disk tracer ($\theta_{\rm disk}$) and the jet/floor tracer ($\theta_{\rm
floor}$), to diagnose any inflow-outflow asymmetry during the eruption.


\bsp	
\label{lastpage}
\end{document}